%% file: paper.tex

\input My.sty
\elevenpoint  

\input title
\input 1a

\input 2a

\input 3a

\input 4a

\input 5a

\input 6a

\input 7a

\input 8a

\input 9a

\input 10a

\input 11a

\input 12a

\raggedright

\subhead{References}
\input refsa

\bye

%% file: title.tex
\def\title#1{\centerline{\bf #1}}
\def\author#1{\centerline{#1}}
\def\address#1{\centerline{#1}}
\def\thanks#1{\f{#1}}

\bigskip

\title{A New Graph Grammar Formalism for Robust Syntactic Pattern Recognition}

\bigskip

\author{Peter~Fletcher}

\address{School of Computer Science and Mathematics,}
\address{Keele University,}
\address{Keele,}
\address{Staffordshire, ST5 2BG, U.K.}
\address{E-mail: p.fletcher@keele.ac.uk}

\bigskip

\thanks{Please see also the accompanying paper
\i{Mathematical Theory of Networks for Syntactic Pattern Recognition}.}

\bigskip

\thanks{This research did not receive any specific grant from funding agencies 
in the public, commercial, or not-for-profit sectors.}

\bigskip

\thanks{There are no conflicts of interest.}

\bigskip

\subhead{Abstract}
I introduce a formalism for representing the syntax of recursively structured 
graph-like patterns. It does not use production rules, like a conventional 
graph grammar, but represents the syntactic structure in a more direct and 
declarative way. The grammar and the pattern are both represented as networks, 
and parsing is seen as the construction of a homomorphism from the pattern to 
the grammar.
The grammars can represent iterative, hierarchical and nested recursive 
structure in more than one dimension.

This supports a highly parallel style of parsing, in which all aspects of pattern
recognition (feature detection, segmentation, parsing, filling in missing
symbols, top-down and bottom-up inference) are integrated into a single
process, to exploit the synergy between them.

The emphasis of this paper is on underlying theoretical issues, but I also give
some example runs to illustrate the error-tolerant parsing of complex
recursively structured patterns of 50--1000 symbols, involving variability in
geometric relationships, blurry and indistinct symbols, overlapping symbols,
cluttered images, and erased patches.

\bigskip

\subhead{Keywords}
graph grammar, syntactic pattern recognition, graph parsing, 
error-tolerant parsing

\newpage

%% file: 1a.tex
\subhead{1. Introduction}
The ability to represent and process recursive symbol structures is often seen
as fundamental to cognition, and its characteristic features of productivity,
systematicity and compositionality have been especially celebrated by the
\ic{classical} school of cognitive science [24,51]; and yet
such symbolic processing is often \i{brittle}, i.e., unable to cope gracefully
with noisy input, random variation, contradictory or incomplete data, and
unexpected input. In contrast, nonsymbolic or subsymbolic methods
[96], such as statistical pattern recognition and neural
networks, are much more robust but have great difficulty representing
recursively structured data. In the 1980s and 1990s this contrast was drawn in
adversarial terms [51,53], but since then much effort has
been devoted to reconciling the two by implementing symbolic processing in
subsymbolic architectures [61,87,20].

I approach the problem from the other direction: the aim of my work is to make
symbolic processing more robust. I choose as a suitable problem domain
\i{geometric pattern recognition}, the problem of recognising structured spatial
configurations. This is not computer vision: we are not concerned with
perspective, binocular vision, motion, occlusion, colour, texture, shadow or
illumination. We are concerned with issues that are common to pattern
recognition in general, not any particular sense modality. The raw data from
which the pattern is recognised I shall call the \i{image}, for want of a better
word, although I intend it to be thought of as a spatial domain, not necessarily
a visual one; where necessary I shall assume that it is two-dimensional.

Formal grammars will be used to specify the recursive structure of a class of
patterns. By a \i{grammar} I mean any formal system that specifies how patterns
can be composed out of parts, allowing for iteration and nested recursive
structure; it may be expressed in the form of a set of production rules, a
transition diagram, an accepting automaton, or by any other method. Since we are
not limited to sequential patterns we need graph grammars rather than string
grammars. A \i{graph grammar} is a formalism for representing the syntactic
structure of systems of symbols connected by binary relations. The conventional
type of graph grammar generates patterns by repeatedly applying production
rules; I shall review these grammars and their difficulties in \S2 and shall
instead introduce a different kind of formalism that rejects production rules
and represents syntactic structure directly and declaratively using
\i{networks}, with parsing seen as a homomorphism between networks.

We begin by making a basic distinction between tokens and types. A symbol
\i{token} is an occurrence of a symbol somewhere in the image; every symbol
token has a \i{type}. E.g., a page of text may consist of hundreds of letter
\i{tokens}, each with its own position, size, orientation, and relations to
nearby letter tokens, but there is only ever a fixed set of 52 letter \i{types},
which have no position. The system of symbol tokens present in an image (with
all their relationships) is called a \i{pattern}; the \i{grammar} is a system of
symbol types, specifying the permitted structural and geometric relationships in
patterns.

The grammars should be capable of representing complex patterns (of up to 1000
symbol tokens) related by iteration (not just in one dimension but
two-dimensional iterations such as square grids), hierarchical structure, and
nested recursive structure. Most syntactic pattern recognition research uses
extremely limited grammars: e.g., a fixed compositional hierarchy (possibly
allowing for absent parts) [68]; or AND-OR trees [81]; or
special-purpose representations designed for a particular application, such as
recognising the facades of buildings [93]; limited types of
easy-to-parse graph grammars are used by
[47,95]; specialised grammars are used for
music notation [42,83] and for mathematical formulae
[66,64,15,22,104].

The recognition algorithm must be able to tolerate all kinds of error and
variability:

\blob variability in the geometric relationships between neighbouring symbol
tokens, or between parts and wholes;

\blob blurry and indistinct symbol tokens, and cluttered images where the
lowest-level symbol tokens cannot be detected purely by bottom-up processing;

\blob overlapping symbol tokens (where the image cannot be segmented by
tracing boundaries or finding connected components);

\blob missing patches in the image, small blank regions where the
recognition algorithm must fill in the missing symbol tokens or parts of symbol
tokens using top-down information.

\medskip\f These requirements are extremely challenging for any existing
pattern recognition algorithm. In my view the brittleness of syntactic pattern
recognition is caused by the imposition of an unwarranted sequential order on
what is conceptually a parallel process. In the first place, many graph parsing
algorithms impose a sequential order on the graph and then traverse it in that
order (see [48, \S2] for a survey); a single error at an early
stage may cause the whole parse to go astray. My algorithm will examine all
parts of the image in parallel. Secondly, recognition should not involve taking
early irrevocable decisions: the algorithm should generate alternative
interpretations of (parts of) the image, which are then extended and combined in
parallel, with all but one eliminated by the end. Thirdly, most recognition
algorithms subdivide the task into a fixed sequence of phases. Recognising a
pattern requires decisions on how the image is segmented into symbol tokens, the
type of each symbol token present, which parts of the image are pixel noise
(i.e., not part of any symbol token), the structural relationships between the
symbol tokens, and the position, orientation, size and degree of stretching and
shearing of each symbol token. It is widely recognised that these tasks are
inter-dependent and that the synergy between them should be exploited. Tu \i{et
al} [98] point out the need for segmentation, object detection and
recognition, top-down parsing, and bottom-up parsing to work together; Tombre
[97] argues that segmentation, recognition and higher-level semantic
analysis need to interact; Casey \& Lecolinet [21], in a
survey of character segmentation, point out the inter-dependence of decisions on
segmentation, shape similarity and contextual acceptability; see also
[82,59]. Yet, despite all this, algorithms still
separate these tasks and perform them in sequence. Many image recognition
algorithms begin with a special segmentation method to identify the foreground
object, find its bounding box, and transform it to a standard position, before
the main parsing (identification of parts) begins, and there is commonly
postprocessing after parsing [78,82]. For highly
specialised applications, where the structure present is highly predictable,
such as recognition of music scores, a top-down approach may be successful
[83]. Other algorithms identify the lowest-level features first,
constructing a graph from them, and then parse the graph; for example, Celik \&
Yanikoglu's mathematical formula recognition system [22] achieves
very low recognition accuracy because any failure to recognise symbol tokens at
the OCR stage cannot be corrected at the parsing stage. As Flasi\'nski \etal\
emphasise [46], the conversion of the numeric input
representation into a symbolic representation in terms of pattern primitives can
lose essential information. Better results can be achieved if the recognition of
pattern primitives can be guided by grammatical information from the
higher-level stages [104]. Few algorithms can cope with overlapping
objects, because bottom-up approaches to segmentation are inapplicable (a
partial exception is [81]).

In my algorithm there is no division into phases; all aspects of recognition are
integrated. During the recognition process the pattern is built up
incrementally; symbol tokens are created, linked to other symbol tokens, merged,
split and pruned; many rival pattern-fragments are in play simultaneously,
competing to attract parts and to grow at one another's expense. Bottom-up and
top-down influences apply simultaneously; indistinct, missing or overlapping
symbol tokens can be \ic{hallucinated} using grammatical context.

The symbol tokens are arranged in spatial configurations, so some mathematical
formalism is needed for representing geometric relationships and how they are
permitted to vary in a pattern. In \S5 I shall review the range of
representational methods used and propose a more general concept of a
\i{fleximap}, an elastic affine transformation relating one symbol token to
another. The recognition process should be invariant under affine
transformations, to accommodate different points of view on the same pattern;
most existing invariant pattern recognition techniques assume that the image has
already been segmented and apply only to a single object (with little or no
occlusion).

Finally, the recognition process must also cope with symmetry; this is a
neglected topic in the literature. A line segment has two-fold symmetry; the
algorithm must understand that it may occur either way round as part of a
complex symbol token and that the two ways are equivalent. A hexagonal grid has
six-fold symmetry; the algorithm must recognise the six symmetry variants of a
pattern as equivalent, not as rival interpretations. The symmetries of each
symbol type must therefore be represented explicitly in the grammar.

This paper builds on my previous work [48], where regular graph
grammars describing two-dimensional geometric patterns were parsed and learned
from positive examples. Here I am using more powerful grammars, using the full
geometry of the plane rather than a discrete grid, and adding error-tolerance.
However, this paper considers only recognition of patterns, not learning of
grammars.

The emphasis of this paper is on the new grammatical formalism. For reasons of
space I give only a summary of the mathematical theory; see [49]
and [50] for full details and proofs.

Section 2 reviews graph grammars in general and my reasons for being
dissatisfied with the conventional approach based on production rules.

My new type of graph grammar is introduced informally in \S3 and formally in
\S4, with the spatial aspects covered in \S5. The problem of pattern recognition
is specified in \S6. The algorithm for recognition is developed in \S\S7--9. The
core parts of the theory are provably correct, but peripheral parts rely on
heuristic arguments (particularly in \S7.4).

Three example grammars are constructed in \S10, illustrating various types of
iteration and recursion. Example runs with these three grammars (combined into a
single grammar) are described in \S11. These runs are intended purely as proof
of concept and are not experiments: the emphasis of this paper is on
investigating fundamental qualitative issues underlying all pattern recognition,
in particular the difficulty of combining recursive representational
capabilities with error-tolerance; I make no claims of applicability.
Conclusions are drawn in \S12.

%% file: 2a.tex
\subhead{2. Graph grammars}
There is a well-established theory of grammars for representing \ic{strings}
(sequences of symbols). String grammars contain \i{production rules} of the form
$L \to R$, where $L$ and $R$ are strings, called the \i{left-hand side} and the
\i{right-hand side} of the rule. Define a relation $\derives$ between strings as
follows: for any strings $S$ and $T$, $S \derives T$ iff $S$ can be turned into $T$ by
replacing a substring matching $L$ in $S$ by $R$; in that case we say that the
rule $L \to R$ has been \i{applied} to $S$ to yield $T$. To \i{parse} a string
$S$ is to find a sequence of production rule applications $S_0 \derives S_1
\derives S_2 \derives \cdots \derives S$, which is called a \i{derivation} of
$S$ (where $S_0$ is the \i{start symbol}). The \i{language generated by} the
grammar is the set of all strings $S$ that have derivations.

To generalise all this to non-sequential patterns we replace the strings $L$
and $R$ by graphs, in which the nodes, and perhaps also the edges, bear labels.
This gives us a \i{graph grammar}, which generates a language of graphs. (The
node labels are symbol types and a labelled node in the generated graph is a
symbol token.) However, several complications and perplexities arise in doing
so:

\item{(i)} the complexity of the concept of applying a production rule,
i.e, the $\derives$ relation;

\item{(ii)} the lack of a well-defined Chomsky hierarchy for graph grammars;

\item{(iii)} the difficulty of combining good expressive power with the
existence of parse trees;

\item{(iv)} the imperspicuity of graph grammars;

\item{(v)} the intractability of parsing;

\item{(vi)} the difficulty of coping with noise, i.e., errors in the graph to
be parsed.

\medbreak\f
I shall take these issues in turn.

\subsubhead{2.1 The complexity of production rule application}
The concept of \ic{applying} a production rule to a graph is not as obvious as
in the string case. To apply the rule $L \to R$ to a graph $G$ we must identify
an isomorphic copy of $L$ in $G$ and replace it with an isomorphic copy of $R$,
giving a new graph $H$. This may be expressed mathematically by graph
endomorphisms (injective homomorphisms) $g:L \to G$ and $h:R \to H$. Thus $g(L)$
is removed from $G$ and replaced by $h(R)$, giving $H$. But how exactly is
$h(R)$ to be joined to $G\setminus g(L)$? This is called the \i{embedding
problem}. There are three approaches in the literature. (I shall consider only
\i{sequential} grammars, not \i{parallel} grammars, here.)

\medbreak

\i{First approach: the context subgraph}. In this method there is a subgraph
of $L$ that is shared with $R$, or in bijection with a subgraph of $R$; this
subgraph is called $K$, the \i{context subgraph}. This can be expressed by
endomorphisms $l:K\to L$ and $r:K\to R$. Often the bijection between the
subgraphs $l(K)$ and $r(K)$ is expressed by marking the nodes of $l(K)$ and
$r(K)$ with unique numeric labels. Then, when a rule is applied, $g(L)$ is
removed from $G$ and $h(R)$ is added to $G\setminus g(L)$ in such a way that
$h(r(K))$ occupies the same position as its isomorphic copy $g(l(K))$ did
before.

This can be expressed neatly in category-theoretic terms as a pair of push-outs
[35,36]:
$$
\def\mapleft#1{\smash{\mathop{\longleftarrow}\limits^{#1}}}
\def\mapright#1{\smash{\mathop{\longrightarrow}\limits^{#1}}}
\def\mapdown#1{\big\downarrow\rlap{$\vcenter{\hbox{$\scriptstyle#1$}}$}}
\matrix{
L & \mapleft l & K & \mapright r & R \cr
\mapdown g & & \mapdown d & & \mapdown h \cr
G & \mapleft{c_1} & D & \mapright{c_2} & H \cr
}$$
where all the arrows are graph endomorphisms. (Think of $D$ as the part of $G$
that is unchanged by the rule application (either because it is not involved in
the operation or because it is the context part $g(l(K))$), and think of
$c_1,c_2$ as inclusion endomorphisms.)

Given a rule $L \to R$ and a graph $G$, the rule can only be applied to $G$ if a
$D$ exists that forms a push-out with $K$, $L$ and $G$. This will be the case
iff there are no edges in $G\setminus g(L)$ incident to nodes in $g(L\setminus
l(K))$: this is called the \i{dangling-edge condition}.

In the most general formulation $l,r,g,d,h,c_1,c_2$ are not required to be
injective, i.e., they are simply graph homomorphisms. In this case, the
\i{identification condition} must also hold for the rule to be applicable to
$G$: this states that whenever two nodes or edges $x,x'$ in $L$ have
$g(x)=g(x')$ then $x$ and $x'$ are in $l(K)$.

(I have been assuming so far that the rule $L \to R$ is applied from left to
right, i.e., that we interpret $G \derives H$ as meaning that $G$ is transformed
into $H$; but it is also possible to apply the rule from right to left, i.e., to
regard it as transforming $H$ into $G$; corresponding dangling-edge and
identification conditions must be imposed to guarantee the existence of $D$
forming a push-out with $R$, $K$ and $H$.)

Examples of this approach are hyperedge replacement grammars
[56,33]; positional grammars [25], which
use plex-like structures with attaching points; Lichtblau's grammar
[69]; reserved graph grammars [106] and variants of
them such as spatial grammars [62].

Ferrucci \etal\ [44] consider this type of grammar restrictive,
because the context subgraph $K$ is fixed. The dangling-edge and identification
conditions are also restrictive. There are some slight extensions that allow
some variation in $K$. Layered graph grammars [84] and breeze graph
grammars [67] allow the nodes in $K$ to have \ic{wildcard} labels, each
of which stands for one of a definite set of possible labels. In edge-based
graph grammars [90,91], an edge in $K$ may bear an asterisk,
meaning that it represents any number of edges. In Blosetin's grammar
[15] $K$ may contain \ic{set nodes}, each of which matches
against as many nodes in $G$ as possible. Other variants, which involve
something similar to a context subgraph, are
[101,102,25,88].

\medbreak

\i{Second approach: connection relations}. In this approach we still have the
endomorphisms $g:L\to G$ and $h:R\to H$, but no context subgraph $K$. When
$g(L)$ is removed from $G$, edges of $G\setminus g(L)$ that are incident to
nodes of $g(L)$ must also be removed and replaced by new edges incident to nodes
of $h(R)$. How this is done is specified by a \i{connection relation}.

The advantage of this method is that it gives more powerful grammars than the
\ic{context subgraph} method -- at least, this appears to be true in general and
has been proved for particular classes of grammar [28].

A disadvantage is that descriptions of how the connection relation is used to
connect the new edges are complicated [60,86] and often
informal and incomplete [102,66]. This
does not lend itself to a neat mathematical theory.

A second disadvantage is that application of a production rule may not be
reversible, or not reversible in a unique way. In [4] the rules
can be applied left-to-right or right-to-left, but the two directions are not
inverses of each other. This is a problem for bottom-up parsing: the definition
of a language involves left-to-right application, but bottom-up parsing involves
right-to-left application.

The commonest type of grammar of this kind is the \i{node-replacement grammar}
[41], in which the left-hand side $L$ of each production rule
is just a single node, $n$. Most commonly the nodes bear labels and the labels
control the connection of new edges: this is called a \i{node-label controlled}
(NLC) grammar [40]. The edges may bear labels too and may be
directed. Rule application works as follows: $g(n)$ is replaced by $h(R)$, and
if there is an edge $e$ in $G$ between $g(n)$ and a neighbour node $u$, it is
replaced by new edges between $h(v)$ and $u$, for some nodes $v$ in $R$,
depending on the labels of $u$ and $v$ (and $e$'s label and direction if they
exist), as directed by the \i{connection relation} that is provided with the
grammar.

A slightly more general type of grammar is the \i{neighbourhood-controlled
embedding} (NCE) grammar [41]. Here, the connection of new
edges depends on the identity of the node $v$ in $R$, not just its label.
Relation grammars [43,44] are essentially equivalent to
NCE grammars, presented in a different format. NCE grammars generate the same
class of languages as NLC grammars, but that is not true when they are combined
with some other extensions [40].

Alternatives to node-replacement grammars are \i{edge-label controlled grammars}
(where the left-hand side $L$ of a rule is one edge and its incident nodes)
[75] and Adachi \etal's context-sensitive NCE grammars, where $L$ can
be a graph of any size [4,5,3].

The most general type of production rule of this kind is Nagl's
\i{set-theoretic approach} [79], in which new edges can be connected
to $G\setminus g(L)$, not just to neighbour nodes of $g(L)$.  A further
generalisation is [14], where more general types of graph
operation are used in place of production rules, the application of these
operations being controlled by logical formulae and a tree grammar.

The \ic{context subgraph} and the \ic{connection relation} approaches should be
seen as alternatives. Adachi \etal~[4] use both in tandem, but in
later work [5] abandon the context subgraph without affecting the
generative power of the grammars.

\medbreak

\i{Third approach: implicit edges.} The problem we are considering is how to
connect $h(R)$, the new part of the graph, to $G\setminus g(L)$, the old part,
by edges. The simplest solution is to do without edges entirely. The graph $G$
is simply a set of nodes, with geometric attributes such as position and size. 
The grammar contains \ic{constraints} between symbol types, which are essentially
edges; if two nodes in $G$ are related in a way that fits the
constraints, there is implicitly a geometric relationship between them, but it
is not represented explicitly as an edge. In the production rules $L$ is a
single node. The geometric attributes of the node in $g(L)$ are related to
those of the nodes in $h(R)$; this means that new implicit geometric
relationships are produced when the rule is applied, but there is no need for a
context subgraph or a connection relation.

Examples of this approach are picture layout grammars [55] and
constraint multiset grammars [23,76]. They are described in
terms of \ic{multisets}, but they are best thought of as graphs with implicit
edges.

\subsubhead{2.2 The lack of a well-defined Chomsky hierarchy}
For string grammars there is a clear classification into levels according to
generative power: type 0 (unrestricted grammars), type 1 (grammars where the
left-hand side of each rule is no longer than the right-hand side, and parsing
is guaranteed to halt), type 2 (context-free grammars, expressing iterative and
nested structure), and type 3 (regular grammars, expressing iterative structure
only). These types can be characterised in various ways (in terms of production
rules, recognition automata, or alternative notations such as regular
expressions), and are unaffected by minor differences in formulation.

Unfortunately the same cannot be said for graph grammars. One may still define
type 0 grammars, having no restrictions on the production rules; there are
inequivalent kinds of type 0 grammar, such as context-sensitive NCE grammars
[3], edge-based graph grammars [90], reserved graph
grammars [106], and the \ic{set-theoretic approach} [79];
reserved graph grammars are more expressive than edge-based grammars
[109], whereas the set-theoretic approach seems to be the most
expressive type. There are several concepts of type 1 grammar; the aim is
always to impose a restriction that ensures that parsing is guaranteed to halt.
The simplest restriction is that the number of nodes on the left-hand side of
each rule is less than or equal to the number of nodes on the right-hand side
[4]; or alternatively that the number of nodes and edges on the
left is less than or equal to the number of nodes and edges on the right
[35]. Layered graph grammars [84,3] impose a
more complicated condition of the same kind, with node labels distinguished by
\ic{layer}. Reserved graph grammars did the same at first [106], but
more recent work drops the layering and reverts to the simpler condition
[110].

There is a variety of inequivalent concepts of \i{context-free} grammar
[39]. Some
authors say that a context-free grammar is one where each rule has a single
(labelled) node as left-hand side
[60,106,102,91,52,72,25].
However Lange \& Welzl [65] point out that this is not really
context-free in the full sense: \ic{\dic{context-sensitivity} is hidden in the
embedding mechanism of NLC grammars}. Hence context-freedom is often identified
with a confluence or finite Church-Rosser property [16].
Definitions of this vary. The general idea is that \ic{the result of a
derivation is independent of the order in which the productions are applied}
[40]. One version is that if $G \derives_p H_1 \derives_q K_1$
and $G \derives_q H_2 \derives_p K_2$ (where \ic{$\derives_p$} means \ic{derives
by using production rule $p$}) then $K_1$ is isomorphic to $K_2$. Another
version is that if $G \derives_p H_1$ and $G \derives_q H_2$, where $p\ne q$,
then $G \derives_p H_1 \derives_q K$ and $G \derives_q H_2 \derives_p K$ for
some $K$. These are left-to-right versions; right-to-left versions can be
defined by reversing the arrows.

Courcelle argues from a more abstract perspective that \ic{context-free} should
be defined to mean confluent and associative [26].

These notions are heavily dependent on the surrounding formalism. In the
\ic{set-theoretic approach} [79] and boundary NLC grammars
[86], the context-free languages coincide with the
context-sensitive ones, though this is not true for the grammars.

There are a few concepts of \i{regular} graph grammar in the literature
[27,54,79,2,1,103,48]. 
These are heavily dependent on the details of the definition, especially on whether
the right-hand side is allowed to be disconnected. There is no unique notion.

There has been little work to characterise classes of graph language by
accepting automata. Some classes of NCE grammars (linear NCE and boundary NCE
grammars) can be characterised by accepting automata, but this does not seem
possible for arbitrary NCE grammars [17].

There has also been a little work on algebraic characterisations of classes of
graph language. Courcelle [28] gives formal characterisations,
using monadic second-order logic,
of the set of languages generated by edge-labelled edge-directed NCE grammars
and the set of languages generated by hyperedge-replacement grammars. Bauderon
\etal\ [6,9] claim to provide an algebraic
formulation of production rule application in NLC grammars as a pullback. But
this cannot be true, as their graphs have no node labels. There is a kind of
node labelling mechanism implicit in their \ic{unknown} homomorphisms
[6, definition~3], but these labels are freely changed every
time a production rule is applied, so this is not at all like what is normally
understood as an NLC grammar. In their later work
[7,8] an elaborate mechanism is specified for
updating these labelling homomorphisms every time a rule is applied; this,
along with the pullback, specifies how rule application works. There is also an
element of parallel application of rules. It is now acknowledged that this
generates a larger class of languages than node-replacement grammars.

\subsubhead{2.3 Perspicuity, parse trees, expressive power}
In a context-free string grammar the production rules (e.g., $Sentence \to NP\
verb\ NP$) express something about the structural possibilities of the language.
After parsing a sentence, a parse tree is produced, which makes explicit the
grammatical structure of the sentence. It is the central function of grammars to
do this. The parse tree provides a declarative view of the structure of the
sentence, abstracting away from the operational details of the derivation.
A parse tree may also be thought of as an algebraic expression that is
evaluated to produce the sentence [34].
Unfortunately, with more complex types of grammar, including most graph
grammars, the production rules are merely engines for generating sentences and
do not convey any grammatical information. Parse trees are often not possible,
so there is no representation of the grammaticality of the sentence more
abstract than the sequence of production rule applications.

When is a parse tree possible in a graph grammar? If the production rules have
a single node on the left-hand side then one may construct a tree of the nodes
involved in the derivation; however, the edges will cut across between the
branches in a tangle [55,29]. Useful parse trees can be
produced if the grammar is confluent. This is done for hyperedge-replacement
grammars [33], constraint multiset grammars [23], and
relation grammars [43,44]. In [60] no
confluence condition is imposed and the parse trees are dependent on the order
of application of the production rules, so they do not express the syntactic
structure in a fully abstract way. Parse trees are also possible for the graph
expansion grammars of [34], which are essentially an extension of
hyperedge replacement grammars.

Sometimes a nested graphical representation of syntactic structure is used,
which is equivalent to a parse tree [102,17]. I
believe this only works well if the grammar is confluent.

Thus it seems that parse trees are possible only for context-free graph
grammars. Unfortunately, these grammars are surprisingly weak in their
expressive power. One would expect context-free grammars to be able to represent
iterative and nested structure, but, as Zou \etal\ say, \ic{context-free graph
grammars have difficulty in specifying a large portion of graphical VPLs [visual
programming languages]} [110]. Even in the weakest sense of
context-free, \ic{Not all, nor even the majority, of VL [visual language]
formalisms are context-free even in this loose sense} [102].
Even an iterative pattern in two dimensions, which one would expect to be
captured by a regular graph grammar, is beyond the reach of context-free graph
grammars: Schuster [89,16] showed that the set of
quadratic graphs (rectangular grids) cannot be generated by any graph grammar
with the finite Church-Rosser property.

Bauderon \etal's pullback grammars can generate the set of square grids
[7], but this is a very much more powerful type of grammar, with
an elaborate mechanism for changing labels at every derivation step, and
depending essentially on parallel application of rules; so the rules themselves
do not fully express the grammatical knowledge.

This is the key limitation of graph grammars: to obtain decent expressive power
we have to go beyond the context-free grammars, but then we lose the ability to
represent syntactic structure through parse trees. Thus we have a purely
operational concept of what it means to belong to the language. There is no
description of how the graph fits the grammar simpler than the sequence of
production rule applications.

\subsubhead{2.4 Parsing algorithms}
Graph grammars are very powerful mechanisms, and so parsing is necessarily
intractable. Type 0 grammars in the \ic{set-theoretic approach} [79]
and the \ic{algebraic approach} [35] and for edge-label controlled
grammars [75] can generate any recursively enumerable set of graphs,
so the membership problem (deciding whether a given graph is in the language) is
undecidable.

Type 1 restrictions, which all amount to saying that the left-hand side is no
larger than the right-hand side, make the membership problem decidable
[35].

Even in the case of context-free grammars the membership problem is still intractable. NLC
grammars can generate PSPACE-complete graph languages [16].
Hyperedge-replacement grammars [33], linear
edge-replacement grammars [33], confluent edge-labelled
edge-directed NCE graph grammars [41], and even regular NLC
grammars [1] can generate NP-complete graph languages. This
is unlike context-free string grammars, which can be parsed in cubic time using
the standard Cocke-Younger-Kasami algorithm. The reason is that a string has
quadratically many substrings, whereas a graph has exponentially many
subgraphs [13].

Some improvement in efficiency is possible by
clever optimisation to reduce backtracking [111]; the worst-case
time-complexity however remains exponential in the size of the input graph,
for a fixed grammar. 
Hence much research effort is devoted to identifying subclasses of grammar that
can be efficiently parsed. Polynomial-time parsing is possible if

\blob the grammar has the finite Church-Rosser property,

\blob the grammar generates only connected graphs,

\blob the grammar generates graphs of bounded degree.

\medskip\f
If any of these three conditions is dropped then a grammar exists for which the
membership problem is NP-hard [16]. Note however that even
where parsing is possible in polynomial time the degree of the polynomial is
very high, so tailor-made efficient algorithms are still needed for particular
classes of grammar [86,43].

If the grammar is confluent in the right-to-left direction then bottom-up
parsing without backtracking, called \i{selection-free} parsing, is possible;
this gives polynomial time complexity [110,106,72]. 
This is a very restrictive condition, however.

Precedence graph grammars can be parsed in linear time [35].  Graph
languages generated by evaluating algebraic expressions generated by tree
grammars can parsed in polynomial time if certain restrictions are imposed
[14,12]. Other fast parsing algorithms for restricted
classes of graph grammar are
[31,32,13,54,103].

\subsubhead{2.5 Coping with noise}
Graph parsing algorithms have little ability to cope with errors in the graph
(false or missing nodes, and errors in labels, edges or structure).

Many graph parsing algorithms are based on exhaustive enumeration of all
possibilities, either by backtracking or by carrying forward a set of possible
parses, e.g., algorithms based on the Cocke-Younger-Kasami algorithm
[86,55,16] or on the Earley algorithm
[43,101,18,19,69]. With noise
the possibilities proliferate and become unmanageable.

Most algorithms work by traversing the graph, i.e., by visiting each node or
edge one at a time
[69,1,25,101].
This is unsuitable for non-Eulerian graphs and is error-intolerant, since an
error in the first node visited may throw the process onto the wrong track, and
the erroneous node cannot be corrected later because it is never revisited.

In addition, many pattern recognition systems work in a fixed sequence of
phases, or in a fixed direction (top-down or bottom-up), where
misinterpretations made in one phase cannot be corrected in later phases.

Some systems have a limited ability to handle simple types of error. 
\i{Error-correcting grammars} can, in principle, model the errors using production
rules and make the correction of error part of the parsing process; however,
these have only been used for the simplest sorts of error. Skomorowski's
grammar [95] can correct errors in the
label of a node or edge, but not structural errors. In Fahmy \& Blostein's
music recognition system [42], the symbols can have ambiguous
labels; one node is created for each possible value of the label. Thus
labelling errors (and stray dots) can be corrected. Liu \& Yang's
[72] production rules include \ic{uncertain edges}, which are
excess edges that it is anticipated may occur in the graph. If such edges
occur they are removed during bottom-up parsing. S\'anchez \etal's grammar
[88] recognises tessellations of rectangles and octagons. There
are production rules that add special \ic{inserted} nodes where rectangles or
octagons are expected at the boundary of the tessellation; at the end of parsing
other production rules either remove each \ic{inserted} node or relabel it as a
\ic{cut} symbol or a \ic{split} symbol.  Mas \etal's grammar [76]
recognises hand-drawn diagrams; it includes production rules for eliminating
errors where one stroke has been misconstrued as two.

Error-correcting grammars are only suitable for simple, explicitly anticipated
types of error. It would not be feasible to use them for general error
correction, where any symbol or relation may be spurious, missing, mislabelled,
misinterpreted as two, or conflated with another.

Instead of modelling the errors in the grammar, an alternative is an
error-tolerant parser, which finds the closest match between the graph and what
the grammar would accept (for example [74], which matches a tree
against a tree automaton). There are also error-tolerant parsing algorithms for
matching a graph against a fixed set of graphs (rather than a grammar) by building up
partial matches or by using edit distance [77,85].

%% file: 3a.tex
\subhead{3. Networks and homomorphisms}

\subsubhead{3.1 Requirements}
From the literature survey in the previous section I draw the following
conclusions for my own work.

\blob I require a grammatical formalism that can represent iterative and
nested structure (including iterations in two dimensions such as tessellations).

\blob The syntactic structure in an image should be represented explicitly and
declaratively, in a structure called a \i{pattern}, like the way a parse tree
represents the structure of a string, rather than by a sequence of production
rule applications. I shall do this by abandoning production rules altogether.
The grammar and the pattern will each be a graph-like structure called a
\i{network}, and parsing will be a matter of constructing a homomorphism from
the pattern to the grammar.

\blob The grammatical formalism should be amenable to mathematical treatment.
There should be no complicated semi-formal prose definitions. All definitions
will be expressed algebraically.

\blob All phases of pattern recognition should be simultaneous and
interdependent. Pattern recognition involves both constructing the pattern from
the given image and parsing the pattern. Top-down and bottom-up parsing occur
simultaneously. We break free completely from sequential ideas drawn from string
parsing: there is no concept of traversal. Error-tolerant parsing must work
opportunistically, starting with the most clear-cut parts of the image and
recognising the ambiguous parts later in the light of top-down evidence. Many
alternative interpretations of the image will be held in the same pattern and
worked on in parallel, with one interpretation chosen by the end of parsing.

\medskip\f
In this section I am considering only the structural aspects of the pattern: the
geometric relationships will be considered in \S5.

\subsubhead{3.2 Networks}
Consider the example in figure~1. The raw data is called the \i{image}. The
pattern recognition system recognises the presence of one symbol token, called
$A1$, of type $A$, and three symbol tokens, $line1$, $line2$, $line3$, of type
$line$, which are its parts.

\midinsert

\medbreak

\percentsize{100}\Figure{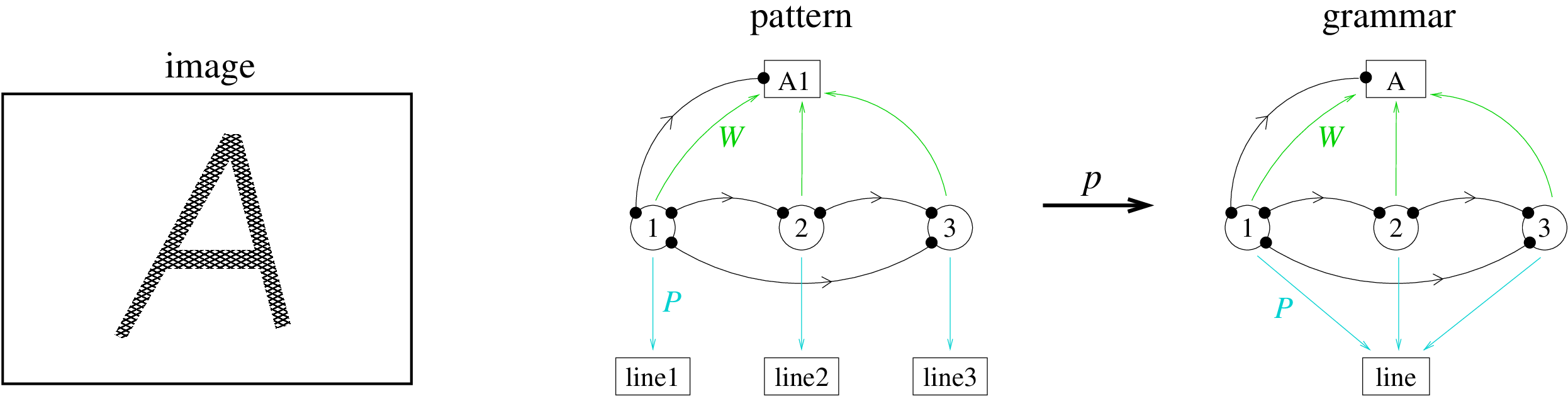}

\smallskip

\longcaption{Figure 1. An image, the corresponding pattern, and the grammar
against which it is parsed. Rectangles are symbols, circles are nodes, small
filled discs are hooks, and black curves are edges. The $W$ and $P$ functions
are depicted by green and blue arrows.}

\medbreak

\endinsert

The symbol tokens are connected together to make the pattern. Part-whole
relationships are represented by \i{node tokens}. Whenever one symbol token,
e.g., $line1$, is a part of another, $A1$, there is a node $n$ between them. Two
functions $W,P$ represent the relationship: $W(n) = A1$ and $P(n) = line1$. The
node represents the role that the part plays in the whole; for example, node
1 = left stroke, node 2 = cross-bar, node 3 = right stroke. (In all my figures,
symbols are depicted by rectangles and nodes by circles; the $W$ and $P$
functions are depicted by green and blue arrows.) When two parts of a
whole are related to each other they are called \i{siblings}, and their
relationship is represented by an \i{edge token} between the nodes. For example,
the left stroke of an $A$ must be joined at its top to the right stroke, and so
there is an edge token between node 1 and node 3. In this example all three
parts are one another's siblings. (In the figures edge tokens are depicted by
black curves, sometimes directed.) Actually, an edge token does not run between two node tokens
but between two \i{hook tokens}, which are attached to node tokens. (Hook tokens
are depicted in the figures as small filled discs.) Note that there is also an
edge token between node token 1 and the symbol token $A1$.

The grammar is similarly constructed, except that it is composed of symbol
types, node types, hook types and edge types. Parsing consists of constructing a
homomorphism $p$ from the pattern to the grammar, under which each symbol, node,
hook and edge token maps to its type, preserving incidence relations. The entire
recognition process is, given an image and a grammar, to construct the pattern
and the homomorphism.

This illustrates the main features of networks, albeit in an uninteresting
example. Figure~2 shows an example with iteration (the image is omitted). An
\i{alkane} is a hydrocarbon molecule consisting of a carbon chain of arbitrary
length surrounded by hydrogen atoms. Propane is the case where there are three
carbon atoms. Diagrammatic conventions are as before; under $p$, each
symbol token or node token maps to the symbol type or node type with the same
alphanumeric label, except that \i{propane} maps to \i{alkane}. (These symbol
and node labels are just a diagrammatic convention for conveying how $p$ works;
they are not really there in the network.)

\midinsert

\bigbreak

\percentsize{75}\Figure{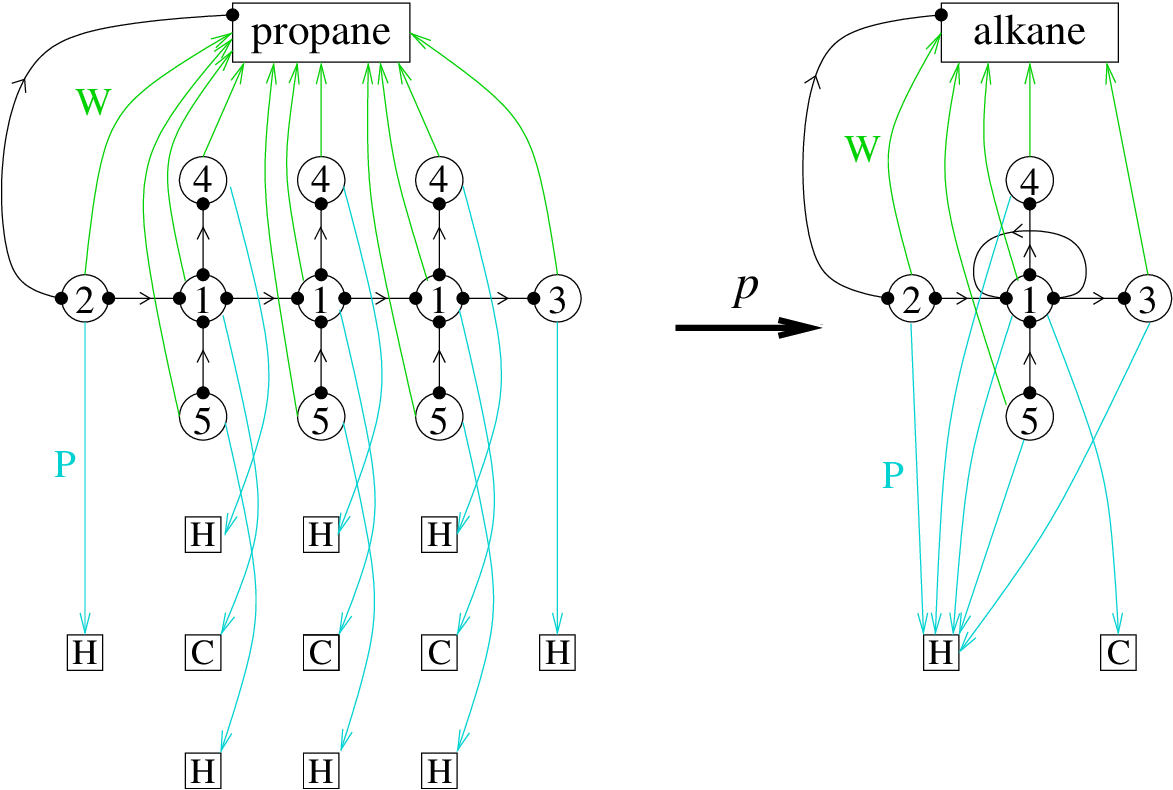}

\smallskip
\caption{Figure 2. On the right, the grammar for alkanes. On the left, one
possible alkane pattern.}

\medbreak

\endinsert

There is a rule that each hook token in the pattern must have exactly one
incident edge token. As a consequence, the \i{propane} token must be connected
to one \ic{2} node token, which must be connected to one \ic{1} node token. Each
\ic{1} node token must be connected to one \ic{4} and one \ic{5} token, and to
either a \ic{2} or a \ic{1} token (by its left hook), and to either a \ic{3} or
a \ic{1} token (by its right hook). This permits unbounded iteration of \ic{1}
node tokens, and hence any number of carbon atoms. Thus, in the grammar, a hook
with several incident edges represents a one-from-$n$ choice, and allows for
grammatical alternatives and iteration, which can be in one or more dimensions.

During the recognition process, while the pattern is under construction, there
may be any number of edge tokens incident to a hook token (either because no
edge token has been added yet or because several alternatives are being
considered in parallel). Thus alternative interpretations for various parts of
the image may be entertained in parallel, in the same pattern, without being
confused. By the end, one incident edge token must be chosen for every hook
token.

\subsubhead{3.3 Subsymbols}
One further type of grammatical constraint is needed. Consider figure~3, which
shows (a) a hexagonal grid image, (b) part of the pattern corresponding to the
portion of the image encircled in red, with symbol tokens (the grid itself and
the component lines) omitted, and (c) the relevant portion of the grammar. (This
is a simplified version of the hexagonal grid grammar of \S10.1.) As usual, each
node and edge in the pattern is mapped under $p$ to the node and edge with the
same label in the grammar.

\midinsert

\medbreak

\percentsize{100}\Figure{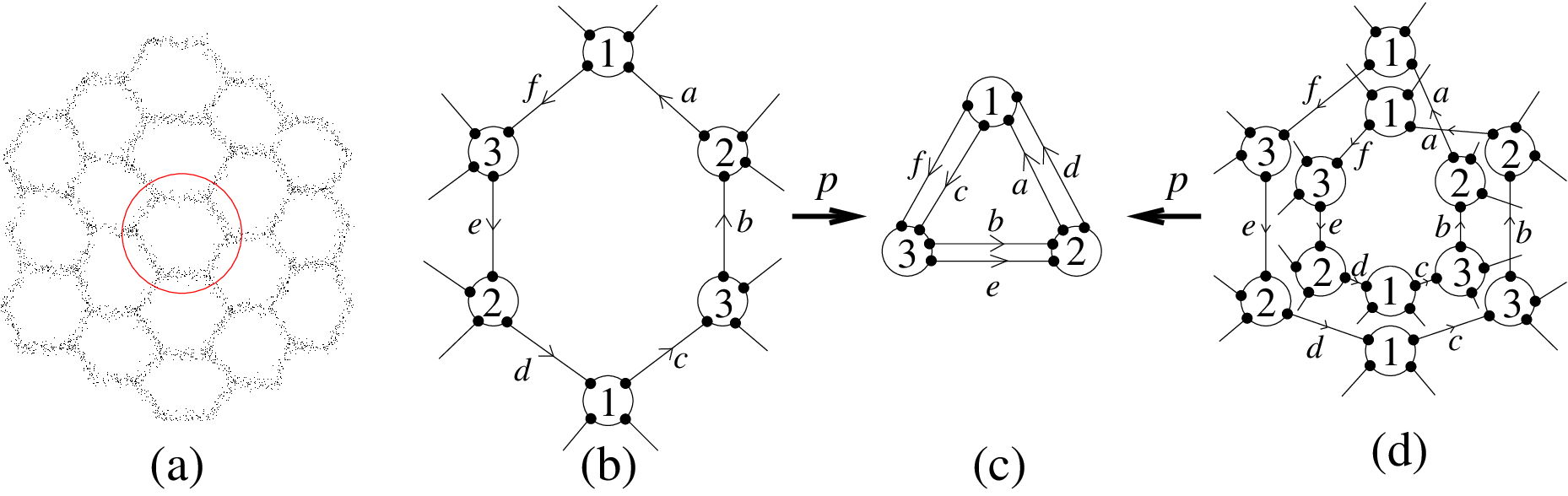}

\smallskip

\longcaption{Figure 3. (a) An image. (b) Part of the pattern (without symbol
tokens). (c) The relevant part of the grammar. (d) An unwanted pattern.}

\medbreak

\endinsert

There is a problem with this. The grammar appears to allow \ic{nonstandard
models}, unintended patterns that could map homomorphically to the grammar just
as well (see (d)). To cure this problem we need to treat a hexagon (i.e., the
portion of the hexagonal grid shown in figure~3(b)) as a symbol in its own
right. The hexagon is said to be a \i{subsymbol} of the hexagonal grid, meaning
that all the parts of the hexagon (the six line symbol tokens) are parts of the
hexagonal grid. Figure~4 shows, in part~(a), the relevant portions of the
pattern: the $hexagon1$ symbol token and all its nodes, hooks and edges; the
$hex1$ hexagonal grid symbol token (only the relevant nodes, hooks and edges are
shown); and the six parts of $hexagon1$, the symbol tokens $line1$--$line6$,
which are also parts of $hex1$. Part~(b) of the figure shows the grammar. As
usual, under the parsing homomorphism $p$, each node and edge token in the
pattern maps to the node or edge type with the same label in the grammar.

\midinsert

\medbreak

\percentsize{100}\Figure{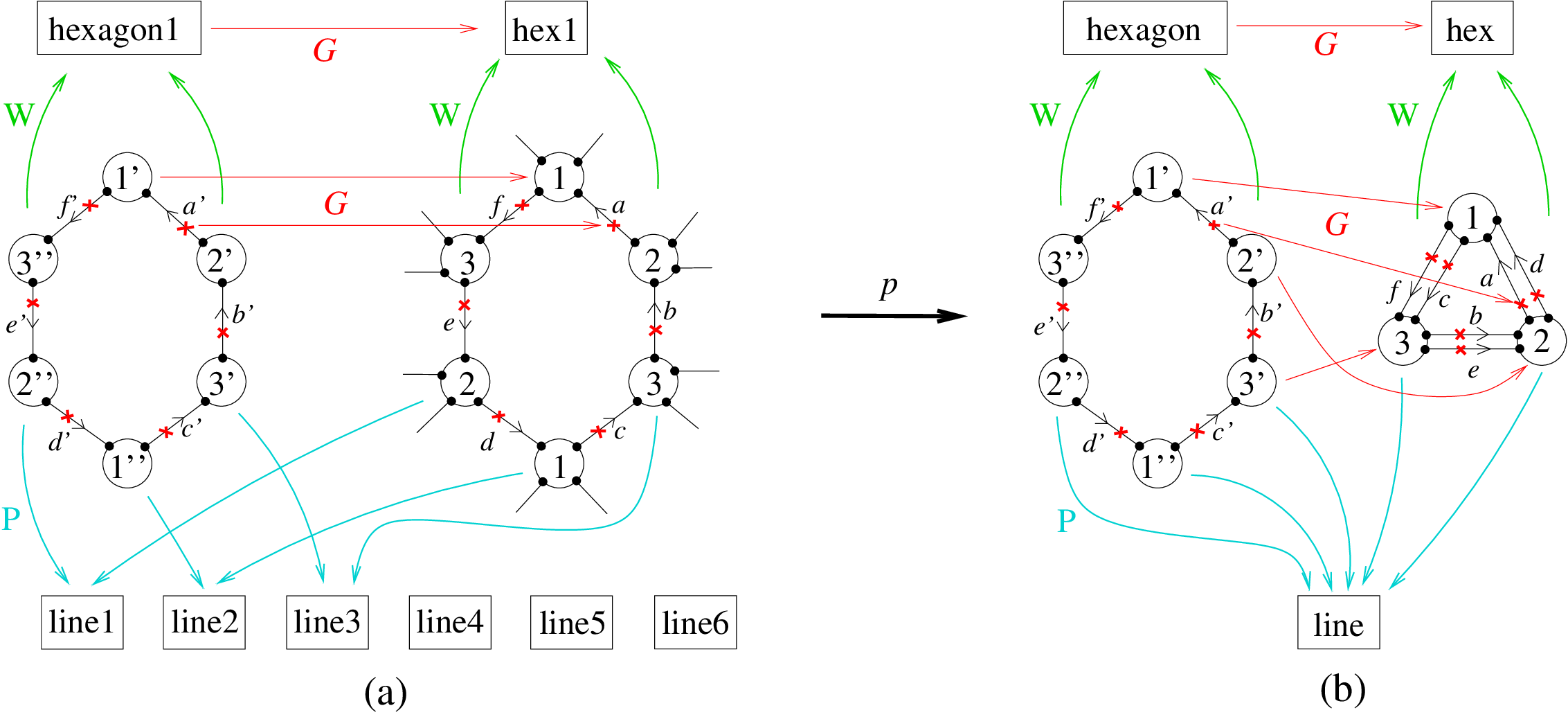}

\longcaption{Figure 4.(a) Part of a pattern: a hexagonal grid symbol token
$hex1$ and some its nodes, hooks, edges; and a subsymbol $hexagon1$, with all
its nodes, hooks and edges and parts $line1$--$line6$. (b) The grammar,
including the hexagonal grid symbol type $hex$ and its subsymbol type $hexagon$.
Only a few of the $W$ and $P$ arrows are shown. The gluing relation $G$ is shown
in red (only a few of the gluings are shown). Facets are shown as small red
crosses.}

\medbreak

\endinsert

To indicate the subsymbol relationship, $hexagon1$ is \ic{glued} to $hex1$ by a
gluing relation $G$; formally we write $G(hex1,hexagon1)$. We say $hexagon1$ is
a \i{subsymbol} of $hex1$ and $hex1$ is a \i{supersymbol} of $hexagon1$. The
corresponding nodes are also glued: $G(1,1'), \allowbreak G(1,1''), \allowbreak
G(2,2'), \allowbreak G(2,2''), \allowbreak G(3,3'), \allowbreak G(3,3'')$. We
say node $1'$ is a \i{subnode} of node $1$ and node $1$ is a \i{supernode} of
node $1'$. Corresponding hooks and edges are also glued:
$G(a,a'),G(b,b'),\ldots$ . This is done in both the pattern and the grammar; the
gluings in the pattern must match those in the grammar. Some of the gluings are
shown in red in the figure.

One final syntactic feature is needed. The grammar must specify where $hexagon$
subsymbols should occur in a $hex$ grid. This is done by giving some of the
edge types one or more \i{facets} (these are depicted as small red crosses in
the figure). Some of the facets in the grammar are glued together in pairs. The
edge tokens in the pattern have facets similarly. There is a new grammatical
rule: in the pattern, each facet must be glued to \i{one} other facet (in a way
consistent with the grammar). When two facets are glued this forces their edges
to be glued, and hence their incident hooks are glued, hence the nodes to which
they are attached are glued, hence the whole symbol tokens above them are glued.

As another example, the three lines meeting at a vertex or junction of the hexagonal grid
also form a subsymbol. There are two sorts of junction subsymbol: those in which
the three lines are oriented towards the junction and those in which the three
lines are oriented away from the junction, so we need two subsymbol types,
$junctionA$ and $junctionB$. Figure 5 shows the $junctionA$ subsymbol type and one
token $junctionA$-$1$ of that type. We give the edges $a$--$f$ in the grammar and
pattern a second facet: this means that each such edge will be part of two
subsymbols: a junction and a hexagon. The $junctionA$ edges $a',b',c'$ only need one
facet. As before, in the pattern each facet must be glued to one other facet (in
a way consistent with the grammar). This will force edge tokens $a',b',c'$ to be
glued to $a,b,c$, respectively, in the pattern. This will force their incident
hook tokens to be glued; this will force the node tokens $1',2',3'$ to be glued
to $1,2,3$; and so $junctionA$-$1$ will be glued to $hex1$.

\midinsert

\medbreak

\percentsize{100}\Figure{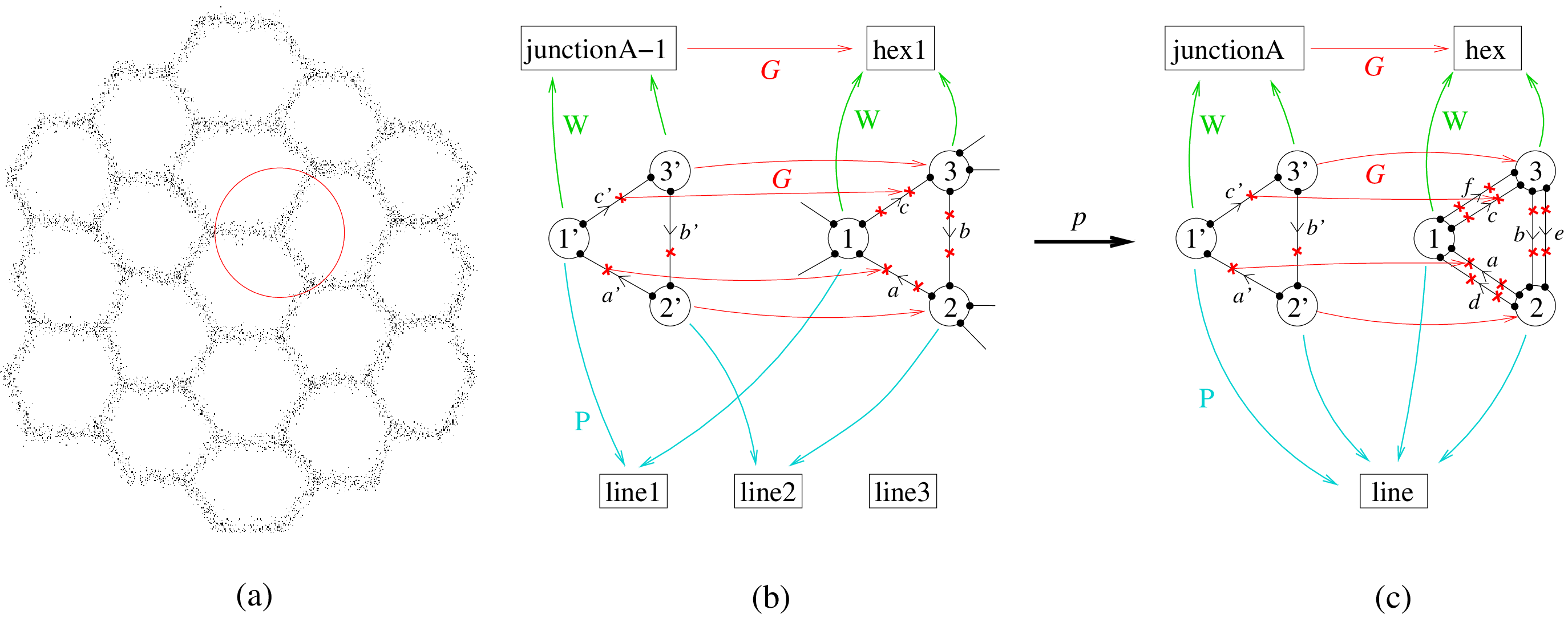}

\longcaption{Figure 5. (a) The image, with the relevant portion encircled. (b)
The pattern: the $junctionA$-$1$ subsymbol token and the relevant portion of
the $hex1$ symbol token. (c) The grammar: the $junctionA$ and $hex$ symbol
types. A few of the $P$ and $W$ arrows and a few of the gluings are shown.
Every node or edge with a primed label is glued to the one with the unprimed
label. Every node or edge in the pattern is mapped under $p$ to the one with
the same label in the grammar.}

\medbreak

\endinsert

Further example grammars are given in \S10.
In the next section the concepts of this section will be defined formally.

%% file: 4a.tex
\subhead{4. The structural part of the theory}

\subsubhead{4.1 Preliminaries}
We shall deal with sets, functions and binary relations. I use standard
set and function notation.

\defn The \i{domain} $\dom(R)$ and \i{range} $\ran(R)$ of a relation $R$ are
defined by
$$ \dom(R) = \comp{a}{\exists b\ R(b,a)}, \qquad
\ran(R) = \comp{b}{\exists a\ R(b,a)}. $$

\defn A relation $R$ is said to be \i{on} a set $A$ iff $\dom(R) \subseteq A$
and $\ran(R) \subseteq A$.

\defn For any set $A$, the \i{identity} relation $id_A$ is defined by
$$ \forall a,b\quad (id_A(a,b) \ \iff \ a=b\in A). $$

\defn The empty relation $\bot$ is defined by $\forall a,b \ \neg\, \bot(a,b)$.

\defn For any relations $R$ and $S$,
$$\eqalign{
R = S \ &\iff \ \forall a,b\ (R(a,b) \iff S(a,b)) \cr
R \subseteq S \ &\iff \ \forall a,b\ (R(a,b) \Rightarrow S(a,b)) \cr 
R \subset S \ &\iff \ R \subseteq S \and R \ne S.
}$$

\defn For any relations $R$ and $S$, the relations $R \cap S$, $R \cup S$ and
$R\setdiff S$ are defined by
$$\forall a,b\ \left\lbrace\, \eqalign{
(R \cap S)(a,b) &\iff R(a,b) \wedge S(a,b), \cr
(R \cup S)(a,b) &\iff R(a,b) \vee S(a,b), \cr
(R \setdiff S)(a,b) &\iff R(a,b) \wedge \neg S(a,b).
}\right.$$

\defn For any relations $R$ and $S$, the \i{composed} relation $R \circ S$ is
defined by
$$ \forall a,c\ \ 
((R \circ S)(a,c) \ \iff \ \exists b\ (R(a,b) \wedge S(b,c))).
$$\ignore{
(In all expressions, the composition operator $\circ$ has higher syntactic
precedence than $\cap$, $\cup$ and $\setdiff$.)}

\defn For any relation $R$, the \i{inverse} relation $R\inv$ is defined
by
$$ \forall a,b\ \ (R\inv (b,a) \ \iff \ R(a,b)). $$

\defn The \i{graph} of a function $f$ is the relation $\overline f$ defined by
$$ \forall a,b\ \ (\overline f(b,a) \ \iff \ a \in \dom(f) \wedge b=f(a)). $$
Note that, for any $f,g$, $\overline{f \circ g} = \overline f \circ \overline
g$.

\defn A relation $R$ is \i{finite} iff there are finitely many pairs $(a,b)$
such that $R(a,b)$ holds.

\defn The \ic{not-equal-to} relation $NE$ is defined by $NE(a,b) \iff a \ne b$.

\defn A finite relation $R$ is \i{acyclic} iff
$$ \NOT\, \exists R^*\ (\bot \ne R^* \subseteq R \, \and\, R^* \subseteq (R^*
\circ NE \, \cap\, NE \circ R^*)). $$

\def\mapleft#1{\,\smash{\mathop{\leftarrow}\limits^{#1}}}
\def\mapright#1{\smash{\mathop{\rightarrow}\limits^{#1}}\,}

\medskip\f
(Informally, $R$ is acyclic iff every non-empty subrelation $R^*$ has an element
of valency 1, i.e., an element related to just one element by $R^*$ or to just
one element by ${R^*}\!\inv$. This is equivalent to the non-existence of a
finite cyclic  sequence $x_1,\ldots x_n,x_1$, with $n$ even and $n>2$, where the terms
are all different and are related by $x_1 \mapright R x_2 \mapleft R x_3
\mapright R x_4 \mapleft R x_5 \mapright R \cdots \mapright R x_n \mapleft R
x_1$.)

\defn If $f:X\to Y$ and $R$ is a relation on $X$ then $R$ is \i{connected}
relative to $f$ iff, for any set $Z$ and function $g:X\to Z$ such that
$\overline g \circ R \subseteq \overline g$, there exists a function $i:f(X)\to
Z$ such that $i \circ f = g$.

\medskip\f
(Informally, $R$ is connected relative to $f$ iff, for any $x,x'\from X$ such
that $f(x)=f(x')$, there exists a finite sequence $x=x_0,x_1,\ldots x_n=x'$ such
that, for each $i\from\{1,\ldots n\}$, $R(x_{i-1},x_i)$ or $R(x_i,x_{i-1})$
holds.)

\defn If $f:X\to Y$ and $R$ is a relation on $X$ then $R$ is \i{minimal}
relative to $f$ iff
$$ \forall R^*\subseteq R\ \ (\overline f \circ R^* = \overline f \circ R
\Rightarrow R^* = R). $$

\medskip\f
(Informally, $R$ is minimal relative to $f$ iff, for any $x\from X$ and $y\from
Y$, there exists at most one $x'\from X$ such that $f(x')=y$ and $R(x',x)$.)

\medskip\f
I shall also use some basic concepts of category theory, namely pullbacks, sums
and coequalisers.

\subsubhead{4.2 Networks and homomorphisms}

\defn A \i{network} is a 12-tuple $\network{}$, where $\Sigma,N,H,E,K$ are
disjoint finite sets; $W:N\cup\Sigma\to\Sigma$, $P:N\to\Sigma$, $A:H\to
N\cup\Sigma$, $F,S:E\to H$ and $C:K\to E$ are functions such that $\forall
\sigma\from\Sigma\ W(\sigma)=\sigma$; and $G$ is a relation on $\Sigma \cup N
\cup H \cup E \cup K$ such that

\item{(1)} $id_\Sigma \circ G = G \circ id_\Sigma$, $id_N \circ G = G \circ id_N$,
$id_H \circ G = G \circ id_H$, $id_E \circ G = G \circ id_E$, $id_K \circ G = G
\circ id_K$;

\item{(2)} 
$\overline W \circ G = G \circ \overline W$, \ $\overline P \circ G
\subseteq \overline P$, \ $\overline A \circ G \subseteq G \circ \overline A$, \
$\overline F \circ G \subseteq G \circ \overline F$, \ $\overline S \circ G
\subseteq G \circ \overline S$, \ $\overline C \circ G \subseteq G \circ
\overline C$;

\item{(3)} $G_H$ and $G_H\inv$ are minimal relative to $A$;
$G_K$ and $G_K\inv$ are minimal relative to $C$; $G_E$ and $G_E\inv$ are minimal
relative to $F$ and $S$ (where $G_H = G \circ id_H$, $G_K = G \circ id_K$ and
$G_E = G \circ id_E$);

\item{(4)} $G \circ G = \bot$.

\medskip\f
The elements of $\Sigma,N,H,E,K$ are called \i{symbols}, \i{nodes}, \i{hooks},
\i{edges} and \i{facets}, respectively. $G$ is called the \i{gluing relation};
$G(x,y)$ means that $y$ is a subsymbol, subnode, subhook, subedge or subfacet
of $x$, i.e., $x$ is a supersymbol, supernode, superhook, superedge or
superfacet of $y$.
The functions $W,P,A,F,S,C$ express the incidence relations: a node $n$
connects a part $P(n)$ to a whole $W(n)$; a hook $h$ is attached to a node (or
possibly a symbol)
$A(h)$; an edge $e$ runs from its \i{first} hook $F(e)$ to its \i{second} hook
$S(e)$; a facet $k$ belongs to the edge $C(k)$. 

Condition~(1) means the gluing relation $G$ may be
considered as the disjoint union of a relation $G_\Sigma = G \circ id_\Sigma$ on
symbols, a relation $G_N = G \circ id_N$ on nodes, a relation $G_H = G \circ
id_H$ on hooks, a relation $G_E = G \circ id_E$ on edges, and a relation $G_K =
G \circ id_K$ on facets. 

Condition~(2) says that $G$ preserves incidence: e.g., if a hook $h_1$ is
glued to a hook $h_2$ then the node $A(h_1)$ is glued to the node $A(h_2)$.

Condition~(3) means that a hook is glued to at most one hook of any given
node; a facet is glued to at most one facet of any given edge; and an edge is
glued to at most one edge of any given hook (for each edge direction).

Condition~(4) says that a subsymbol, subnode, subhook, subedge or subfacet
cannot be also be a supersymbol, supernode, superhook, superedge or superfacet.

\defn If $\N_1=\network{_1}$ and $\N_0=\network{_0}$ are networks, a
\i{homomorphism} $p:\N_1\to\N_0$ is a function from $\Sigma_1 \cup N_1 \cup H_1
\cup E_1 \cup K_1$ to $\Sigma_0 \cup N_0 \cup H_0 \cup E_0 \cup K_0$ such that

\item{(1)} $p(\Sigma_1) \subseteq \Sigma_0$, \quad $p(N_1) \subseteq N_0$, \quad
$p(H_1) \subseteq H_0$, \quad $p(E_1) \subseteq E_0$, \quad $p(K_1) \subseteq
K_0$;

\item{(2)} $W_0 \circ p = p \circ W_1$, \quad $P_0 \circ p = p \circ P_1$, \quad
$F_0 \circ p = p \circ F_1$, \quad $S_0 \circ p = p \circ S_1$;

\smallskip

\item{(3)}
$\bigrectangleA{N_0\cup\Sigma_0}{N_1\cup\Sigma_1}{H_0}{H_1}
{p|_{N_1\cup\Sigma_1}}{A_0}{A_1}{p|_{H_1}}$ and
$\bigrectangleA{E_0}{E_1}{K_0}{K_1}{p|_{E_1}}{C_0}{C_1}{p|_{K_1}}$ \ are pullbacks in
the category of sets;

\medskip

\item{(4)} $\overline p \circ G_1 = G_0 \circ \overline p$;

\item{(5)} $G_1$ is minimal relative to $p$.

\medbreak\f
Condition~(1) means that $p$ maps symbols to symbols, nodes to nodes, hooks to
hooks, edges to edges, and facets to facets.

Condition~(2) means that $p$ preserves the $W,P,F,S$ incidence functions.

Condition~(3) means that $p$ maps the hooks of any node $n$ bijectively onto
the hooks of $p(n)$, and maps the facets of any edge $e$ bijectively onto the
facets of $p(e)$.

Condition~(4) means that $p$ preserves gluing (i.e., if $G_1(x,y)$ then
$G_0(p(x),p(y))$), and if $p(x)$ is a subsymbol, subnode, subhook, subedge or
subfacet then so is $x$.

Condition~(5) says that the gluings in $\N_1$ are induced by those in $\N_0$,
i.e., two things are glued in $\N_1$ only if they are forced to be by
condition~(4).

\medbreak

The composition of two homomorphisms is a homomorphism and the inverse of a
bijective homomorphism is a homomorphism [50, theorems~15,16].

\defn If $\N=\network{}$ is a network, a \i{subnetwork} of $\N$ is a 12-tuple
$\network{'}$, where

$\Sigma' \subseteq \Sigma, \quad N' \subseteq N, \quad H' \subseteq H, \quad E'
\subseteq E, \quad K' \subseteq K$

$W(N') \subseteq \Sigma', \quad P(N') \subseteq \Sigma', \quad H' =
A\pre{N'\cup\Sigma'}, \quad F(E') \subseteq H', \quad S(E') \subseteq H', \quad
K' = C\pre{E'}$

$W' = W|_{N'\cup\Sigma'}, \quad P' = P|_{N'}, \quad A' = A|_{H'}, \quad F' =
F|_{E'}, \quad S' = S|_{E'}, \quad C' = C|_{K'}$

$G \circ id_{\Sigma'\cup N'\cup H'\cup E' \cup K'} \subseteq id_{\Sigma'\cup
N'\cup H' \cup E' \cup K'} \circ G, \quad G' = id_{\Sigma'\cup N'\cup H' \cup E'
\cup K'} \circ G \circ id_{\Sigma'\cup N'\cup H'\cup E' \cup K'}$.

\medskip\f
If $\N'$ is a subnetwork of $\N$ then $\N'$ is a network and the inclusion 
function from $\N'$ into $\N$ is a homomorphism [50, theorem~24].

A \i{proper} subnetwork of $\N$ is a subnetwork $\N'$ such that $\N'\ne\N$.

\subsubhead{4.3 Semi-definite and definite networks}
The grammar is required to be a \i{semi-definite network} and the pattern to be a
\i{definite} network. (While the pattern is under construction it is not
definite.) First we need the concept of a \i{minimal} gluing relation.

\defn If $\N=\network{}$ is a network then its gluing relation $G$ is
\i{minimal} relative to $\N$ iff
$$ id_K \subseteq G_K \circ G_K\inv \cup G_K\inv \circ G_K $$
where $G_K = G \circ id_K$; and, for any relation $G^* \subseteq G$ such that

\blob $\overline W \circ G^* \subseteq G^* \circ \overline W$, \ $\overline P
\circ G^* \subseteq \overline P$, \ $\overline A \circ G^* \subseteq G^* \circ
\overline A$, \ $\overline F \circ G^* \subseteq G^* \circ \overline F$, \
$\overline S \circ G^* \subseteq G^* \circ \overline S$, \ $\overline C \circ
G^* \subseteq G^* \circ \overline C$,

\blob $id_K \subseteq G^*_K \circ {G^*_K}\inv \cup {G^*_K}\inv
\circ G^*_K$ (where $G^*_K = G^* \circ id_K$),

\medskip\f
we have $G^* = G$.

\medskip\f
(Informally, the condition that $G$ be minimal relative to $\N$ means that every
facet is glued to another facet and, subject to this constraint, $G$ is as 
small as it can be.)

\defn A network $\network{}$ is \i{semi-definite} iff

\item{(1a)} $W \circ A \circ F = W \circ A \circ S$,

\item{(3)} $G_N$ is acyclic,

\item{(4a)} $H = F(E) \cup S(E)$,

\item{(5a)} $G_K \circ G_K\inv \subseteq id_K$,

\item{(6)} $G$ is minimal relative to $\N$,

\item{(7a)} $\overline F \circ G = G \circ \overline F$ and
$\overline S \circ G = G \circ \overline S$,

\item{(8a)} $\forall R\ \ (\overline A\inv \circ R = (\overline F \cup \overline S)
\circ R \ \Rightarrow\ G_{\Sigma} \circ R \subseteq \overline W \circ G_N \circ
R \ \cup \ id_\Sigma \circ \overline A \circ G_H \circ \overline A\inv \circ
R)$.

\defn A network $\network{}$ is \i{definite} iff

\smallskip

\item{(1b)} $E \doublearrow{A \circ F}{A \circ S} N\cup\Sigma \buildrel W \over
\rightarrow \Sigma$ is a coequaliser diagram in the category of sets,

\smallskip

\item{(2b)} $G_N$ is connected relative to $P$,

\item{(3)} $G_N$ is acyclic,

\item{(4b)} $E \buildrel F \over \rightarrow H \buildrel S \over \leftarrow E$ is a
sum diagram in the category of sets,

\item{(5b)} $id_K = G_K \circ G_K\inv \cup G_K\inv \circ G_K$,

\item{(6)} $G$ is minimal relative to $\N$.

\medskip\f
Condition~(1a) says that the two nodes at the ends of any edge belong to the same
whole. The stronger condition~(1b) says that the nodes and edges belonging to
any whole form a \i{connected} graph.

Condition~(2b) means that two nodes share the same part only when they are
glued together (directly or indirectly). (This condition prevents the same
symbol from being interpreted as part of two unrelated wholes.)

Condition~(3) means that there is no cyclic sequence of gluings. This is a
technical condition for ensuring that condition~(2b) holds in the pattern at
the end of recognition [50, \S3.6].

Condition~(4a) says that every hook has at least one incident edge. The
stronger condition~(4b) says that every hook has exactly one incident edge.

Condition~(5a) says that every facet has at most one superfacet.
The stronger condition~(5b) says that every facet has exactly one subfacet or superfacet.

Condition~(6) says that every facet is glued to another facet, but the gluings
are minimal subject to this constraint.

Condition~(7a) says that, if any hook is a subhook, then all its incident
edges are subedges.

Condition~(8a) means roughly that, whenever a subsymbol is
glued to a supersymbol, there are sufficiently many nodes and
hooks belonging to the subsymbol that are glued to nodes and hooks belonging to
the supersymbol. This is another
technical condition for ensuring that condition~(2b) holds in the pattern at
the end of recognition (it is used directly in
[50, theorem~22]).

Every definite network is semi-definite [50, theorem~18].

\subsubhead{4.3 The recognition problem}
We can now make our first formal statement of the recognition problem:
given a semi-definite network $\N_0$ (the \i{grammar}) and an image, 
the task is
construct a definite network $\N_1$ (the \i{pattern}) and a homomorphism
$p:\N_1\to\N_0$ (the \i{parse}).

This statement will be refined in \S6.3.

%% file: 5a.tex
\subhead{5. The geometric part of the theory}
This section is concerned with the spatial aspects of the pattern, in particular
the following five issues:

\blob the geometric relationships between one symbol token and another;

\blob the \i{embedding} or \ic{pose} of a symbol token in the image, i.e., the
transformation that maps the symbol token into the image at a certain position,
orientation and size;

\blob how to represent the variability in the embeddings and relationships;

\blob invariance of the whole pattern under affine transformations;

\blob symmetries of the pattern.

\medskip\f
I shall review the available approaches in the literature and introduce my own
concept of a \i{fleximap}, a variable affine transformation.

\subsubhead{5.1 Representing geometric relationships and embeddings}
Some authors represent the relation between one symbol token and another using
qualitative relationships (i.e., special named relationships) such as horizontal
and vertical neighbourhood [94,15,63]; \i{above},
\i{below}, \i{before}, \i{after}, \i{overlap}, etc.~[100,104];
closeness [42]; or topological relationships such as \i{hinged},
\i{butting}, \i{collinear}, \i{parallel}, \i{attached}, \i{concentric},
\i{radial}, \i{contained} [71] or \i{parallel}, \i{incidence},
\i{perpendicular}, \i{intersection}, \i{neighbour}, \i{inclusion}
[76].

A more versatile approach is to represent relationships between symbol tokens
in terms of geometric attributes, such as vanishing points and orientation
angles [57]; distance and relative size [100]; relative
angles [47]; distance and relative angle
[70]; relative position, angle and size [92,22];
internal angles and ratios of lengths [108,107]; curvature
and direction of bending [73]; closeness of attachment points,
ratios of widths, curvature and angles of tilt [80]; and
perpendicular distance from an endpoint of a line to another line
[11].

Embedding transformations are sometimes represented as sequences of simple
transformations, e.g., as $\i{translation} \cdot \i{rotation} \cdot
\i{dilation}$ [11] (where the dot denotes function
composition), or $\i{stretch} \cdot \i{rotation} \cdot \i{shear}$
[98], or via a singular value decomposition as $\i{translation}
\cdot \i{rotation} \cdot \i{dilation} \cdot \i{stretch} \cdot \i{rotation}$
[10].

\subsubhead{5.2 Representing variability}
Since symbol tokens are not likely to be perfectly positioned, some way is
needed to allow for variability in their embeddings and relationships. The
crudest way of allowing for variability is by rounding of geometric parameters
(such as position, angle and size) into a small discrete set of possible values.
A small variation in position, angle or size is tolerated, provided it does not
cross over a rounding boundary [100,47]. A
better-behaved approach is to allow geometric parameters to vary in a certain
range [15,58]. These are \ic{hard} constraints: they
either hold or do not.

Other authors use \ic{soft} constraints: deviations of the geometric parameters
from their nominal values are penalised by a \ic{penalty} or \ic{affinity
score} [42]. Sometimes fuzzy conditions are used [80].
Soft constraints have the advantage that they allow several alternative
interpretations of a symbol token to be scored and a choice made between them.
Most commonly, a quadratic penalty function is used to penalise deviations of
the attributes from their nominal values
[99,57,108,11,30]; an
\ic{energy} or \ic{match} function is calculated by summing the penalties of
all geometric relationships along with other penalty terms (and in some papers
the energy is exponentiated to give a Gaussian probability density
[81]).  The energy can then be minimised, giving the optimal match
of the symbols with the image. The advantage of using an energy function is
that a high deviation in one symbol-pair relationship can be traded off against
low deviations in other relationships to give the overall best match. In this
way the pattern recognition process becomes sensitive to context.

Particularly interesting is the use of \ic{springs} to connect pairs of symbol
tokens [45]. This allows variability in the
relative position of the symbols; a different device is used for controlling
relative orientation and length. This is similar to deformable templates,  
such as [105], in which a human eye is modelled with parabolic
curves.

Each of these methods represents the variability of a relationship in a fixed
way, for a particular purpose. For example, in [99] the variability is always analysed
into rotations around the centre point, dilations from the centre point, and
translations. What if one wanted a different decomposition, such as rotation or
dilation about a different point? For example, in figure~6 we have three
\ic{line} symbol tokens, which are parts of an \ic{A} symbol token. We may want
to allow the crossbar part $\sigma_3$ to rotate around its end-point, which is
40\% up the length of $\sigma_2$, while $\sigma_3$ may also dilate relative to
$\sigma_2$ (keeping one end in contact with $\sigma_2$); $\sigma_3$ may be
translated up or down $\sigma_2$ from the nominal 40\% position; it may also
overshoot or undershoot $\sigma_2$ by a tiny amount. Each of these kinds of
variation are to be penalised by different coefficients. The total penalty
should be a weighted sum of the squares of these variations.

\midinsert

\medbreak

\percentsize{25}\Figure{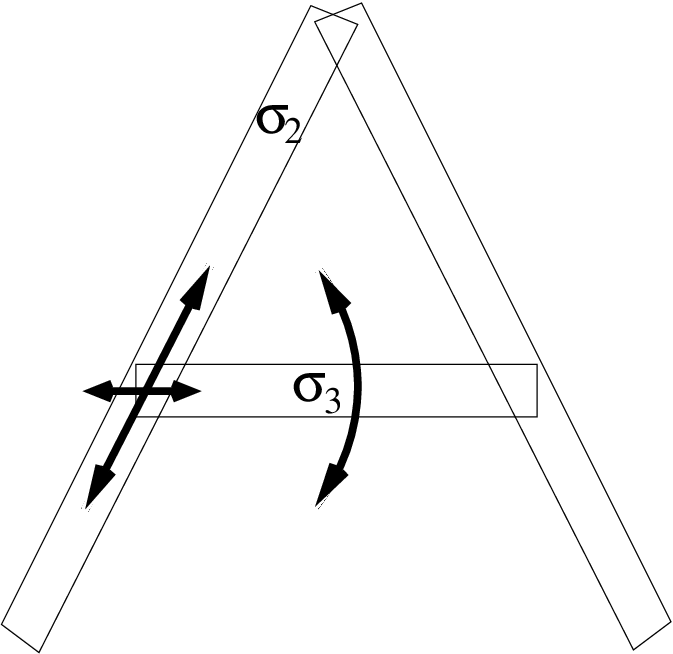}
\caption{Figure 6. The relationship between two symbol tokens, and some of its
dimensions of variability}

\medbreak

\endinsert

This is accomplished by my concept of a \i{fleximap}, a variable affine
transformation in which variations in relative position, orientation, size,
stretching and shearing are treated in a uniform formalism. Assume that every
symbol type has a \i{template}, depicting an ideal token of that type, at a
standard position, size and orientation. Every symbol token has an
\i{embedding}, an affine transformation mapping from the plane of the template
into the image plane. Figure~7 shows the templates for the \ic{$A$} and
\ic{$line$} symbol types, and embeddings for one symbol token $\sigma_1$ of type
$A$ and three symbol tokens $\sigma_2,\sigma_3,\sigma_4$ of type $line$. The
embeddings are called $u(\sigma_1),u(\sigma_2),u(\sigma_3), u(\sigma_4)$.

\midinsert

\medbreak

\percentsize{80}\Figure{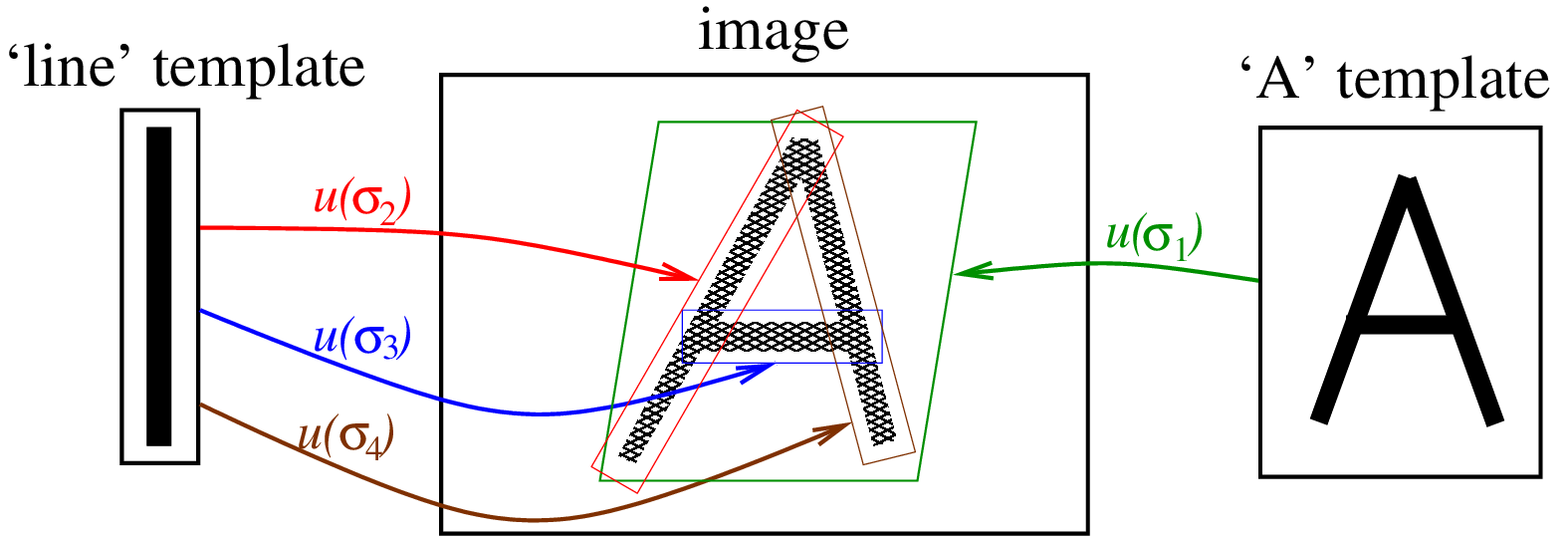}
\longcaption{Figure 7. This shows an image and the templates for two symbol
types $line$ and $A$. In colour are shown the embedding transformations for four
symbol tokens, $\sigma_1$ (of type $A$) and $\sigma_2,\sigma_3,\sigma_4$ (of
type $line$).}

\medbreak

\endinsert

Now consider the geometric relationship between $\sigma_2$ and $\sigma_3$: this
is defined as $u(\sigma_3)\inv \cdot u(\sigma_2)$, which is a member of $\G$,
the group of affine transformations (if the image is a plane this is a
six-dimensional Lie group).

The relationship between $\sigma_2$ and $\sigma_3$ may be expressed in the form
$$ u(\sigma_3)\inv \cdot u(\sigma_2) = F\cdot \exp(A) \eqno(1) $$
where $F \from \G$ is the \i{nominal} or \i{ideal} value of the relationship,
$A$ is a member of $\A$, the six-dimensional Lie algebra of $\G$, and
$\exp:\A\to\G$ is the exponential function. If you would like a concrete
representation, think of a point $x$ in the plane, an affine transformation
$G\from\G$, and a member $A\from\A$ of the Lie algebra in matrix form using
projective coordinates:
$$ x = \mat{x_1 \cr x_2 \cr 1}, \quad
G = \mat{g_{11} & g_{12} & s_1 \cr g_{21} & g_{22} & s_2 \cr 0 & 0 & 1}, \quad
A = \mat{h_{11} & h_{12} & t_1 \cr h_{21} & h_{22} & t_2 \cr 0 & 0 & 0}. $$
Then $G$ may be applied to $x$ by matrix multiplication, and exponentiation is
defined by
$$ \exp(A) = \sum_{n=0}^\infty \frac{A^n}{n!}. $$
$A$ is called the \i{deviation} of $u(\sigma_3)\inv \cdot u(\sigma_2)$ from its
nominal value. This deviation is quadratically penalised using a metric tensor.

\defn A \i{metric tensor} is a function $g:\A \to (\A \to \R)$ such that

\blob $\forall c_1,c_2 \from\R\ \forall A,B_1,B_2\from\A\ \, g(A)(c_1B_1+c_2B_2)
= c_1g(A)(B_1) + c_2g(A)(B_2)$,

\blob $\forall A,B \from\A\ \, g(A)(B) = g(B)(A)$,

\blob $\forall A\from\A\ (A \ne 0 \Rightarrow g(A)(A) > 0)$.

\medbreak\f
A fleximap is simply the combination of the nominal transformation and the
metric tensor.

\defn A \i{fleximap} is a pair $(F,g)$, where $F\in\G$ and $g$ is a metric
tensor.

\medskip\f
The \i{penalty} of the affine transformation $u(\sigma_3)\inv \cdot u(\sigma_2)$
relative to the fleximap $(F,g)$ is defined as $g(A)(A)$, where $A$ is given by
equation~(1) above. To state it in general,

\defn For any fleximap $\tau = (F,g)$ and any affine transformation $G$,
define the \i{penalty} $E_\tau(G)$ of $G$ relative to $\tau$ by
$$ E_\tau(G) = g(A)(A) $$
where $G = F \cdot \exp(A)$.

\medskip\f
If we choose a basis $(a_1,\ldots a_6)$ for $\A$, $A$ may be expressed as $A =
\sum_{i=1}^6 A^i a_i$, where $A^1,\ldots A^6$ are real coefficients, and $g$ may
be represented as a positive-definite, symmetric real $6\times 6$ matrix
$(g_{ij})$, with $g(A)(A) = \sum_{i,j=1}^6 g_{ij} A^i A^j$.

By a suitable choice of basis $(a_1,\ldots a_6)$, $g$ may be put into
diagonal form, \newline $g = diag(g_1,g_2,g_3,g_4,g_5,g_6)$, where
$g_1,g_2,g_3,g_4,g_5,g_6 > 0$. Then the penalty is $g(A)(A) = \sum_{i=1}^6 g_i
(A^i)^2$. We can choose $g$ so that $a_1,\ldots a_6$ are the six desired
dimensions of variation ($a_1 = $ rotation about a certain point, $a_2 = $
dilation about a certain point, $a_3 = $ translation in a certain direction,
etc.), and $g_1,g_2,g_3,g_4,g_5,g_6$ are the penalties for each of these types
of deviation. Hence $g(A)(A)$ is a weighted sum of squares of deviations, as
desired.

Thus the deviation $A$ may be decomposed into any six (linearly independent)
dimensions of variation that we like, and each dimension may be penalised as
much as we like. For the full and rigorous theory of fleximaps see
[49, \S3,\S5].

In the grammar, relationships between siblings are represented by edges; so each
edge type will have a fleximap $\tau$ describing the geometric relationship. In the
pattern, each edge token has the actual value of the relationship, e.g.,
$u(\sigma_3)\inv \cdot u(\sigma_2)$. The penalty is calculated for the actual
relationship relative to the fleximap, and equals 
$E_\tau(u(\sigma_3)\inv \cdot u(\sigma_2))$.

Relationships between part and whole, e.g., between $\sigma_2$ and $\sigma_1$ in
figure~7, are represented by nodes; each node type in the grammar will have a
fleximap constraining this relationship. A penalty is calculated for the actual
relationship $u(\sigma_1)\inv \cdot u(\sigma_2)$ relative to the fleximap.

All these penalties are summed (multiplied by $-1$), along with other penalty
terms, to give the total \i{match function} (see \S6.2,\S7.2), which is to be
maximised.

\subsubhead{5.3 Invariance of the whole pattern under affine transformations}
Whichever method is used to represent geometric transformations, some allowance
needs to be made for invariance of the pattern under change of frame of
reference. Qualitative spatial relationships such as \i{leftOf}, \i{aboveOf} are
invariant under translation but not rotation; they are appropriate for
applications where the image has already been normalised for orientation and
spatial directions have a special significance, such as recognition of
mathematical formulae [15] or musical scores [42].

Topological qualitative relationships [71] are invariant under
homeomorphisms. The qualitative relationships in [76] are a mixture of
topological invariants, affine invariants and similarity invariants.

Where geometric relationships are expressed in terms of angles and ratios of
lengths [100,92,22,108,107] they are invariant under similarities;
others are invariant under isometries [11].

My fleximap formalism is invariant under affine transformations. If the whole
pattern is subjected to an affine transformation $f$ then the embedding
$u(\sigma)$ of every symbol token $\sigma$ is transformed by $u(\sigma) \mapsto
f \cdot u(\sigma) $, so the relationship $u(\sigma_2)\inv \cdot u(\sigma_1)$
between two symbol tokens $\sigma_1$ and $\sigma_2$ is unchanged, so the penalty
value is unchanged.

\subsubhead{5.4 Symmetry}
The final issue to be considered is symmetry of the symbols and the whole
pattern. A hexagonal grid pattern, for example, has six-fold rotational
symmetry: every interpretation of the image has six symmetry variants. A pattern
recognition system must recognise that these are equivalent, rather than
treating them as six alternative, competing hypotheses, all of precisely equal
merit. Thus symmetries must be represented explicitly. Every symbol type has a
symmetry group, which we assume to be finite.

For example, the $line$ symbol type has two symmetries: the identity
transformation and rotation by $\pi$. Look back at figure~7. The symbol token
$\sigma_1$ of type $A$ expects its three line parts to be oriented in a certain
way relative to it; for example, it may expect the crossbar to be oriented
left-to-right; yet the line token $\sigma_3$ that plays the role of the crossbar
may actually be oriented the other way round. If so, then if we calculated the
penalty of $u(\sigma_3)\inv \cdot u(\sigma_1)$ against the fleximap we would get
a spuriously high value. We need a way of saying that rotations by $\pi$ are to
be disregarded when calculating the penalty. To allow for this, it is convenient
to give node tokens embeddings as well as symbol tokens. Recall that the
part-whole relationship between $\sigma_3$ and $\sigma_1$ is represented by a
node token $n$ with $P(n) = \sigma_3$ and $W(n) = \sigma_1$. We give $n$ an
embedding $u(n)$, which is the same as $u(\sigma_3)$ except that it is oriented
the expected way round, considered as a part of $\sigma_1$. That is,
$u(\sigma_3) = u(n) \cdot s$, where $s$ is either the identity or rotation by
$\pi$. It is $u(\sigma_3)\inv \cdot u(n)$, rather than $u(\sigma_3)\inv \cdot
u(\sigma_1)$, that we take as the part-whole relationship in the calculation of
the penalty relative to the part-whole fleximap.

To specify a symmetry we must provide

\blob an automorphism $a:\N_0\to\N_0$ of the grammar (i.e., an isomorphism onto
itself),

\blob an affine transformation $s(\sigma)$ to be applied to every symbol type
$\sigma$ (it must be one of $\sigma$'s symmetries),

\blob an affine transformation $s(n)$ to be applied to every node type $n$
(it must be one of $P(n)$'s symmetries),

\medskip\f
such that the grammar is unchanged when the automorphism is applied to it and
all fleximaps are transformed using the $s(\sigma)$ and $s(n)$ functions. The
symmetry itself is the pair $(a,s)$. (See [49, \S6.4,\S7.5] for
precise details.) For example, in one symmetry the hexagonal grid symbol type
$hex$ (see \S3.3) is rotated by $\frac\pi3$, some of its nodes are rotated by
$\pi$ (but $line$ is not), and the automorphism maps $hex$ to $hex$, $line$ to
$line$, nodes 1,2,3 to nodes 2,3,1, and edges $a,b,c,d,e,f$ to $b,c,a,e,f,d$,
respectively. Such symmetries of the grammar can then be applied to the pattern
$\N_1$ with a parse $p:\N_1\to\N_0$ and a set of embeddings; a \i{local}
symmetry of $\N_1$ consists of applying a separate symmetry to each symbol,
producing a new parse and a new set of embeddings.  For example, a symmetry can
be applied to a single token of type $hex$ and its subsymbols, nodes and edges,
leaving everything else unchanged [50, \S4.3].

\subsubhead{5.5 Embedding tokens and types}
It is convenient to bundle together all the embedding transformations into a
single mathematical object, the function $u:\Sigma_1\cup N_1 \to \G$ (where
$\N_1=\network{_1}$ is the pattern). This is called an \i{embedding token} for
$\N_1$. It is subject to the constraint
$$ \forall n,n^* \from N_1\ (G_1(n,n^*) \Rightarrow u(n)=u(n^*)) $$
i.e., if two nodes are glued then they have the same embedding.

It is also convenient to bundle together all the fleximaps that constrain
these embedding transformations
into a single mathematical object, called an \i{embedding
type} for $\N_1$. This is a sextuple $v_1=(sub_1,\allowbreak con_1,\allowbreak
rel_1,\allowbreak symm_1,\allowbreak tem_1, \allowbreak in_1)$, consisting of functions
\newline ${sub_1:\comp{(\sigma,\sigma^*) \from
\Sigma_1\times\Sigma_1}{G_1(\sigma,\sigma^*)} \to \Flex}$, $con_1:N_1\to\Flex$,
$rel_1:E_1\to\Flex$, $symm_1:\Sigma_1\to Sub(\G)$, $tem_1:\Sigma_1\to Tem$, and
$in_1:\Sigma_1\to\Flex$, where $\Flex$ is the set of all fleximaps, $Sub(\G)$ is
the set of all subgroups of $\G$, and $Tem$ is the set of all templates. These
are interpreted as follows.

For any supersymbol $\sigma$ and subsymbol $\sigma^*$,
$sub_1(\sigma,\sigma^*)$ is the fleximap defining the variable geometric
relationship between them.

For any node $n$, $con_1(n)$ is the fleximap defining the geometric
relationship between the part $P(n)$ and the whole $W(n)$, or, more
accurately, the relationship between the node and the
whole, as explained in the previous subsection.

For any edge $e$, $rel_1(e)$ is the fleximap defining the geometric
relationship between the nodes $A(F(e))$ and $A(S(e))$.

For any symbol $\sigma$, $symm_1(\sigma)$ is the symmetry group of $\sigma$,
and $tem_1(\sigma)$ is its template; $in_1(\sigma)$ is a fleximap $(I,g)$ whose
nominal part $I$ is just the identity affine transformation, but whose metric
$g$ is the \i{inertial} metric of $\sigma$, which is used to determine how 
$\sigma$ moves in response to forces (see \S7.3).

But where does this embedding type $v_1$ for $\N_1$ come from? We are given an
embedding token $v=(sub,\allowbreak con,\allowbreak rel,\allowbreak
symm,\allowbreak tem, \allowbreak in)$ for the grammar $\N_0$, which captures
all the geometric information associated with the grammar. We can transfer it
to the pattern using the homomorphism $p$, giving an embedding type $v_1 = v
\circ p$ for $\N_1$, defined by
$$ v \circ p
= (sub_1,con \circ p,rel \circ p,symm \circ p,tem \circ p,in \circ p). $$
where
$$ \forall \sigma,\sigma^*\from\Sigma_1 \quad (G_1(\sigma,\sigma^*) \
\Rightarrow\
sub_1(\sigma,\sigma^*) = sub(p(\sigma),p(\sigma^*))). $$
$v_1$ is called the \i{induced} embedding type on $\N_1$. It is $v_1$ that we
compare with the embedding token $u$, as just described.

Embedding tokens are further constrained by a symmetry condition. If $u$ is 
the embedding token for the pattern $\N_1$ and $v_1=v \circ
p=(sub_1,\allowbreak con_1,\allowbreak rel_1,\allowbreak symm_1,\allowbreak
tem_1, \allowbreak in_1)$ is the embedding type for $\N_1$, the \i{symmetry condition} for $\N_1,u,v_1$ is 
$$ \forall n\from N_1 \quad u(P_1(n))\inv \cdot u(n) \in symm_1(P_1(n)). $$
This states
that each node $n$ is embedded in the image plane in the same way as the part
$P_1(n)$, up to a symmetry of $P_1(n)$ (as explained in the previous
subsection).  This symmetry condition will be
imposed throughout the recognition process.

%% file: 6a.tex
\subhead{6. Templates and the definite match function}

\subsubhead{6.1 Templates}
So far I have said nothing about the relationship between the image and the
pattern. It is only at this point that we need to make any assumptions about
the nature of the raw data, which I have called the \ic{image}. Assume in this
section that the image is a rectangular array of monochrome pixels.  Formally,
an \i{image} is a function $I:\R^2 \to \closeopen{0,\infty}$. For any point
$p\in\R^2$, $I(p)$ is the image intensity at the pixel $p$. The domain of $I$
is called the \i{image plane}.

Recall figure~7 in \S5.2. Formally, a template is a differentiable
function $T:\R^2 \to \closeopen{0,\infty}$ such that the set
$\comp{x\from\R^2}{T(x) > 0}$ is bounded. The domain of $T$ is called the
\i{template plane}. Each token
$\sigma$ of this type has an embedding transformation $G=u(\sigma)$, mapping from
the template plane to the image plane.

We shall define a measure $\rho_{I,T}(G)$, called the \i{correlation
function}, of how well a template $T$ matches
the image (transferred into the template plane) $I\circ G$,
and will seek to choose $G$ to maximise it. Define
$$ \rho_{I,T}(G) = |det(G)| \int T(u) \, (I(G(u))-I_0) \d^2u =
\int T(G\inv(x)) \, (I(x)-I_0) \d^2x $$
where $I_0$ is a positive real constant associated with $T$. The integrals are
over the whole of $\R^2$, or equivalently over a large enough region to include
in its interior all the points where $T$ is non-zero.

Each symbol type has a template (this is provided as part of the embedding type
$v$, see \S5.5).  This is transferred across to the pattern when we form $v
\circ p$, so now we can say that each symbol token $\sigma$ has a template,
$tem(\sigma)$, which is equal to the template of its symbol type.

There is one complication. We wish to prevent two identical symbols from
forming at the same place in the image, or more generally to discourage two or
more symbols from claiming credit for the same patch of the image. If two or
more symbol tokens' templates overlap in the image plane, i.e., if there are
points $x$ in the image plane where $tem(\sigma)(u(\sigma)\inv(x)) > 0$ for two
or more symbol tokens $\sigma$, then the point $x$ should become \i{saturated}
and the contribution it makes to the correlation function should be reduced to
penalise the overlapping symbol tokens. For this it is necessary to calculate
the saturation $sat(x)$ of each point $x$, which is roughly the sum of
$tem(\sigma)(u(\sigma)\inv(x))$ over each symbol token $\sigma$. (However, this
needs modification for subsymbols to avoid double-counting.) The formal
definition of saturation follows.

Suppose we are given a pattern $\N_1=\network{_1}$, an embedding token $u$ for
$\N_1$, and a function $tem:\Sigma_1\to Tem$, then $sat:\R^2\to\R$, is defined by 
$$ \forall x\from\R^2\quad sat(x) = \sum_{\sigma\in\Sigma_1} (1-g_\sigma) \cdot
tem(\sigma)(u(\sigma)\inv(x)). $$
where $g_\sigma =
\bigl|\comp{\sigma^*\from\Sigma}{G_1(\sigma^*,\sigma)}\bigr|$.
We call $sat(x)$ the \i{saturation at} the image point $x$.

The correlation function is now modified to take account of saturation:
$$ \rho_{I,T,sat}(G)
= |det(G)| \int w(sat(G(u))) \, T(u) \, (I(G(u))-I_0) \d^2u
= \int w(sat(G)) \, T(G\inv(x)) \, (I(x)-I_0) \d^2x $$
where $w$ is a weighting function that suppresses the integrand at points $x$
where $sat(x)$ is above 1. A suitable definition of $w$ is
$$ \forall s\from\R\quad w(s) =
\cases{
1 & if $s \le 1$, \cr
(1.6-s)/0.6 & if $1 < s < 1.6$, \cr
0 & if $1.6 \le s$.
}$$

We also define \i{mass}, $m$, of a symbol token, with template $T$, 
and which is embedded in the image $I$ by $G$, by
$$ m = |det(\rep G)| \int T(u) \d^2u = \int T(G\inv(x)) \d^2x. $$
This is used when calculating how much the symbol token moves in response to a
force (see \S7.3 below).

\subsubhead{6.2 The definite match function}
We are now in a position to state the \i{definite match function}, which
measures how well a definite pattern $\N$ matches an image $I$, and how well $\N$'s
embedding token $u$ matches an embedding type $v$ for $\N$. It is the aim of
recognition to construct a definite pattern that maximises the definite match
function.

The definite match function $DM$ is defined by
$$\eqalign{
DM(I, \N, u, v)
=& \sum_{\sigma\in\Sigma} \rho_{I,tem(\sigma),sat}(u(\sigma))
\ - \ \theta |\Sigma\setdiff P(N)|
\ - \sum_{(\sigma,\sigma^*)\mid G(\sigma,\sigma^*)} \kern-15pt
E_{sub(\sigma,\sigma^*)}\bigl(u(\sigma)\inv \cdot u(\sigma^*)\bigr)
\cr&
\ - \ \sum_{n\in N} E_{con(n)}\bigl(u(W(n))\inv \cdot u(n)\bigr)
\ - \ \sum_{e\in E} E_{rel(e)}\bigl(u(A(S(e)))\inv \cdot u(A(F(e)))\bigr)
}$$
where $\N=\network{}$, $v=(sub,con,rel,symm,tem,in)$, 
$\theta$ is a positive real constant and $sat$ is the saturation
function, as defined in the previous section. 

The first term on the right-hand side is a sum over all symbols; $tem(\sigma)$
is the template for $\sigma$, and $\rho_{I,tem(\sigma),sat}(u(\sigma))$
measures the correlation between the template (embedded in the image using
$u(\sigma)$) and the image $I$.

The second term on the right-hand side applies a fixed penalty of $\theta$ for
each \ic{bare} symbol, i.e., each symbol that is not a part of another symbol.
This term encourages the symbols to connect themselves together rather than
remaining separate.

The third term measures how well each subsymbol matches its supersymbol
geometrically. The summation
is over all pairs $(\sigma,\sigma^*)$ such that $\sigma$ is a supersymbol of
$\sigma^*$; $sub(\sigma,\sigma^*)$ is a fleximap that defines what the 
geometric relationship between $\sigma^*$ and $\sigma$
should be; $u(\sigma)\inv \cdot u(\sigma^*)$ is the actual relationship
between them; the quadratic penalty function $E_\tau(G)$ calculates the
penalty for the deviation between an affine transformation $G$ and a 
fleximap $\tau$, as defined in \S5.2.

The fourth term measures how well each part matches its whole geometrically. The summation
is over all nodes $n$; $P(n)$ is the part symbol and $W(n)$ is the whole
symbol. The node $n$ has its own embedding $u(n)$, which equals $u(P(n))$ up
to a symmetry; $u(W(n))\inv \cdot u(n)$ is the actual geometric relationship
between the part (or rather the node) and the whole; $con(n)$ is a fleximap
specifying what the relationship should be.

The final term measures how well each pair of siblings match geometrically.
The sum is over every edge $e$, representing a sibling relationship between two
nodes $A(F(e))$ and $A(S(e))$; $u(A(S(e)))\inv \cdot u(A(F(e)))$ is the actual
relationship and $rel(e)$ is the fleximap specifying what the relationship
should be.

All these fleximaps are provided by the embedding type $v$; the final
component $in$ of $v$ is not used yet.

The $DM$ function is invariant under application of affine transformations to the
symbols' internal frames of reference, under local symmetries, and under affine
transformations of the image of determinant $1$ 
[50, theorems~28-31].

The $DM$ function is used as follows. We are given an embedding type $v$ for
the grammar, we transfer it across to an embedding type $v\circ p$ for the
pattern, and $DM$ allows us to compare this with the embedding token $u$ that
we have constructed for the pattern, while also comparing the pattern $\N_1$
with the image $I$.

\subsubhead{6.3 Full statement of the recognition problem}
We can now refine
our statement of the recognition problem in \S4.3.  Given a semi-definite
network $\N_0$ (the \i{grammar}), an embedding type $v$ for $\N_0$, and an
image $I$, the task is construct a definite network $\N_1$ (the \i{pattern}),
an embedding token $u$ for $\N_1$ (specifying how everything in the pattern is
embedded in the image), and a homomorphism $p:\N_1\to\N_0$ (the \i{parse}),
maximising $DM(I,\N_1,u,v\circ p)$, subject to the symmetry condition for
$\N_1,u,v\circ p$.

%% file: 7a.tex
\subhead{7. Inclusion functions and the indefinite match function}
We now turn to the algorithm for solving the recognition problem. A pattern
$\N_1$ is constructed incrementally. While it is under construction it is
indefinite, meaning that alternative grammatical interpretations co-exist for
different portions of the image. Some alternatives are evaluated as better
than others. Each symbol token $\sigma$ has an \i{inclusion value} $i(\sigma) \from
[0,1]$, which is the algorithm's degree of confidence in $\sigma$, where $i(\sigma)=1$ means that $\sigma$ is definitely correct and
$i(\sigma)=0$ means that $\sigma$ is definitely wrong and will be pruned.
Similarly each node token $n$ and each edge token $e$ has an inclusion value
$i(n)$ and $i(e)$. The function $i$ is called an \i{inclusion function} and
is accompanied by a second inclusion function $j$, explained below. 

By the end of the recognition process, all inclusion values have been driven to
the extremes, $0$ or $1$, and everything with inclusion value
$0$ has been pruned.

\subsubhead{7.1 Definition of inclusion functions}

\defn A pair of \i{inclusion functions} on a network $\N=\network{}$ is defined
as $(i,j)$, where $i:\Sigma\cup N \cup E \to [0,1]$ and $j:\Sigma\cup N\cup H
\cup (K\setdiff dom(G)) \to [0,1]$, such that
$$\eqalignno{
\forall \sigma\from\Sigma \quad i(\sigma)
&= j(\sigma) \ + \sum_{n\in P\preo\sigma} \hskip-3mm (1-g_n)i(n),
\qquad\text{where $g_n = \bigl|\comp{n^*\from N}{G(n^*,n)}\bigr|$} \qquad & (1)
\cr
\forall n \from N \quad i(W(n)) &= j(n) + i(n) & (2)
\cr
\forall h\from H \quad i(A(h)) &= j(h) \ + \sum_{e\in F\preo h} \hskip-3mm i(e)
\ + \sum_{e\in S\preo h} \hskip-3mm i(e) & (3)
\cr
\forall k \from K\setdiff dom(G) \quad
i(C(k)) &= j(k) + \sum_{k^* \mid G(k,k^*)} i(C(k^*)). & (4)
}$$
(The summation notation in (4) means a sum over all $k^*$ such that $G(k,k^*)$.)

This definition may be interpreted informally as follows. (In the following
explanation I shall say \ic{should} or \ic{is correct} to state what will hold or
exist at the end of recognition when the pattern is definite, and a simple
\ic{is} for what is true during recognition when the pattern is indefinite.)

Line (1). Each symbol $\sigma$ should either be a \ic{bare} symbol (with $P\preo\sigma
=\emptyset$) or a part of one larger symbol (with $|P\preo\sigma| = 1$). 
Hence the nodes presently in $P\preo\sigma$ are in competition with one
another; at most one can be correct. $i(\sigma)$ is
interpreted as the degree of confidence in $\sigma$, $j(\sigma)$ is the degree
of confidence that $\sigma$ should be a bare symbol, and $i(n)$
is the degree of confidence in $n$. An exception to this competition is that if
$n$ is a subnode of $n^*$ (i.e., $G(n^*,n)$) then they may both be correct; 
the $1-g_n$ factor allows for this co-existence.

In line (2), $i(W(n))$ is the degree of confidence in $W(n)$, $j(n)$
is the degree of confidence that $W(n)$ is correct but $n$ is not, and
$i(n)$ is the degree of confidence in $n$.

Line (3) expresses the fact that each hook $h$ should have a single edge
incident to it (i.e., $|F\preo h| + |S\preo h| = 1$); hence the edges presently
in $F\preo h$ and $S\preo h$ are in competition with one another. Thus
$i(A(h))$ is the degree of confidence in the node or symbol $A(h)$, $i(e)$ is
the degree of confidence in $e$, and $j(h)$ is the degree of confidence that
$A(h)$ is correct but that none of its present edges is (i.e., the correct edge has
yet to be created).

Line (4) expresses the fact that each facet $k$ that is not itself a sub-facet
should be glued to a single sub-facet. Hence if $k$ is presently glued to several
sub-facets then they are in competition with one another; $i(C(k^*))$ is the
degree of confidence in the edge $C(k^*)$, $j(k)$ is the degree of confidence
that $C(k)$ is correct but that none of $k$'s present sub-facets is.

At the end of recognition all surviving symbols, nodes and edges will have
inclusion values $1$ and the pattern will be definite.

\theorem 32 (from [50]). If $\N=\network{}$ is a definite network and $(i,j)$ is a pair of
inclusion functions on $\N$ satisfying $\forall x\from\Sigma\cup N \cup E\ i(x)
= 1$, then
$$ \forall\sigma\from\Sigma\ j(\sigma) =
\cases{0 & if $\sigma \in P(N)$\cr
1 & otherwise}, \quad
\forall n\from N\ j(n) = 0, \quad
\forall h\from H\ j(h) = 0, \quad
\forall k\from K\setdiff\dom(G)\ j(k) = 0. $$

\subsubhead{7.2 The indefinite match function, $IM$}
We can now generalise the definite match function $DM$ to a function
applicable to indefinite patterns. First we generalise the
(definite) saturation function defined in \S6.1 to the \i{indefinite}
saturation function $sat:\R^2\to\R$, defined by
$$ \forall x\from\R^2\quad
sat(x) = \sum_{\sigma\in\Sigma} (1-g_\sigma) \, i(\sigma) \,
tem(\sigma)(u(\sigma)\inv(x)). $$
(The difference is that every symbol $\sigma$ is now weighted by $i(\sigma)$.)
Then the indefinite match function $IM$ is defined by
$$\eqalign{
& \kern-10pt IM(I, \N, u, v, i, j, B) = \cr
& \sum_{\sigma\in\Sigma}
\bigl( i(\sigma) \rho_{I,tem(\sigma),sat}(u(\sigma)) - j(\sigma) B(\sigma)
\bigr)
-\sum_{(\sigma,\sigma^*)\mid G(\sigma,\sigma^*)} \kern-15pt i(\sigma^*)
E_{sub(\sigma,\sigma^*)}(u(\sigma)\inv \cdot u(\sigma^*))
\cr&
- \sum_{n\in N} i(n) E_{con(n)}(u(W(n))\inv \cdot u(n))
- \sum_{h \in H} j(h) B(h)
- \sum_{k \in K\setdiff\dom(G)} \kern-12pt j(k) B(k)
\cr&
- \sum_{e\in E} i(e)
\bigl( E_{rel(e)}(u(A(S(e)))\inv \cdot u(A(F(e)))) +
E_{in(W(A(F(e))))}(u(W(A(S(e))))\inv \cdot u(W(A(F(e))))) \bigr).
}$$
Here, $I$ is the image, $\N=\network{}$ is the network we are comparing with the image,
$u$ is the embedding token that specifies how $\N$ is embedded in $I$, $v=
(sub,\allowbreak con,\allowbreak rel,\allowbreak symm,\allowbreak
tem, \allowbreak in)$ is
the embedding type against which $u$ is evaluated, $(i,j)$ are the inclusion
functions, and $B$ is called the \i{bareness function}.

The terms of $IM$ are much like the terms of $DM$, except that everything is
weighted by its inclusion value. The fixed penalty $\theta$ levied for each
bare symbol is replaced by a variable penalty $B(\sigma)$; it is weighted by
$j(\sigma)$ because $j(\sigma)$ is the degree of confidence that $\sigma$ is
bare. There is a new penalty $B(h)$ for a bare hook (i.e., a hook with no
incident edges); this is weighted by the degree of confidence $j(h)$ that $h$
is bare. Similarly there is a new penalty $B(k)$ for a bare facet (a facet
that should have a sub-facet but does not); this is weighted by the
degree of confidence $j(k)$ that $k$ is bare.

All these variable penalties are specified by the bareness function
$B:\Sigma\cup H\cup (K\setdiff\dom(G))\to\closeopen{0,\infty}$. 

The final term ($E_{in(W(A(F(e))))}(\ldots)$) is new. Normally, for any edge $e$, we have
$W(A(F(e)))=W(A(S(e)))$, which means that the two nodes at each end of
$e$ belong to
the same whole (this is called a \i{coherent} edge). However, incoherent edges
are temporarily permitted during recognition. They may occur, for example,
with the hexagonal grids in \S10.1. There may be genuine uncertainty about
whether two overlapping hexagonal grids should be merged or kept separate.
Incoherent edges may form between a node of one grid and a node of the other.
If so, this is penalised in $IM$ by using a metric tensor $in(\sigma)$, where $in$ is
taken from $v$ and $\sigma$ is one of the two grids. This term penalises
incoherent edges; it is $0$ if the edge is coherent. As we are about to see,
each term in $IM$ gives rise to forces that change the symbols' embeddings.
This final term exerts a force that draws the two wholes together; eventually, either
they
will come close enough to be merged or the incoherent edges will be
removed. At the end all edges will be coherent. 

The next theorem shows that $DM$ is a special case of $IM$.

\theorem 33 (from [50]). If

\item{(a)} $\N$ is definite,

\item{(b)} $\forall x\from\Sigma\cup N \cup E\ i(x) = 1$,

\item{(c)} $\forall\sigma\from\Sigma\setdiff P(N)\ B(\sigma)=\theta$,

\smallskip\f
then $IM(I,\N,u,v,i,j,B) = DM(I,\N,u,v)$.

\medbreak\f
When we use $IM$ during recognition, the network we take is the
pattern $\N_1$, and the embedding type we take is $v \circ p$, i.e., the given
embedding type $v$ for the grammar $\N_0$ transferred across to $\N_1$ using the
parse $p:\N_1\to\N_0$. So it is $IM(I,\N_1,u,v\circ p,i,j,B)$ that we shall be
maximising. At the end of the recognition process, when the conditions of
theorem~33 will hold, we shall have maximised $DM(I,\N_1,u,v)$, as required.

\subsubhead{7.3 Adjustment of the embedding token}
During recognition the embeddings of all the symbols and nodes are continually
adjusted to maximise $IM(I,\N_1,u,v\circ p,i,j,B)$. Consider making a small
change to the embedding token $u \mapsto u \cdot \Delta u$, where
$$\eqalign{
\forall\sigma\from\Sigma\ \ \Delta u(\sigma) &= \exp(\epsilon V_\sigma) \cr
\forall n\from N\ \ \Delta u(n) &= \exp(\epsilon V_n) \cr
}$$
where the increments $V_\sigma,V_n \in \A$ (recall that $\A$ is the Lie algebra
of the affine group). Then, developing the change in the value of $IM$ to first
order,
$$ IM(I, \N, u \cdot \Delta u, v, i, j, B) =
IM(I, \N, u, v, i, j, B) +
\epsilon \sum_{\sigma\in\Sigma} F_\sigma(V_\sigma) + o(\epsilon). $$
Here $F_\sigma \in \F$, the dual vector space to $\A$. (The theory of $\A$ and
$\F$ is developed fully in [49].) Elements of $\F$ are called
\i{forces}. Note that the change depends only on the $V_\sigma$ increments,
not also on the $V_n$ increments, because they are tightly related by the symmetry
condition.

The forces can be understood informally as follows. Each term in $IM$
generates (through its derivative) a force at a symbol, node or edge, which 
propagates through the network; the forces reaching any symbol $\sigma$ are
summed to give the total force $F_\sigma$ acting on $\sigma$ (see
[49, \S7.6] and [50, \S6.5]).

We then apply gradient ascent, choosing our increment $\Delta u$ to increase
the value of $IM$. To make this well-defined we must specify the cost of
making the adjustment; this requires a choice of an \i{inertial} metric tensor
$in(\sigma)$. The cost is defined as $\frac12\epsilon \sum_{\sigma\in\Sigma} 
m_\sigma \, in(\sigma)(V_\sigma)(V_\sigma)$, where $m_\sigma$ is the mass of
$\sigma$, defined in \S6.1. It is a simple exercise [49, \S7.7]
to show that the optimal choice of increment is
$$ \forall\sigma\from\Sigma\ \ V_\sigma = \frac1{m_\sigma}
in(\sigma)\inv(F_\sigma). $$
Thus the role of the inertial metric $in(\sigma)$ is to convert (covariant)
forces into (contravariant) increments. The increments $V_n$ for nodes $n$ are
then determined by the symmetry condition. This determines the change 
$u \mapsto u \cdot \Delta u$ in the embedding token (up to an arbitrary choice
of the positive constant $\epsilon$). This incremental adjustment, continually
repeated, optimises the embedding of the pattern in the image.

The inertial metrics $in(\sigma)$ are determined by the function $in$,
provided as part of the embedding type. (As we saw in the last section
$in(\sigma)$ also has a secondary use, in the penalty for incoherent edges in the
$IM$ function; this is separate from its primary use here.)

\subsubhead{7.4 How the inclusion functions are determined}
The inclusion functions $i,j$ are continually recalculated during the
recognition process by a simulated annealing process, governed by a
\i{temperature} parameter that varies across the pattern and with time. Where the structure of
the pattern is changing, the temperature is high, and this makes the inclusion
functions take on mid-range values, so several alternative interpretations can
co-exist in parallel; when structural changes stop, the temperature
declines and the inclusion functions are pushed towards $0$ or $1$, and so a
choice is forced between alternatives. Every symbol token, node token or edge
token with inclusion value $0$ is pruned. Eventually everything has inclusion
value $1$ and the pattern becomes definite.

For the purposes of this section only, we reformulate $i,j$ and the constraints 
on them in a vector notation in order to emphasise their linear nature.
Given a pair of inclusion functions $(i,j)$ on a
network $\N=\network{}$, we shall convert $(i,j)$ into the \i{inclusion vector}
$\bold i$ on $\N$, with components $\bold i_x$, for all $x \in X$, where
$$ X = \Sigma \times \set{0,1} \ \cup\ N \times \set{0,1} \ \cup\ H \ \cup\ E
\ \cup\ K\setdiff dom(G). $$
Each component of $\bold i$ represents one value of $i$ or $j$, as follows.
$$\eqalign{
\forall\sigma\from\Sigma \quad \bold i_{(\sigma,0)} &= i(\sigma), \quad
\bold i_{(\sigma,1)} = j(\sigma) \cr
\forall n \from N \quad \bold i_{(n,0)} &= i(n), \quad \bold i_{(n,1)} = j(n) \cr
\forall h \from H \quad \bold i_h &= j(h) \cr
\forall e \from E \quad \bold i_e &= i(e) \cr
\forall k \from K\setdiff dom(G) \quad \bold i_k &= j(k) \cr
}$$
The constraints on $i$ and $j$ in \S7.1 are linear and so can be expressed as 
a set of linear conditions on $\bold i$ of the form
$$ \forall y \from Y \quad \bold c^y \cdot \bold i
= \sum_{x\in X} \bold c^y_x \bold i_x = 0 $$
using a set of vectors $\bold c^y$, for all $y\from Y$, where
$$ Y = \Sigma \,\cup\, N \,\cup\, H \,\cup\, K\setdiff dom(G) $$
and $\bold c^y$ has components $\bold c^y_x$ for $x\from X$.

The $IM$ function is also linear in $\bold i$
(if we disregard the dependence of the saturation function on $\bold i$) and
so can be written in the form 
$$ IM(I, \N, u, v, i, j, B) = \bold i \cdot \bold m = \sum_{x\in X} \bold i_x
\bold m_x $$
where $\bold m$ is a vector whose components $\bold m_x$ depend on
$I,\N,u,v,B$.

The vector $\bold i$ is determined by maximising the expression
$$ E = \sum_{x\in X} \frac{\bold i_x \bold m_x}{T_x}
-\sum_{x\in X} \bigl(
\bold i_x \ln \bold i_x + (1-\bold i_x) \ln (1-\bold i_x)
\bigr) $$
subject to the constraints $\forall y\from Y \ \bold c^y \cdot \bold i = 0$,
where each $T_x$ is a positive number, known as the \i{temperature} of $x$.
To be precise, every symbol, node, hook, edge and facet has a temperature, and 
we define $T_{(\sigma,0)} = T_{(\sigma,1)} = T_\sigma$ and $T_{(n,0)} = 
T_{(n,1)} = T_n$. The solution is
$$ \bold i_x
= \sigmoid\Bigl(
\frac{\bold m_x}{T_x} + \sum_{y\in Y} \lambda_y \bold c^y_x
\Bigr)
$$
where the sigmoid function $\sigmoid:\R \to (0,1)$ is defined by
$ \forall u\from\R\ \sigmoid(u) = \frac1{1+e^{-u}} $
and the $\lambda_y$ parameters are unknown Lagrange multipliers.
By examining the second derivatives it can be determined that this solution is
a local maximum of $E$. 

We can split each $\bold c^y$ vector into two vectors $\bold c^{y+}$ and 
$\bold c^{y-}$
by separating positive and negative components:
$$ \forall y\from Y\ \forall x\from X\quad \bold c^{y+}_x = \max(\bold c^y_x,0),
\quad \bold c^{y-}_x = \max(-\bold c^y_x,0), $$
so that the constraints may be written as $\forall y\from Y\ 
\bold c^{y+} \cdot \bold i = \bold c^{y-} \cdot \bold i$.
Also define
$$ \forall y\from Y\ \ 
C^y = \bigl(\max_{x\in X} \bold c^{y+}_x\bigr) + 
\bigl(\max_{x\in X} \bold c^{y-}_x\bigr). $$
The following iterative algorithm seeks values of $\lambda_y$ satisfying the
constraints.

\bigskip

\indent For each $y\from Y$, initialise $\lambda_y$ to its final value last time
this algorithm was run;

\smallbreak

\indent repeat

\smallbreak

\indent\indent for each $x\from X$, do
$\bold i_x \assign \sigmoid\bigl( \bold m_x/T_x + \sum_{y\in Y} \lambda_y \bold
c^y_x \bigr)$ $\mid$

\smallbreak

\indent\indent for each $y \from Y$, do $\displaystyle\lambda_y \assign
\lambda_y + \frac1{C^y}
\ln\Bigl(\frac{\bold c^{y-}\cdot \bold i}{\bold c^{y+}\cdot \bold i}\Bigr)$

\smallbreak

\indent until equilibrium;

\smallbreak

\indent for each $x\from X$, if $\bold i_x$ is very close to $0$ or $1$,
then round it to $0$ or $1$.

\bigbreak\f
In this algorithm the semicolon means sequential composition, the \ic{$\mid$} 
symbol means parallel composition, and the \ic{for each} loops are
parallel loops. This means that all the assignment statements in the
\ic{repeat} loop body
may be executed concurrently in any fair order. Note that this loop is by far
the most computationally expensive part of the entire recognition process, so the
high degree of parallelism is relevant from the point of view of time
complexity.

I have no proof of convergence for this, but a heuristic argument
[50, \S7.4] shows that it tends to bring the constraints
closer to satisfaction. It can only halt when all the constraints are
satisfied.

Thus the inclusion functions are chosen to maximise $E$. 
In the final stages of recognition, all the temperatures will converge to the
minimum allowed temperature $T_{min}>0$ and so each $\bold i_x$ is likely to
approach $0$ or $1$, and is then rounded to $0$ or $1$, giving
$$ E = \frac{IM(I, \N_1, u, v\circ p, i, j, B)}{T_{min}}. $$
Hence maximising $E$ ultimately maximises $IM(I, \N_1, u, v\circ p, i, j, B)$.
By theorem~33 this maximises $DM(I,\N_1,u,v\circ p)$ for the final pattern.

%% file: 8a.tex
\subhead{8. The structural operations}
This section describes the \i{structural operations} by which the pattern
is incrementally grown during recognition. There are four kinds:

\blob pruning operations,

\blob extension operations,

\blob merging two symbol tokens,

\blob partitioning a symbol token into two.

\medbreak\f
When such an operation is applied to the pattern $\N_1$, the parse
$p:\N_1\to\N_0$, the embedding token $u$ on $\N_1$, and the inclusion
functions $(i,j)$ are updated accordingly.

\subsubhead{8.1 Pruning operations}

\defn Given a network $\N=\network{}$ and a pair $(i,j)$ of inclusion
functions on $\N$, a \i{pruning operation} is a transformation from $\N$ to a
subnetwork $\N'=\network{'}$ such that
$$ \forall \sigma\from\Sigma\setdiff\Sigma'\ i(\sigma) = 0, \quad
\forall n\from N\setdiff N'\ i(n) = 0, \quad \forall e\from E\setdiff E'\ i(e)
= 0. $$
A pruning operation is \i{trivial} iff $\N'=\N$.

\medbreak\f
In practice we may confine ourselves to \i{elementary} pruning operations,
which consist of 

\blob \i{pruning a symbol $\sigma$} -- removing $\sigma$ and its subsymbols,
and all their dependent nodes, hooks, edges and facets;

\blob \i{pruning a node $n$} -- removing $n$ and its subnodes, and all their
dependent hooks, edges and facets;

\blob \i{pruning an edge $e$} -- removing $e$ and its subedges, and all their
facets.

\medbreak\f
A pruning operation, since it only removes things with inclusion value $0$,
does not alter the value of the $IM$ function [50, theorem~41].

\defn A network $\N=\network{}$ is \i{unprunable}, given a pair of
inclusion functions $(i,j)$ on $\N$, iff there is no proper subnetwork
$\N'=\network{'}$ of $\N$ satisfying
$$ \forall \sigma\from\Sigma\setdiff\Sigma'\ i(\sigma) = 0, \quad
\forall n\from N\setdiff N'\ i(n) = 0, \quad
\forall e\from E\setdiff E'\ i(e) = 0. $$

\medskip\f
(In other words, a network is unprunable iff no non-trivial pruning operation is
possible on it, or, equivalently [50, theorem~40], iff no 
elementary pruning operation is possible on it.)

\subsubhead{8.2 Extension operations}

\defn Given a grammar $\N_0$, a pattern $\N_1$, a homomorphism $p:\N_1\to\N_0$, and an
embedding token $u$ for $\N_1$, an \i{extension} of $(\N_1,p,u)$ is a triple
$(\N_1',p',u')$ where $\N_1$ is a subnetwork of $\N_1'$, $p':\N_1'\to\N_0$ is a
homomorphism such that $p = p'|_{\N_1}$, and $u'$ is an embedding token for
$\N_1'$ such that $u = u'|_{\N_1}$.

\defn $(\N_1',p',u')$ is a \i{minimal} extension of $(\N_1,p,u)$ satisfying a
condition $P$ iff

\smallskip

\item{(i)} $(\N_1',p',u')$ is an extension of $(\N_1,p,u)$ satisfying $P$;

\item{(ii)} for any extension $(\N_1'',p'',u'')$ of $(\N_1,p,u)$ satisfying $P$,
$|\N_1'| \le |\N_1''|$,

\smallskip\f
where the cardinality $|\N|$ of a network $\N$ is defined by $|\network{}| =
|\Sigma| + |N| + |H| + |E| + |K|$.

\defn An extension is \i{trivial} iff $\N_1'=\N_1$.

\medbreak\f
Using these notions we can define the extension operations used in the
algorithm. The notation is as usual: we have a grammar $\N_0=\network{_0}$
and an embedding type $v = (sub,\allowbreak
con,\allowbreak rel,\allowbreak symm,\allowbreak tem, \allowbreak in)$ on
$\N_0$, a pattern $\N_1=\network{_1}$ with a parse $p:\N_1\to\N_0$, an embedding
token $u$, a pair of inclusion functions $(i,j)$, and a bareness function
$B$ for $\N_1$; we shall produce an extended network $\N_1'=\network{_1'}$
with a new parse $p':\N_1'\to\N_0$ and embedding token $u'$. The
threshold $\theta$ is the positive constant used in the $DM$ function (see
\S6.2);
the thresholds $\theta_0,\theta_1,\theta_2,\theta_3,\theta_4$ are 
used only in the extension operations and may be made dependent on parameters
such as temperature.

In the extension operations the following additional conditions, referred to
collectively as \i{the extension conditions}, will be imposed.

\blob $\forall\sigma\from\Sigma_1\ ({P_1'}\preo\sigma \not\subseteq N_1 \imp 
j(\sigma) > \theta_0 \and B(\sigma) < \theta)$,

\blob $\forall h\from H_1\ ({F_1'}\preo h \cup {S_1'}\preo h
\not\subseteq E_1 \imp j(h) > \theta_0)$,

\blob $\forall e\from E_1'\ (A_1'(F_1'(e)) \notin N_1 \OR
A_1'(S_1'(e)) \notin N_1 \imp W_1'(A_1'(F_1'(e))) = W_1'(A_1'(S_1'(e))))$,

\blob $\forall n \in N_1'\setdiff N_1 \ \
E_{con(p'(n))}(u'(W_1'(n))\inv \cdot u'(n)) < \theta_1$,

\blob the symmetry condition for $\N_1',u',v \circ p'$.

\medskip\f
The \i{extension operations} are as follows (see figure~8 below). For each one,
the extension conditions are checked at the end and the operation is cancelled
if they do not hold.

\medskip\f
(a) (Joining two symbols.) Given symbols $\sigma_1,\sigma_2 \from \Sigma_1$, an
edge $e_0 \from E_0$, and affine transformations $s_1 \from symm(p(\sigma_1))$
and $s_2 \from symm(p(\sigma_2))$ such that

\blob $P_0(A_0(F_0(e_0))) = p(\sigma_1)$ and $P_0(A_0(S_0(e_0))) = p(\sigma_2)$,

\blob $E_{rel(e_0)}(s_2\inv \cdot u(\sigma_2)\inv \cdot u(\sigma_1) \cdot s_1)
< \theta_2$,

\medskip\f
construct a minimal extension $(\N_1',p',u')$ of $(\N_1,p,u)$ (if one exists)
such that $\N_1'$ contains nodes $n_1,n_2$ and an edge $e$ for which

\blob $p'(e)=e_0$, $A_1'(F_1'(e))=n_1$, $A_1'(S_1'(e)) = n_2$,
$P_1'(n_1) = \sigma_1$ and $P_1'(n_2) = \sigma_2$,

\blob $u'(n_1) = u(\sigma_1) \cdot s_1$ and $u'(n_2) = u(\sigma_2) \cdot s_2$.

\medskip\f
(b) (Joining three symbols.) Given symbols
$\sigma_1,\sigma_2,\sigma_3\from\Sigma_1$, edges $e_{01},e_{02}\from E_0$,
$s_1\from symm(p(\sigma_1))$, $s_2\from symm(p(\sigma_2))$ and $s_3\from
symm(p(\sigma_3))$ such that

\blob $A_0(S_0(e_{01}))=A_0(F_0(e_{02}))$, $P_0(A_0(F_0(e_{01})))=p(\sigma_1)$,
$P_0(A_0(S_0(e_{01})))=p(\sigma_2)$ and
$P_0(A_0(S_0(e_{02})))\allowbreak=p(\sigma_3)$,

\blob $E_{rel(e_{01})}(s_2\inv \cdot u(\sigma_2)\inv\cdot u(\sigma_1) \cdot s_1)
+ E_{rel(e_{02})}(s_3\inv \cdot u(\sigma_3)\inv \cdot u(\sigma_2) \cdot s_2) <
\theta_3$,

\medskip\f
construct a minimal extension $(\N_1',p',u')$ of $(\N_1,p,u)$ (if one exists)
such that $\N_1'$ contains nodes $n_1,n_2,n_3$ and edges $e_1,e_2$ for which

\blob $p'(e_1)=e_{01}$, $p'(e_2)=e_{02}$, $A_1'(F_1'(e_1)) = n_1$,
$A_1'(S_1'(e_1))=n_2=A_1'(F_1'(e_2))$, $A_1'(S_1'(e_2))=n_3$,
$P_1'(n_1)=\sigma_1$, $P_1'(n_2)=\sigma_2$ and $P_1'(n_3)=\sigma_3$,

\blob $u'(n_1)=u(\sigma_1)\cdot s_1$, $u'(n_2)=u(\sigma_2)\cdot s_2$ and
$u'(n_3)=u(\sigma_3)\cdot s_3$.

\medskip\f
(There are also variations of operation~(b), in which the directions of $e_{01}$
and $e_1$ are reversed, or the directions of $e_{02}$ and $e_2$ are reversed.)

\medskip\f
(c) (Extending from a hook.) Given a hook $h\from H_1$, construct a minimal
extension $(\N_1',p',u')$ of $(\N_1,p,u)$ such that

\blob $p'({F_1'}\inv(\{h\})) = F_0\inv(p(\{h\}))$ and $p'({S'_1}\inv(\{h\})) =
S_0\inv(p(\{h\}))$,

\blob $\forall e\from E_1'\setdiff E_1\ \ E_{rel(p'(e))}(u'(A_1'(S_1'(e)))\inv
\cdot u'(A_1'(F_1'(e)))) = 0$.

\medskip\f
(d) (Extending from a facet.) Given a facet $k\from K_1$ such that

\blob $W_1(A_1(F_1(C_1(k))))=W_1(A_1(S_1(C_1(k))))$

\blob $j(k) > \theta_0$,

\medskip\f
construct a minimal extension $(\N_1',p',u')$ of $(\N_1,p,u)$ such that

\blob $\overline{p'} \circ {G_1'}\inv \circ id_{\{k\}} = G_0\inv \circ
\overline p \circ id_{\{k\}}$.

\medskip\f
(e) (Extending from a part to a whole.) Given a symbol $\sigma\from\Sigma_1$,
construct a minimal extension $(\N_1',p',u')$ of $(\N_1,p,u)$ such that

\blob $p'({P_1'}\pre{\{\sigma\}}) = P_0\pre{p(\{\sigma\})}$,

\blob $\forall n\from N_1'\setdiff N_1\ E_{con(p'(n))}(u'(W_1'(n))\inv \cdot
u'(n)) = 0$.

\medskip\f
(f) (Filling in a missing part between two parts.) Given nodes $n_1,n_3\from
N_1$, hooks $h_1,h_3\from H_1$, edges $e_{01},e_{02}\from E_0$ and an affine
transformation $f$ such that

\blob $W_1(n_1)=W_1(n_3)$, $A_1(h_1)=n_1$, $A_1(h_3)=n_3$,
$A_0(S_0(e_{01}))=A_0(F_0(e_{02}))$, $F_0(e_{01})=p(h_1)$ and
$S_0(e_{02})=p(h_3)$,

\blob $E_{rel(e_{01})}(f\inv \cdot u(n_1)) + E_{rel(e_{02})}(u(n_3)\inv \cdot f)
< \theta_4$,

\medskip\f
construct a minimal extension $(\N_1',p',u')$ of $(\N_1,p,u)$ (if one exists)
such that $\N_1'$ contains edges $e_1,e_2$ and a node $n_2$ for which

\blob $p'(e_1) = e_{01}$, $p'(e_2) = e_{02}$, $F'_1(e_1) = h_1$,
$A'_1(S'_1(e_1))=n_2=A'_1(F'_1(e_2))$ and $S'_1(e_2) = h_3$,

\blob if $n_2\notin N_1$ then $u'(n_2)=f$.

\medskip\f
(There are also variations of operation~(f), in which the directions of $e_{01}$
and $e_1$ are reversed, or the directions of $e_{02}$ and $e_2$ are reversed.)

\medskip\f
(g) (Filling in a symbol between part and whole.) Given symbols
$\sigma_1,\sigma_3\from\Sigma_1$, symbols
$\sigma_{01},\sigma_{02},\allowbreak\sigma_{03}\from\Sigma_0$, nodes $n_{01},n_{02}\from
N_0$, $s\from symm(\sigma_{03})$ and an affine transformation $f$ such that

\blob $p(\sigma_1)=\sigma_{01}$, $p(\sigma_3)=\sigma_{03}$,
$W_0(n_{01})=\sigma_{01}$, $P_0(n_{01})=\sigma_{02}=W_0(n_{02})$ and
$P_0(n_{02})=\sigma_{03}$,

\blob $E_{con(n_{01})}(u(\sigma_1)\inv \cdot f) +
E_{con(n_{02})}(f\inv \cdot u(\sigma_3) \cdot s) < \theta_4$,

\smallskip\f
construct a minimal extension $(\N_1',p',u')$ of $(\N_1,p,u)$ such that $\N_1'$
contains a symbol $\sigma_2$ and nodes $n_1,n_2$ for which

\blob $p'(n_1)=n_{01}$, $p'(n_2)=n_{02}$, $W_1'(n_1)=\sigma_1$,
$P_1'(n_1)=\sigma_2=W_1(n_2)$ and $P_1'(n_2)=\sigma_3$,

\blob if $n_1 \notin N_1$ then $u'(n_1)=u'(\sigma_2)=f$,

\blob if $n_2 \notin N_1$ then $u'(n_2)=u(\sigma_3)\cdot s$.

\medskip\f
These extension operations are applied concurrently, in a fair order, controlled
by probabilities; the probability is low in cases where new symbols would be
created (particularly operations~(a) and (e)), to avoid the creation of too many
new symbols.

These operations are depicted in figure~8. As in previous figures, rectangles
represent symbols, circles represent nodes, small filled discs represent hooks, lines
with arrowheads halfway along represent edges, and small crosses represent
facets (shown only for operation~(e)); the $W$ and $P$ functions, which map each
node to the whole and part symbols, are depicted by green and blue arrows. 
For each operation solid lines are used for the symbols, nodes, etc.,
assumed to be present in the pattern before the extension operation; dashed
lines are used for the symbols, nodes, etc., added by the extension operation
(if they are not already present in $\N_1$). Thus, for example, operation~(g)
adds one symbol and two nodes (and their associated hooks), unless a suitable
symbol or suitable nodes already exist in $\N_1$. Note, however, that whenever
an edge is added the appropriate number of superedges must also be added, in
order that $p'$ satisfy the conditions for a homomorphism; and the same applies
to symbols, nodes, etc.; these are not shown in the figure.

\midinsert

\bigskip


\twofigures{30}{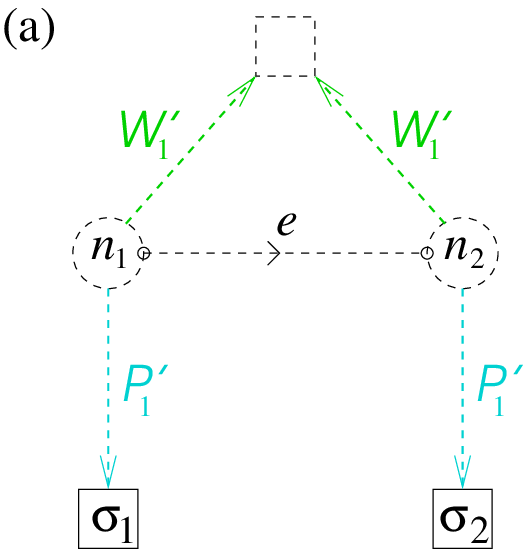}{50pt}{50}{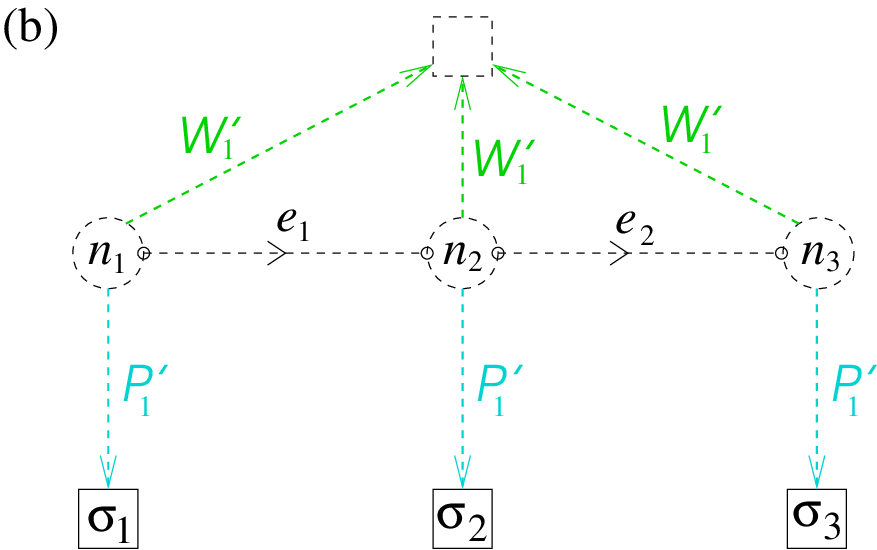}

\bigskip\null\bigskip

\twofigures{35}{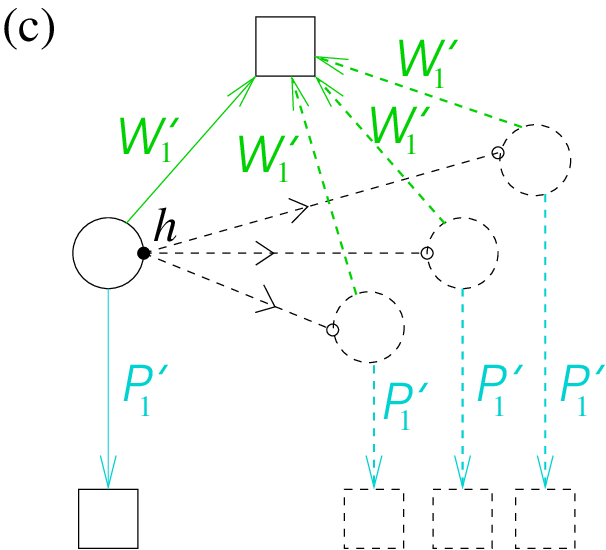}{50pt}{40}{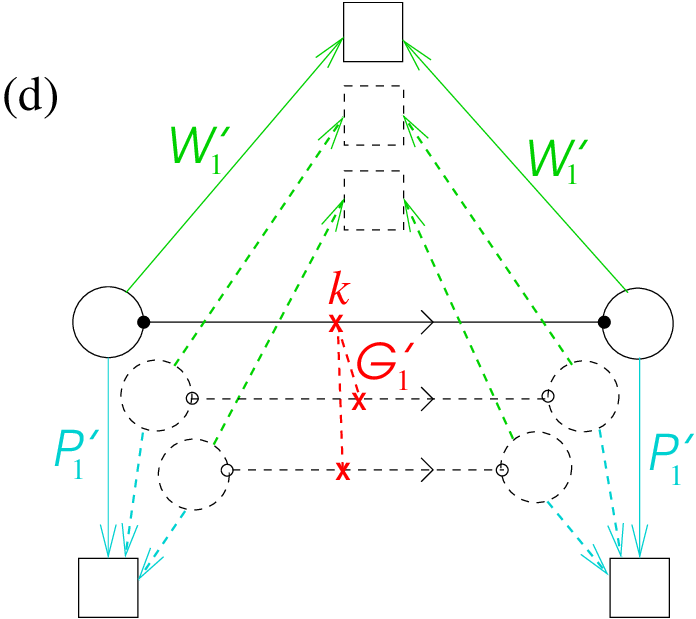}

\bigskip\null\bigskip

{\nobreak

\threefigures{26}{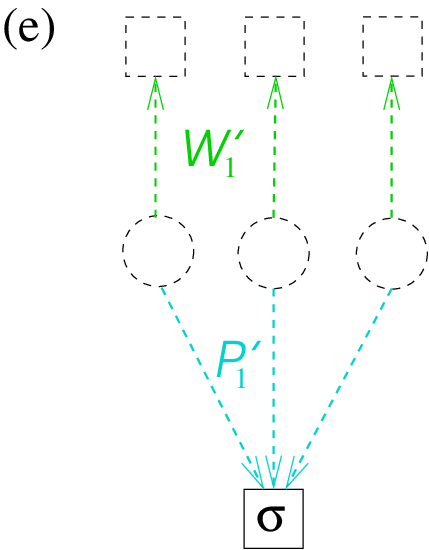}{35pt}{50}{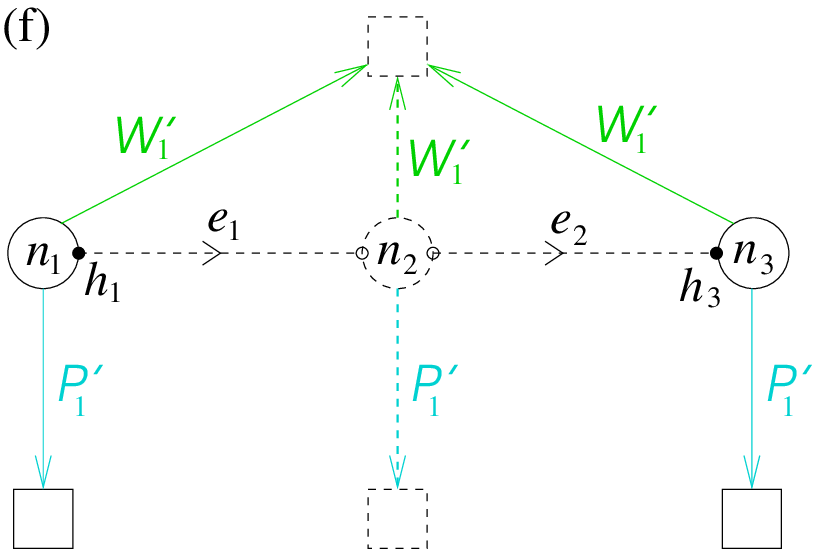}{15pt}{14}{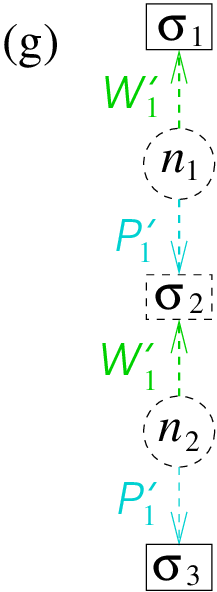}

\bigskip

\longcaption{Figure 8. The extension operations. The figure shows the relevant
parts of the pattern $\N_1'$ after each extension operation. Solid lines depict
what must be present before the operation; dashed lines depict what is added if
not already present.}

}

\endinsert

\defn $(\N_1,p,u)$ is \i{inextendable}, given $i,j,v,B$, iff none of the
extension operations can be applied to it, other than ones giving a trivial
extension.

\subsubhead{8.3 Merging two symbol tokens}
We may merge two symbol tokens of the same type that have similar embedding
transformations (up to a symmetry transformation). Two symbol tokens
$\sigma_1,\sigma_2\from\Sigma_1$ of type $\sigma_0\from\Sigma_0$ are considered
to have similar embeddings up to a symmetry transformation iff there exists
$s\in symm(\sigma_0)$ such that $E_{in(\sigma_0)}(s\inv \cdot u(\sigma_2)\inv
\cdot u(\sigma_1))$ is below a threshold. If this condition holds then the
symmetry $s$ is applied to $\sigma_2$ (this is a local symmetry operation; the
other symbols are unchanged); $\sigma_1$ and $\sigma_2$ are replaced by a single
symbol; and the nodes and edges of $\sigma_1$ and $\sigma_2$ are pooled.
(See [50, \S8.6] for the formal definition.)

\subsubhead{8.4 Partitioning a symbol token into two}
For any $\sigma_0\from\Sigma_1$, if the nodes and symbol in
$W_1\preo{\sigma_0}$ can be partitioned into two disjoint non-empty subsets
$T_1,T_2$,
such that there is no edge between any element of $T_1$ and any element of
$T_2$, then $\sigma_0$ may be replaced by two symbols, $\sigma_1,
\sigma_2$, with $\sigma_1$ getting the nodes of $T_1$ and $\sigma_2$ getting
the nodes of $T_2$. The subsymbols of
$\sigma_0$ are glued to $\sigma_1$ or $\sigma_2$ as appropriate (or
duplicated if necessary). The nodes, hooks, edges and facets above $\sigma_0$
must be duplicated. This operation is called \i{partitioning $\sigma_0$}
[50, \S8.7].
It is roughly the inverse of the merging operation.

\defn In a network $\network{}$, a symbol $\sigma_0\from\Sigma$ is 
\i{partitionable} iff there exist sets $T_1,T_2$ such that 
$T_1 \cup T_2 = W\preo{\sigma_0}$ and $T_1 \cap T_2 = \emptyset$ and
$T_1,T_2 \ne \emptyset$
and 
$F\pre{A\pre{T_1}} \cap S\pre{A\pre{T_2}} = \emptyset$ and 
$S\pre{A\pre{T_1}} \cap F\pre{A\pre{T_2}} = \emptyset$.
The network $\network{}$ is \i{partitionable} iff at least one symbol in
$\Sigma$ is partitionable; otherwise it is \i{unpartitionable}.

\medbreak\f
If no more partitions are possible then we are one step forward in attaining a
definite pattern, as the following theorem shows.

\theorem 46 (from [50]). If $\N=\network{}$ is an unpartitionable network satisfying the
condition $W \circ A \circ F = W \circ A \circ S$ then
$E \doublearrow{A \circ F}{A \circ S}
N\cup\Sigma \buildrel W \over \rightarrow \Sigma$ is a coequaliser
diagram in the category of sets.

%% file: 9a.tex
\subhead{9. The whole recognition process}
There are a few details to add, and then I shall summarise the whole
recognition algorithm. As usual, the grammar is $\N_0=\network{_0}$, the pattern
is $\N_1=\network{_1}$, and the parse is $p:\N_1\to\N_0$.

\subsubhead{9.1 Line operations}
To get the process started there are some operations that introduce symbol
tokens of type $line$ into the pattern (and possibly $bar$ symbol tokens too
-- see \S10). These are as follows.

\item{(a)} Create a line. A random search is used to find an initial embedding 
$u(\sigma)$ with a high value of $\rho_{I,T,sat}(u(\sigma))$

\item{(b)} Randomise a line's embedding, when it is bare and its inclusion value falls too low.

\item{(c)} Remove a bare line, if its temperature falls below a threshold.
This operation has a small probability. 
The purpose of this is to tidy up the pattern by removing excess lines.

\item{(d)} Glue two lines together, end to end: the two lines are replaced by
one; their bars are connected together, end to end.

\subsubhead{9.2 How temperature is determined}
Every symbol token, node token, hook token, edge token and facet token has a
temperature. These change continually by the
following three processes. (The notation $X \spread
Y$ means the assignment statement $Y \assign \max(X,Y)$, and $X \sspread Y$
means $X \spread Y$ and $Y \spread X$. Typical values for the spreading
parameters are $\beta=0.3$ and $\gamma=0.2$.)

\medskip

\item{1.} Every symbol, node, hook, edge or facet token created by a line
operation or an extension operation is given a high initial temperature.

\medskip

\item{2.} Temperature is spread through the pattern by applying the following
operations periodically.

\blob For every hook $h\from H_1$, do $T_{A_1(h)} \spread T_h$; $\beta T_h \spread
T_{A_1(h)}$.

\blob For every edge $e\from E_1$, do $T_{F_1(e)} \sspread T_e$; $T_{S_1(e)}
\sspread T_e$.

\blob For every node $n\from N_1$, do

\indent\indent $T_{P_1(n)} \spread T_n$; $\gamma T_n \spread T_{W_1(n)}$;

\indent\indent if $n\in\dom(G_1)$ then
$\gamma T_{W_1(n)} \spread T_n$ else $\gamma T_n \spread T_{P_1(n)}$.

\blob For every pair $k,k^*\from K_1$ such that $G_1(k,k^*)$, do $T_k \sspread
T_{k^*}$.

\blob For each $k\from K_1$, do $T_{C_1(k)} \sspread T_k$.

\medskip

\item{3.} Periodically each temperature $T$ declines by the formula
$$ T \assign a + \eta T $$
where the constant $\eta$ is slightly below $1$ and the constant $a$ is small
and positive.

\medskip\f
Consequently active regions of the pattern will have high temperature; regions
that have settled down will have temperature converging to $T_{min}=a/(1-\eta)$.

\subsubhead{9.3 How the bareness function $B$ is determined}
The purpose of the bareness function is to prevent the algorithm from getting
stuck by doing the same operations repeatedly. The bareness values
$B(\sigma),B(h),B(k)$ increase monotonically to the maximum $\theta$,
inhibiting repeated extension operations at the same place.

For each $h\from H_1$, $B(h)$ is initially $0$ and increases by a fixed amount
every time an extension operation is applied that adds edges to $F_1\preo h \cup
S_1\preo h$. This makes the algorithm more and more unwilling to remove all the
edges and leave the hook bare.

For each $k\from K_1$, $B(k)$ is initially $0$ and increases by a fixed amount
every time an extension operation is applied that adds facets to $\comp{k^*\from
K_1}{G_1(k,k^*)}$. This prevents the algorithm from repeatedly adding and
removing sub-facets to $k$ forever.

For each $\sigma\from\Sigma_1$, if $P_0\pre{p(\{\sigma\})} = \emptyset$ then
$B(\sigma)$ is initially set to $\theta$, and never changes thereafter. If 
$P_0\pre{p(\{\sigma\})} \ne \emptyset$ then $B(\sigma)$ is initially set to a 
positive value $\theta_0 < \theta$; $B(\sigma)$ is increased by a fixed amount 
every time extension operation (a), (b), (e) or (g) is applied that adds nodes to $P_1\preo\sigma$.
The increment is chosen so that $\theta-\theta_0$ is a multiple of the
increment. Ultimately every symbol token $\sigma$ has $B(\sigma)=\theta$
[50, theorem~47]. This prevents these structural operations from 
being repeated indefinitely.

\subsubhead{9.4 Summary of the entire recognition process}
The input is an image $I$, a semi-definite network $\N_0$ (the grammar),
and an embedding type $v$ for $\N_0$ (defining all the geometric relationships
in the grammar).

The recognition process is a sequence of steps, called \i{cycles}. In each
cycle,

\blob $i$ and $j$ are recalculated (\S7.4);

\blob $u$ is adjusted by one step (subject to the symmetry condition) to increase
$IM(I, \N_1, u, v\circ p, i, j, B)$ (\S7.3);

\blob all temperatures spread and decline a little (\S9.2);

\blob structural operations are applied to $\N_1$ if the conditions are
satisfied (elementary pruning
operations, extension operations, merging two symbol tokens, and partitioning
a symbol token into two); every new symbol, node, hook, edge or facet is given
a high temperature (\S9.2), and some bareness values are increased
after an extension operation (\S9.3);

\blob line operations are applied to $\N_1$ (\S9.1).

\medskip\f
The algorithm halts when

\blob no further structural operations are possible (except for trivial
extensions);

\blob the temperatures have declined very close to the minimum $T_{min}$.

\subsubhead{9.5 The outcome of the recognition algorithm}
When recognition has finished the following have been constructed:

\blob a pattern $\N_1$,

\blob a parsing homomorphism $p:\N_1\to\N_0$,

\blob a pair of inclusion functions $(i,j)$ on $\N_1$,

\blob an embedding token $u$ for $\N_1$,

\blob a bareness function $B$ for $\N_1$.

\medskip\f
The condition
$$ \forall\sigma\from\Sigma_1\ B(\sigma) \le \theta, \text{ with equality if }
P_0\pre{p(\{\sigma\})} = \emptyset $$
will hold, because it holds all the time. The condition 
$$ \forall x\from \Sigma_1 \cup N_1 \cup E_1 \ i(x) \in \{0, 1\} $$
is likely to hold, because all temperatures have reduced to a very low value
$T_{min}$. Also, the condition
$$ W_1 \circ A_1 \circ F_1 = W_1 \circ A_1 \circ S_1 $$
is likely to hold. This is because the final term in the definition of $IM$
penalises incoherent edges (recall the discussion of this in \S7.2). The
penalty is large and its effect is amplified when temperature is very low; 
an incoherent edge will either pull the two wholes together until they are 
merged or be pruned.

If we assume that all three of these conditions hold we can apply the following
theorem.

\theorem 48 (from [50]). If

\item{(a)} $\N_1=\network{_1}$ is a network satisfying
$W_1 \circ A_1 \circ F_1 = W_1 \circ A_1 \circ S_1$,

\item{(b)} $p:\N_1\to\N_0$ is a homomorphism, where $\N_0=\network{_0}$ is a
semi-definite network,

\item{(c)} $(i,j)$ is a pair of inclusion functions on $\N_1$ satisfying
$\forall x\from \Sigma_1 \cup N_1 \cup E_1 \ i(x) \in \{0, 1\}$,

\item{(d)} $u$ is an embedding token for $\N_1$ and $v$ is an embedding type
for $\N_0$,

\item{(e)} $B$ is a bareness function for $\N_1$ satisfying
$\forall\sigma\from\Sigma_1\ B(\sigma) \le \theta, \text{ with equality if }
P_0\pre{p(\{\sigma\})} = \emptyset$,

\item{(f)} $\N_1$ is unprunable, given $(i,j)$,

\item{(g)} $(\N_1,p,u)$ is inextendable, given $i,j,v,B$,

\item{(h)} $\N_1$ is unpartitionable,

\smallskip\f
then

\smallskip

\item{(1)} $\N_1$ is definite,

\item{(2)} $\forall x\from \Sigma_1\cup N_1\cup E_1\ i(x) = 1$,

\item{(3)} $IM(I,\N,u,v,i,j,B) = DM(I,\N,u,v)$.

\proof This uses theorems from [50].

Theorem~46, using hypotheses~(a,h), tells us that

\item{(i)} $E_1 \doublearrow{A_1 \circ F_1}{A_1 \circ S_1}
N_1\cup\Sigma_1 \buildrel W_1 \over \rightarrow \Sigma_1$ is a coequaliser
diagram in the category of sets. 

\smallskip\f
Theorem~42, using hypotheses~(c,f), gives 

\item{(j)} $\forall e\from E_1\ i(e)=1$. 

\smallskip\f
So theorem~45, using hypotheses~(a,c,g) and (j), gives

\item{(k)}
$\forall h\from H_1\ j(h)=0$ and $\forall k\from K_1\setdiff\dom(G_1)\ j(k)=0$.

\smallskip\f
Then theorem~44, using hypotheses~(c,f) and (i,k), gives conclusions~(1,2).

\smallskip\f
Theorem~47, using hypotheses~(a,e,g) and (1,2), then gives

\item{(l)} $\forall\sigma\from\Sigma_1\setdiff P_1(N_1)\ B(\sigma) = \theta$.

\smallskip\f
Conclusion~(3) then follows by theorem~33, using (l) and (1,2). \qed

\medbreak\f
Thus the outcome is a definite pattern. The algorithm has sought to maximise 
$IM(I,\N,u,v,i,\allowbreak j,\allowbreak B)$, and hence to maximise $DM(I,\N,u,v)$ at the end. The
symmetry condition is enforced throughout. The recognition problem is solved.

\subsubhead{9.6 What can go wrong}
The core of the theory is provably correct, but the more peripheral parts of
the algorithm are supported only by heuristic arguments (which I have
indicated throughout by use of the word \ic{likely}).

\item{(i)} The algorithm for determining the inclusion functions (\S7.4) is not
guaranteed to halt. It finds only a local maximum of $E$, not a global
maximum.

\item{(ii)} Even for low temperature, maximising $E$ is not precisely the
same as maximising $IM(I, \N_1, \allowbreak u, \allowbreak v\circ p, 
\allowbreak i, \allowbreak j, B)$.

\item{(iii)} The monotonic raising of the bareness function may cut off possible
structural extensions prematurely.

\item{(iv)} Symbol tokens may be merged or not merged, or partitioned or not
partitioned, wrongly. (Errors in merging can be corrected by partitioning and
vice versa.)

\item{(v)} The whole recognition algorithm is not guaranteed to halt.

\item{(vi)} It is not guaranteed that very low temperature will force all
inclusion values to $0$ or $1$.

\item{(vii)} Incoherent edges may survive to the end.

\item{(viii)} Some portions of the image may have been overlooked and not
covered by the pattern; perhaps they are noise, but perhaps not. 

\medbreak\f
In practice it is only (iii), (iv), (v) and (viii) that matter.

%% file: 10a.tex
\subhead{10. Example grammars}
I shall give three examples of grammars, designed to illustrate various aspects
of iteration, recursion, and the use of subsymbols to enforce syntactic
constraints. First, it is useful to distinguish between a \ic{line} (a
one-dimensional line segment) and a \ic{bar} (a thin rectangle, with a positive
width). I shall use both lines and bars together in my grammars. A line plays a
role as part of a larger symbol, such as a hexagonal grid; a bar gives a
finer-grained representation of the line. We introduce two symbol types, $line$
and $bar$, where $line$ has $bar$ as its sole part. The geometric relationship
between the line and the bar may stretch width-wise, to adapt the width of the
bar to the image; this variation does not interfere with the variation in the
geometric relationship between the line and the rest of the pattern. In the
simplest implementation of this idea, each line token has one bar token as its
sole part. However, instead I shall allow a line token to be made up of one or
more bar tokens, joined end-to-end (see the grammar in figure~10): this is 
useful for representing bent or broken lines.  

\subsubhead{10.1 Hexagonal grids ($hex$)}
The first example illustrates two-dimensional iteration (which is something that
conventional graph grammars can represent only using context-sensitive
production rules, see \S2.3). The grammar for the hexagonal grid symbol type
$hex$, with its subsymbols $hexagon$, $junctionA$ and $junctionB$, was
introduced in \S3.3, in simplified form. I omitted to say there that there are
also edges between the $hexagon$, $junctionA$ and $junctionB$ subsymbols and
each of their nodes, which ensure that each node occurs once per symbol token.
More importantly, I made no provision for what happens at the boundary of the
grid. To cope with this we insert invisible \i{dummy} symbol tokens at the
boundary of the grid, where the next \i{line} would be if the grid continued
(figure~9).

\midinsert

\medskip

\twofigures{55}{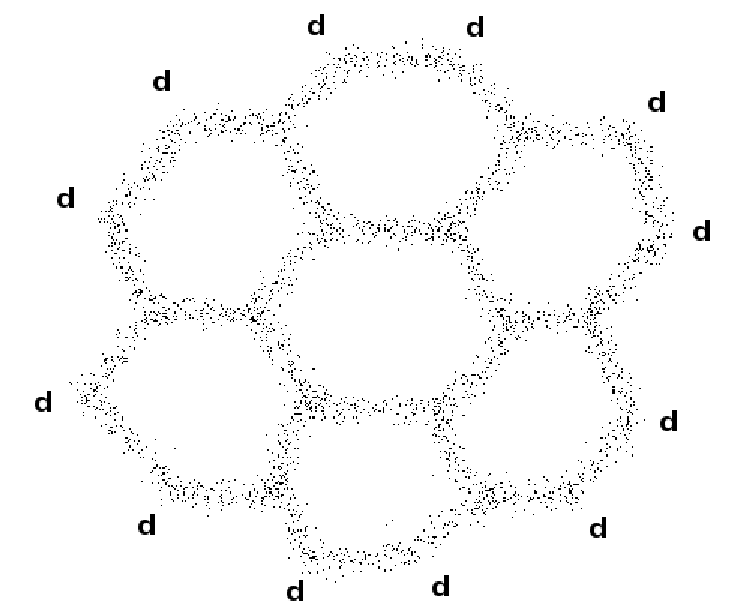}{-125pt}{70}{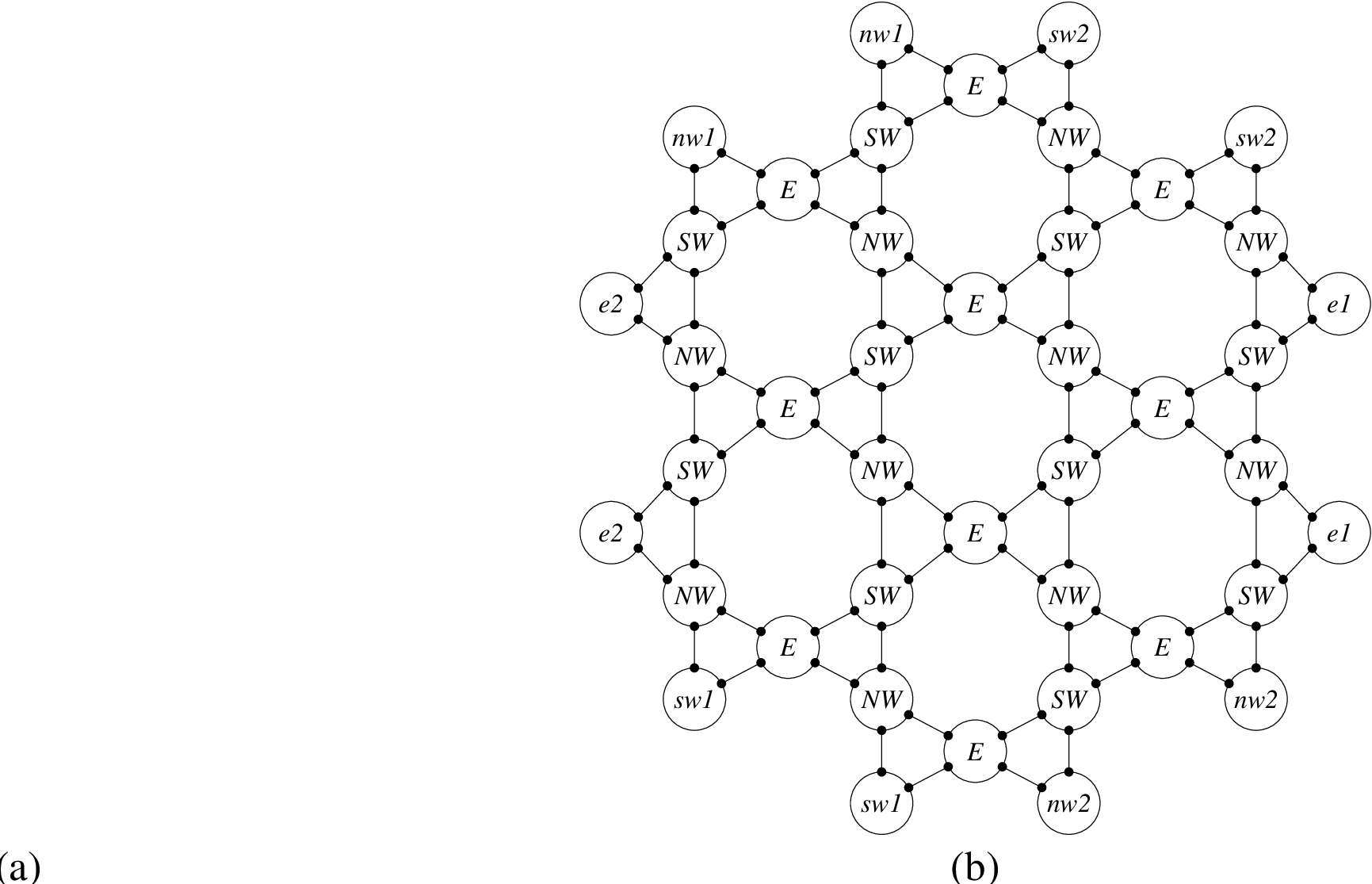}

\medskip

\longcaption{Figure 9. (a) A hexagonal grid image with $dummy$ symbol tokens
(marked by \ic{d}) inserted at the boundary; (b) The corresponding pattern (only
nodes, hooks and edges shown).}

\medskip

\endinsert

The $hex$ grammar is then modified accordingly (figure~10). I have labelled the
node types $E$ (for east-oriented line), $NW$ (northwest-oriented line), $SW$
(southwest-oriented line), $e1$ (dummy on the east boundary), $e2$ (dummy on the
west boundary), etc.; figure 9(b) shows the pattern corresponding to figure 9(a)
(showing only the nodes, hooks and edges of the $hex$).

\midinsert

\medskip

\percentsize{70}\Figure{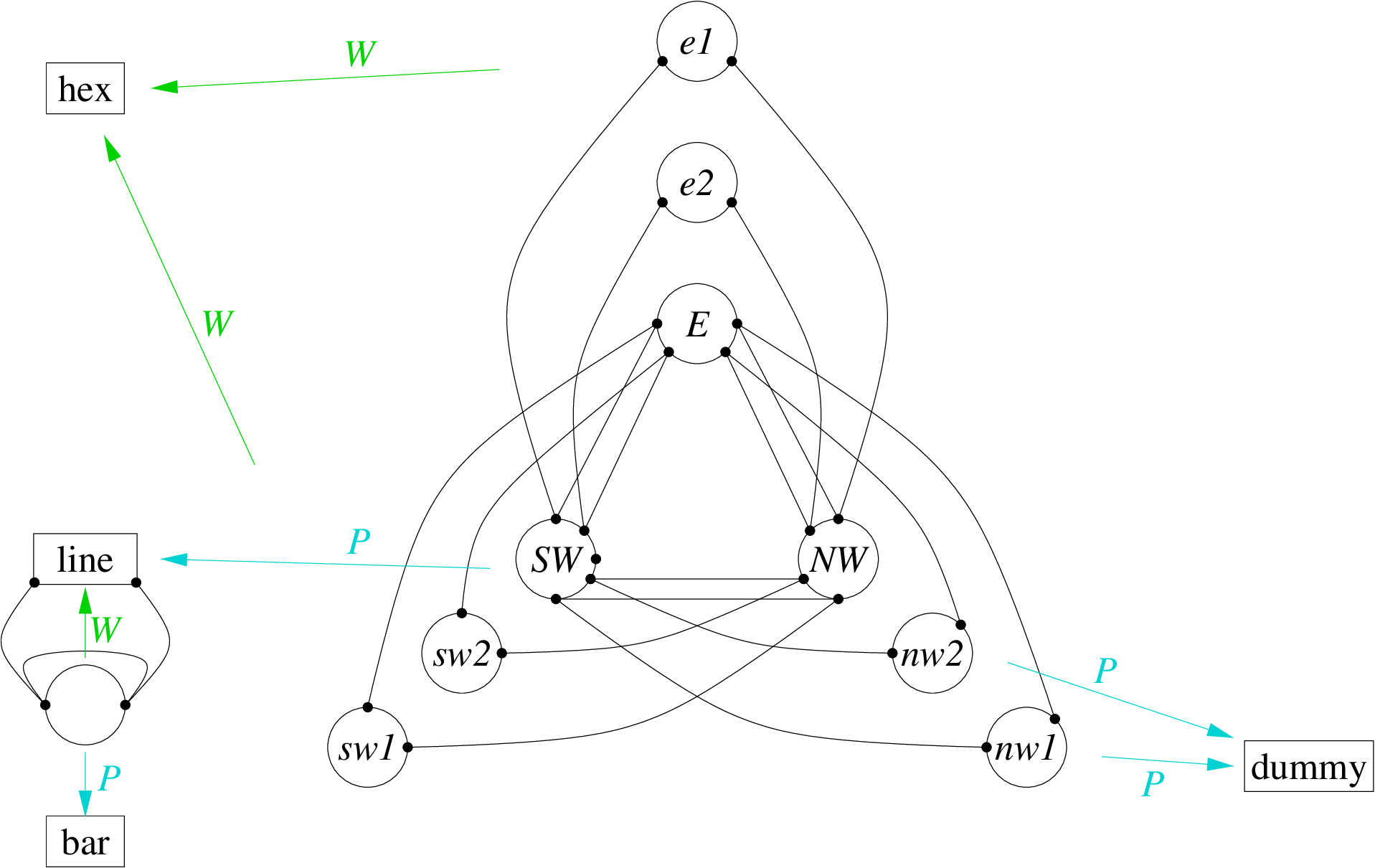}

\medskip

\caption{Figure 10. The $hex$ grammar with the $dummy$ and $bar$ symbol
types added.}

\medskip

\endinsert

The $junctionA$ and $junctionB$ subsymbols are modified similarly to allow for
the possibility that one of the three lines is a dummy. Several variants of the
$hexagon$ subsymbol are also needed, to provide for the cases of a hexagon
occurring on the boundary or at a corner of the grid.

\subsubhead{10.2 Crosses}
The next example involves crosses -- see figure~11 for an example image. It consists
of a central square, with four \i{trunks} radiating from it, each with the same
number of smaller lines called \i{twigs}. The grammar is shown in figure~12. On
the left is the grammar for the symbol type $cross$; the nodes $N,E,S,W$
represent the lines of the square, the nodes $NT,ET,ST,WT$ represent the four
trunks, and the nodes $Nt,Et,St,Wt$ represent the twigs. To enforce the
constraint that every trunk has the same number of twigs we use a subsymbol
$8twigs$, which includes two adjacent twigs of each trunk. Figure~13 shows the
pattern corresponding to the image in figure~11. The presence of $8twigs$ tokens
in the pattern is governed by the facets. The facets $\alpha$--$\delta$ in a
$cross$ token must be glued to corresponding sub-facets in an $8twigs$ token:
this is what forces there to exist the required number of $8twigs$ subsymbol
tokens, and this forces the number of twigs on each trunk to be equal.

\midinsert

\medskip

\percentsize{60}\Figure{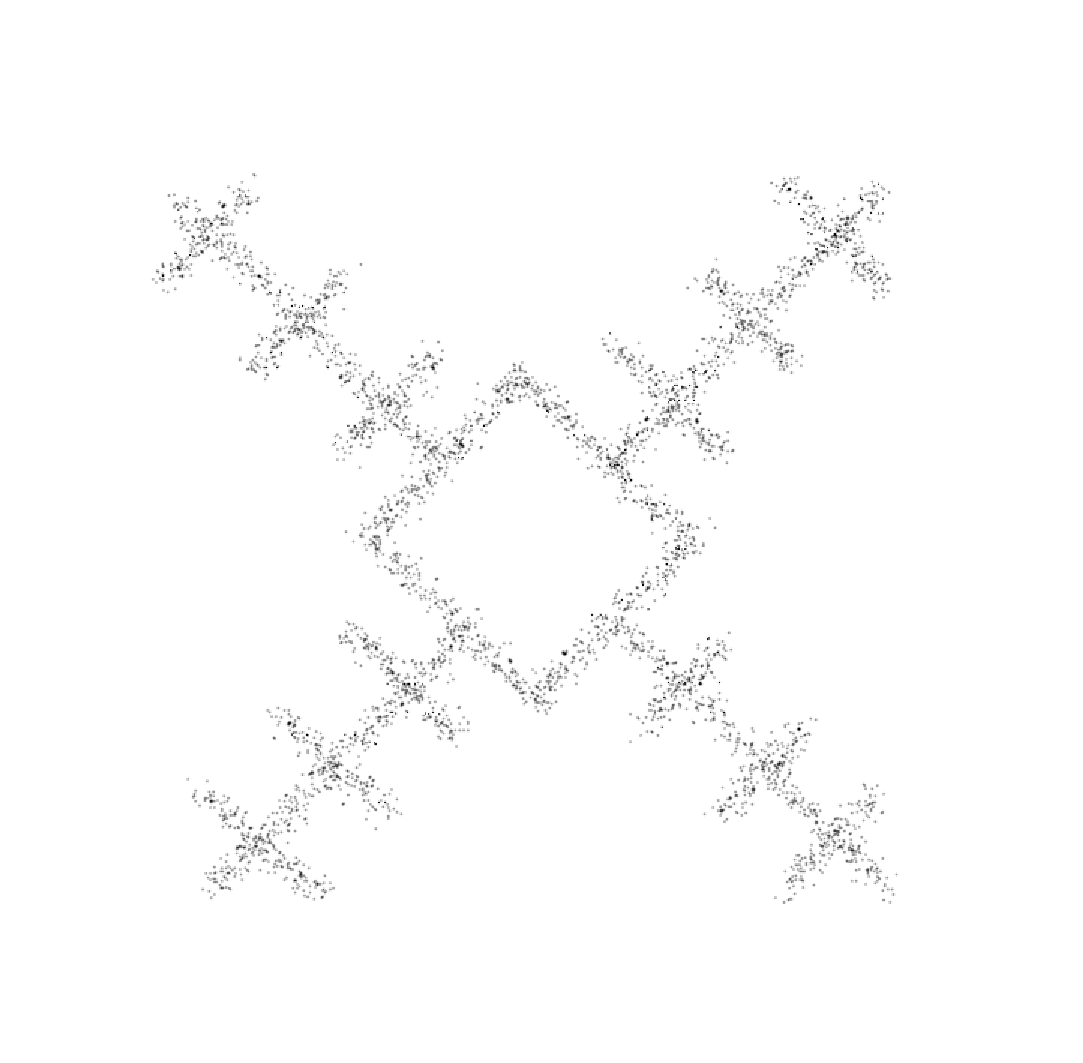}

\caption{Figure 11. A image showing an example of a cross.}

\medskip

\endinsert

\midinsert

\medskip

\percentsize{90}\Figure{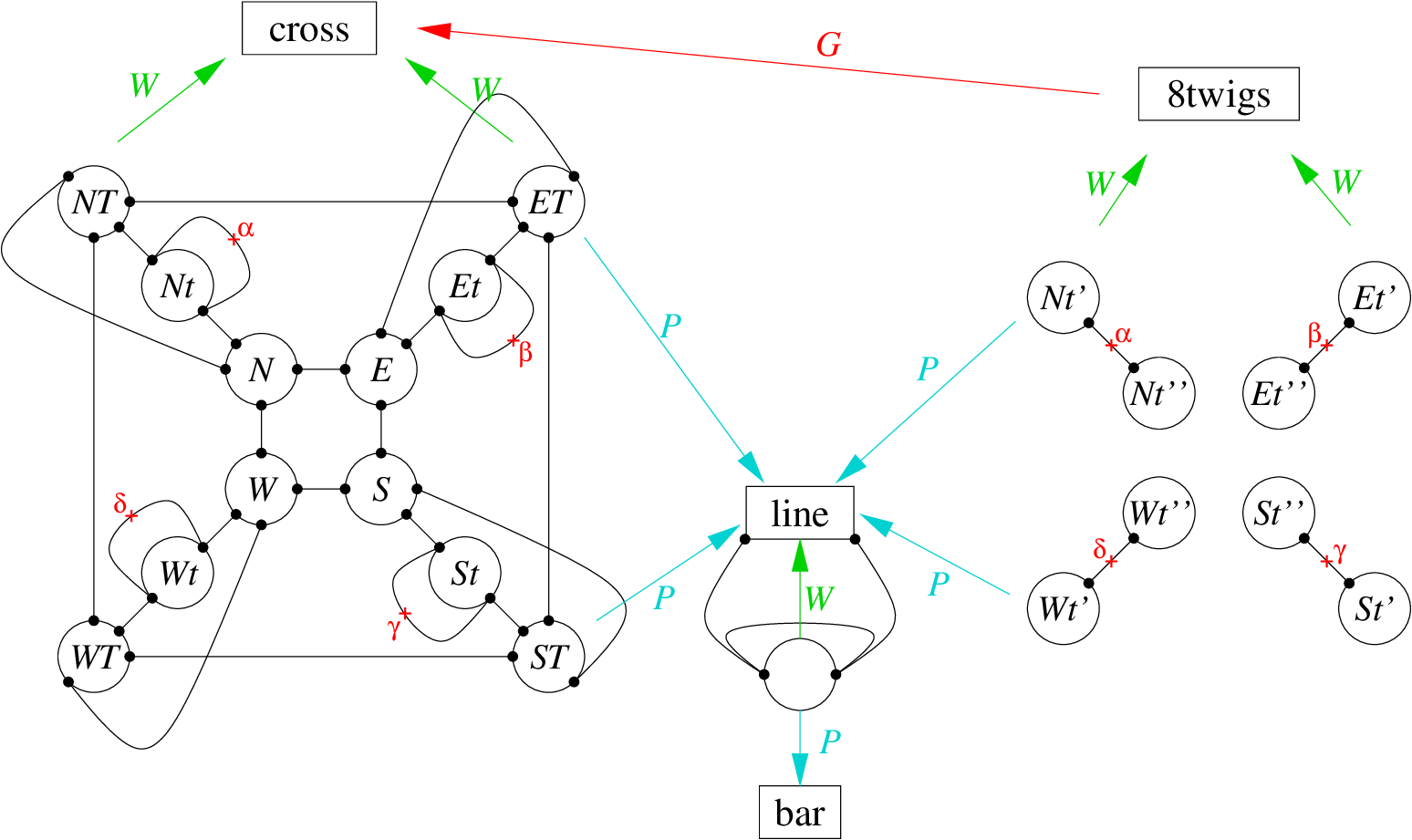}

\longcaption{Figure 12. The $cross$ grammar. Only a few $W$ and $P$ arrows
are shown. $8twigs$ is glued to $cross$; nodes $Nt'$ and $Nt''$ are glued to
node $Nt$, etc.. The facets are shown as red crosses: $\alpha$ is glued to
$\alpha$, $\beta$ to $\beta$, $\gamma$ to $\gamma$, and $\delta$ to $\delta$. }

\medskip

\endinsert

\midinsert

\medskip

\percentsize{100}\Figure{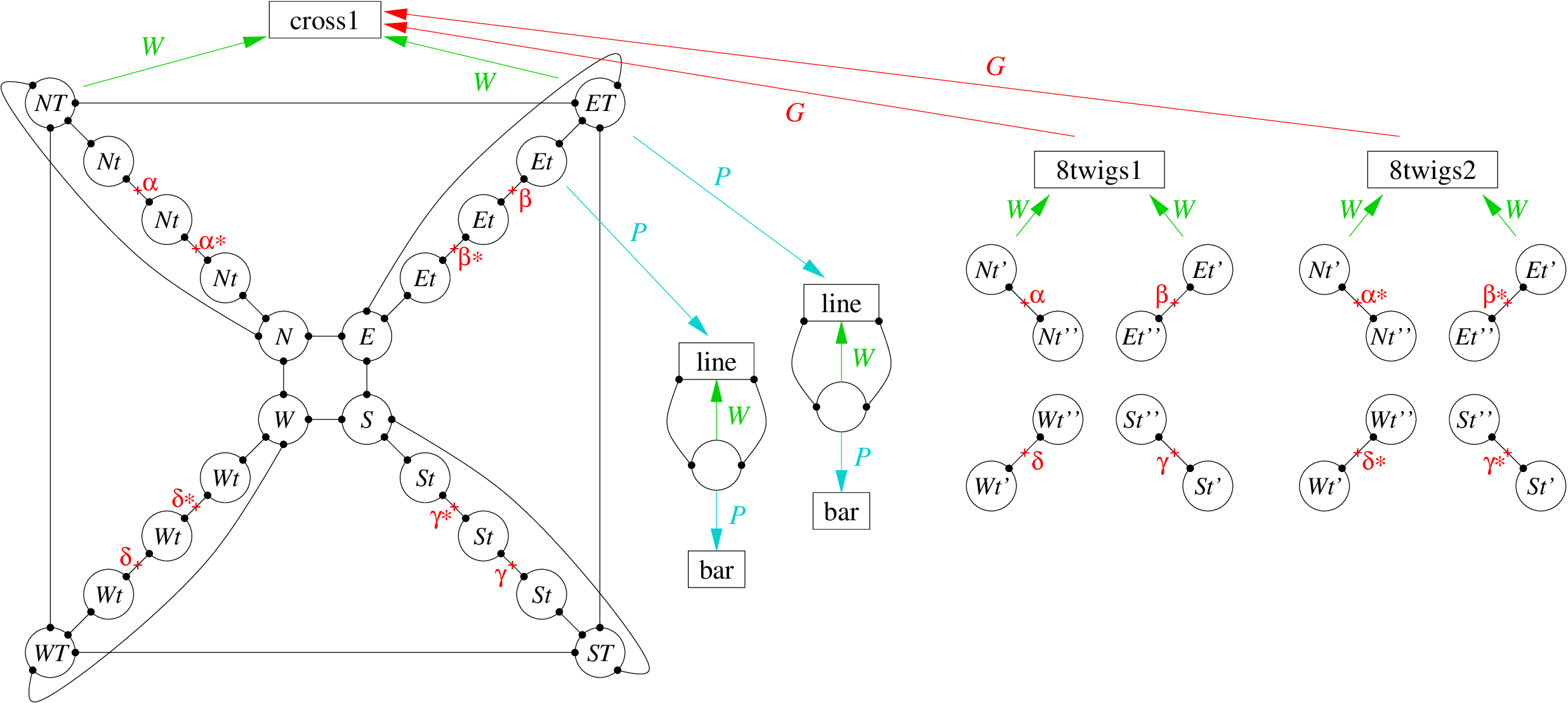}

\longcaption{Figure 13. The pattern corresponding to the $cross$ image in
figure~11. Only two of the $line$ tokens are shown. Nodes $Nt',Nt''$ are glued
to node $Nt$, and so on. Facets
$\alpha,\beta,\gamma,\delta,\alpha*,\allowbreak\beta*,\allowbreak
\gamma*,\allowbreak \delta*$ are glued to facets
$\alpha,\allowbreak\beta,\allowbreak\gamma,\allowbreak\delta,
\allowbreak\alpha*,\beta*,\gamma*,\delta*$, respectively.}

\medskip

\endinsert

(Again I am omitting edges from the symbols to the nodes, to avoid cluttering
the diagrams. There are edges from $cross$ to $N,E,S,W,NT,ET,ST,WT$ and from
$8twigs$ to each of its nodes. This forces there to be one node token of each type per
symbol token.)

\subsubhead{10.3 Hnests}
The third example involves nested recursion. Figure~14 shows an example of an
$Hnest$, which consists of three lines forming an \ic{H} and two smaller
$Hnests$ nested inside it. Figure 15 shows the grammar. As usual, each node
represents a role that a symbol may play as a part of another. The nodes $L$,
$cb$, $R$ represent a line occurring as left bar, crossbar, and right bar of the
\ic{H}, respectively. The nodes $T$, $B$ represent an $Hnest$ occurring as the
top or bottom nested $Hnest$ of an $Hnest$ (note that the $Hnest$ symbol type is
a part of itself). Every $Hnest$ token either has two or no $Hnest$ parts.
There are no subsymbols.

\midinsert

\medskip

\percentsize{60}\Figure{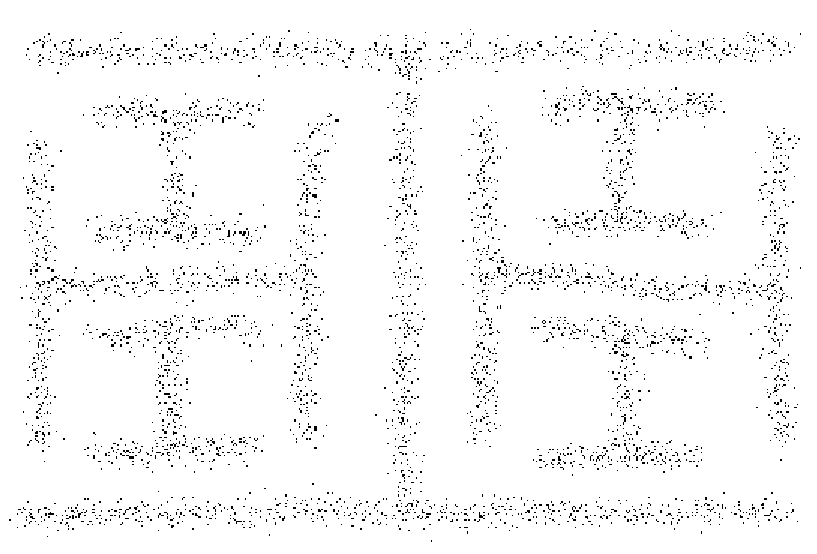}

\caption{Figure~14. An $Hnest$ image.}

\medskip

\endinsert

\midinsert

\medskip

\percentsize{50}\Figure{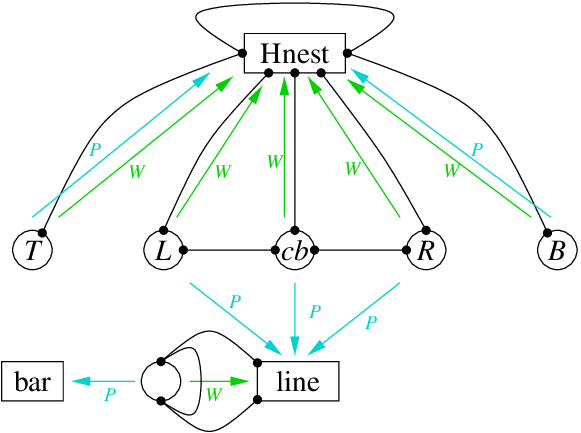}

\medskip
\caption{Figure~15. The $Hnest$ grammar.}

\medskip

\endinsert

Figure~16 shows an example of an $Hnest$ pattern.

\midinsert

\medskip

\percentsize{100}\Figure{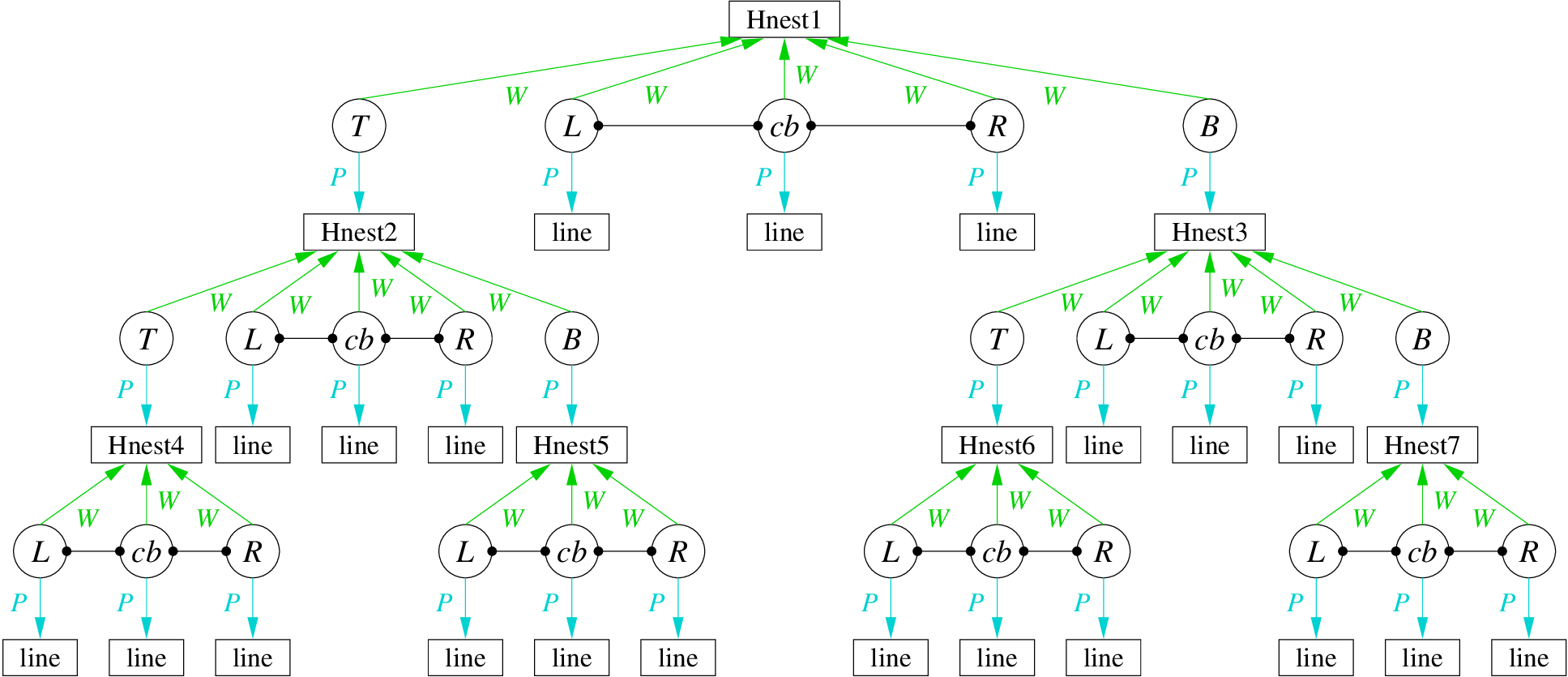}

\medskip
\longcaption{Figure~16. An $Hnest$ pattern. Edges between $Hnest$ tokens and
their nodes are omitted. $Bar$ tokens are omitted.}

\medskip

\endinsert

%% file: 11a.tex
\input table.sty

\subhead{11. Example runs}
I shall provide some examples of the algorithm recognising images according to
the three grammars described in the previous section (combined together into a
single grammar). These example runs are not intended as experiments and I make
no quantitative claims; they merely provide proof of concept. The test images
were designed to illustrate the recursive capabilities and aspects of
error-tolerance listed as desirable in \S1.

Test images were produced as follows. First I drew twenty hexes (with 2--4
hexagons on each side), twenty crosses (with 3--6 twigs on each trunk), and
twenty Hnests (with 1--4 levels of nesting) by hand; these are irregularly
drawn, with bent lines and variable angles and length ratios. The lines are
automatically blurred. A composite image is produced by randomly selecting three
of these sixty simple images and superimposing them, with random translations
(in the range $-300$ to $300$ pixels in each dimension), random rotations (in
the range $0$ to $2\pi$) and random dilations (in the range $0.9$ to $1.3$).
Repeating this produces an indefinitely large supply of composite images.
Corrupted versions are produced by erasing one or two random discs of radius
$50$ pixels in the composite image.

The reasons for using synthetic images rather than natural ones are that they
allow me to emphasise the features I am most concerned with (recursion and
iteration, overlapping symbol tokens, geometric variability, indistinct and bent
lines, and random erasures), while avoiding the need for special tricks, as
often needed with natural images; they avoid issues I am not interested in
(perspective, binocular vision, motion, occlusion, colour, texture, shadow and
illumination); they allow me to calibrate the difficulty of the image set to
match the capabilities of the algorithm; and they allow me to generate an
unlimited number of patterns with consistent statistical properties and to test
the algorithm under controlled conditions.

The above parameters were chosen to give images sufficiently complex and
cluttered to illustrate the desired qualitative features. Varying the number of
simple images makes no difference to the statistical properties of the results,
so I have not gone beyond sixty; the variety in the composite images comes from
the random overlapping of the simple images. Varying the number of composite
images also makes no difference. There is sufficient geometric variability in
the images to ensure that top-down and bottom-up inference must be used in
combination: e.g., a junction between two lines cannot be recognised as
belonging to a hex, a cross or an Hnest merely from the angle or ratio of line
lengths but must be interpreted in the light of grammatical context.

The difficulty of a composite image depends on how much its three component
images overlap, and this may be roughly measured by an \i{overlap index},
calculated as follows. Fit a circumscribing disc to each of the three component
images. The overlap index is defined as the sum of the areas of overlap between
every pair of discs, divided by the sum of the areas of the three discs. It
ranges between $0$ and $1$. Another measure of difficulty is the number of
symbol tokens in the desired pattern.

The recognition algorithm is run on 300 of these composite images and the
results are summarised in tables~1--3, broken down by number of symbols and then
by overlap index. The first six columns give the numbers of composite images for
which the constructed pattern is topologically correct, subdivided by the number
of wrong lines: a line is \ic{wrong} iff it is misplaced or is part of a
misplaced hex, cross or Hnest. The next three columns give the numbers of
composite images for which the constructed pattern is not topologically correct,
subdivided by the number of component images (hexes, crosses or Hnests) that are
not topologically correct (plus the number of spurious hexes, crosses or Hnests,
if any). The final three columns give the numbers of composite images for which
the algorithm gets stuck, i.e., does not halt, again subdivided by the number of
components that are not topologically correct.

\midinsert


\table{
\toprow.
\row symbols, 0, 1, 2, 3, 4, 5+, 1, 2, 3, 1, 2, 3, total.
\Line
\row 50--149, 22, 2, 0, 0, 0, 0, 3, 0, 0, 0, 0, 0, 27.
\row 150--249, 74, 11, 4, 1, 0, 1, 10, 0, 0, 2, 2, 0, 105.
\row 250--349, 27, 3, 3, 2, 0, 0, 3, 0, 0, 1, 0, 1, 40.
\row 350--449, 23, 2, 2, 2, 0, 0, 1, 0, 0, 3, 1, 0, 34.
\row 450--549, 16, 0, 0, 2, 1, 0, 1, 0, 0, 6, 1, 1, 28.
\row 550--649, 20, 4, 4, 2, 2, 0, 7, 0, 0, 2, 2, 0, 43.
\row 650--749, 3, 1, 0, 1, 0, 1, 1, 0, 0, 1, 3, 0, 11.
\row 750--849, 0, 1, 0, 0, 0, 0, 0, 0, 0, 0, 2, 6, 9.
\row 850--949, 0, 0, 0, 0, 0, 0, 0, 0, 0, 0, 1, 0, 1.
\row 950--1049, 1, 0, 0, 0, 1, 0, 0, 0, 0, 0, 0, 0, 2.
\row total, 186, 24, 13, 10, 4, 2, 26, 0, 0, 15, 12, 8, 300.
}

\table{
\toprow overlap.
\row index, 0, 1, 2, 3, 4, 5+, 1, 2, 3, 1, 2, 3, total.
\Line
\row 0--0.2, 13, 1, 0, 1, 1, 0, 1, 0, 0, 0, 0, 0, 17.
\row 0.2--0.4, 90, 4, 1, 3, 2, 0, 4, 0, 0, 3, 3, 2, 112.
\row 0.4--0.6, 72, 11, 9, 5, 1, 1, 16, 0, 0, 8, 8, 3, 134.
\row 0.6--0.8, 11, 8, 3, 1, 0, 1, 4, 0, 0, 4, 1, 3, 36.
\row 0.8--1, 0, 0, 0, 0, 0, 0, 1, 0, 0, 0, 0, 0, 1.
\row total, 186, 24, 13, 10, 4, 2, 26, 0, 0, 15, 12, 8, 300.
}

\caption{Table 1. Results for images with no erased discs.}



\medskip

\table{
\toprow.
\row symbols, 0, 1, 2, 3, 4, 5+, 1, 2, 3, 1, 2, 3, total.
\Line
\row 50--149, 21, 2, 1, 0, 0, 0, 3, 0, 0, 0, 0, 0, 27.
\row 150--249, 64, 15, 2, 4, 1, 2, 9, 0, 0, 7, 1, 0, 105.
\row 250--349, 25, 7, 0, 2, 0, 0, 2, 0, 0, 3, 0, 1, 40.
\row 350--449, 18, 5, 2, 1, 0, 0, 2, 0, 0, 4, 2, 0, 34.
\row 450--549, 12, 3, 0, 0, 1, 1, 1, 0, 0, 7, 2, 1, 28.
\row 550--649, 16, 4, 1, 2, 2, 2, 4, 1, 0, 2, 9, 0, 43.
\row 650--749, 1, 3, 1, 0, 0, 0, 2, 0, 0, 2, 2, 0, 11.
\row 750--849, 0, 0, 0, 0, 1, 0, 1, 0, 0, 2, 0, 5, 9.
\row 850--949, 0, 0, 0, 0, 0, 0, 0, 0, 0, 0, 0, 1, 1.
\row 950--1049, 0, 0, 0, 0, 1, 1, 0, 0, 0, 0, 0, 0, 2.
\row total, 157, 39, 7, 9, 6, 6, 24, 1, 0, 27, 16, 8, 300.
}

\table{
\toprow overlap.
\row index, 0, 1, 2, 3, 4, 5+, 1, 2, 3, 1, 2, 3, total.
\Line
\row 0--0.2, 14, 0, 0, 1, 1, 1, 0, 0, 0, 0, 0, 0, 17.
\row 0.2--0.4, 69, 14, 2, 3, 2, 1, 7, 0, 0, 7, 6, 1, 112.
\row 0.4--0.6, 59, 19, 5, 5, 1, 3, 14, 1, 0, 16, 7, 4, 134.
\row 0.6--0.8, 15, 5, 0, 0, 2, 1, 3, 0, 0, 4, 3, 3, 36.
\row 0.8--1, 0, 1, 0, 0, 0, 0, 0, 0, 0, 0, 0, 0, 1.
\row total, 157, 39, 7, 9, 6, 6, 24, 1, 0, 27, 16, 8, 300.
}

\caption{Table 2. Results for images with one erased disc.}

\endinsert

\midinsert


\table{
\toprow.
\row symbols, 0, 1, 2, 3, 4, 5+, 1, 2, 3, 1, 2, 3, total.
\Line
\row 50--149, 20, 0, 0, 3, 1, 0, 2, 0, 0, 1, 0, 0, 27.
\row 150--249, 59, 17, 7, 1, 1, 1, 10, 0, 0, 4, 5, 0, 105.
\row 250--349, 18, 10, 1, 0, 1, 0, 3, 0, 0, 5, 2, 0, 40.
\row 350--449, 18, 5, 1, 1, 1, 1, 1, 0, 0, 5, 1, 0, 34.
\row 450--549, 10, 3, 0, 3, 0, 0, 2, 0, 0, 4, 5, 1, 28.
\row 550--649, 13, 5, 4, 1, 1, 1, 3, 0, 0, 5, 10, 0, 43.
\row 650--749, 3, 1, 1, 0, 0, 0, 2, 0, 0, 0, 4, 0, 11.
\row 750--849, 1, 0, 0, 1, 0, 0, 0, 0, 0, 1, 2, 4, 9.
\row 850--949, 0, 0, 0, 0, 0, 0, 0, 0, 0, 0, 0, 1, 1.
\row 950--1049, 0, 0, 0, 0, 0, 1, 0, 0, 0, 0, 0, 1, 2.
\row total, 142, 41, 14, 10, 5, 4, 23, 0, 0, 25, 29, 7, 300.
}

\table{
\toprow overlap.
\row index, 0, 1, 2, 3, 4, 5+, 1, 2, 3, 1, 2, 3, total.
\Line
\row 0--0.2, 12, 2, 0, 1, 1, 0, 1, 0, 0, 0, 0, 0, 17.
\row 0.2--0.4, 69, 14, 5, 2, 0, 1, 6, 0, 0, 6, 7, 2, 112.
\row 0.4--0.6, 55, 18, 8, 5, 3, 2, 12, 0, 0, 14, 14, 3, 134.
\row 0.6--0.8, 6, 7, 1, 2, 1, 1, 3, 0, 0, 5, 8, 2, 36.
\row 0.8--1, 0, 0, 0, 0, 0, 0, 1, 0, 0, 0, 0, 0, 1.
\row total, 142, 41, 14, 10, 5, 4, 23, 0, 0, 25, 29, 7, 300.
}

\caption{Table 3. Results for images with two erased discs.}

\endinsert

Tables~2 and 3 show that the algorithm has some ability to restore erased lines
or junctions. However, it tends to restore them in the most grammatically
probable place, rather than where they actually were before they were erased.
This accounts for the increased number of wrong lines. There is also an increase
in the number of cases where the algorithm gets stuck.

The figures show some examples. Figure~17 shows an image consisting of a hex, an
Hnest and a cross, with two erased discs (erased discs are outlined in brown in
the figures). Note that the Hnest is somewhat sheared. The image is successfully
recognised, but with three misplaced lines of an H in one erased disc (this
counts as \ic{3 wrong lines} in table 3). The actual positions of the misplaced
lines are marked in red and the correct positions in blue. Four lines correctly
guessed in the other erased disc are shown in green: one is part of an H and
three are part of a hex.

Figure~18 shows an image consisting of three hexes (I have added an outline to
two of the hexes in the figure as an aid to the reader). This is successfully
recognised. In a version of this image with one erased disc the hex outlined in
red is only partially recognised (this is counted as \ic{stuck -- 1 wrong
symbol} in table~2). In a version with two erased discs the hex outlined in blue
is recognised but the other two are not separated (this is counted as \ic{stuck
-- 2 wrong symbols} in table~3).

Figure~19 shows an image consisting of one hex and two crosses, with one erased
disc. By accident the two crosses are oriented identically and positioned so
that many of their lines overlap, yet the image is successfully recognised. A
version of the image with two erased discs is not successfully recognised: the
twigs of the crosses are not sorted out correctly (this counts as \ic{stuck -- 2
wrong symbols} in table 3).

Figure~20 shows an image consisting of three Hnests, with two erased discs. This
is recognised with two misplaced lines in a single H (this counts as \ic{2 wrong
lines} in table 3). The actual positions of the misplaced lines are marked in
red and the correct positions in blue. The other line in the H (correctly
recognised) is marked in green.

Figure~21 shows an image consisting of one hex and two Hnests, with one erased
disc (the image is truncated in the figure). The hex is incompletely recognised:
the lines marked in blue are missed. This is the commonest type of error for a
hex. (This counts as \ic{topological error -- 1 wrong symbol} in table~2.)

Figure~22 shows an image consisting of one hex and two Hnests, with two erased
discs. The algorithm recognises only part of the smaller Hnest: the H marked in
red is recognised (as an Hnest in its own right), but the parts marked in blue
are overlooked. This is a common type of error for an Hnest. (This counts as
\ic{topological error -- 1 wrong symbol} in table~3.)

Figure~23 shows how a bent or broken line may be represented by two or three
bars joined end to end; four such lines are shown, the bars being marked by red
rectangles. Long bent lines would be hard to recognise if this were not allowed.
The brown circle outlines an erased disc. The whole image is recognised
successfully.

\newpage

\vbox{

\vskip-3pt

\percentsize{84}\figure{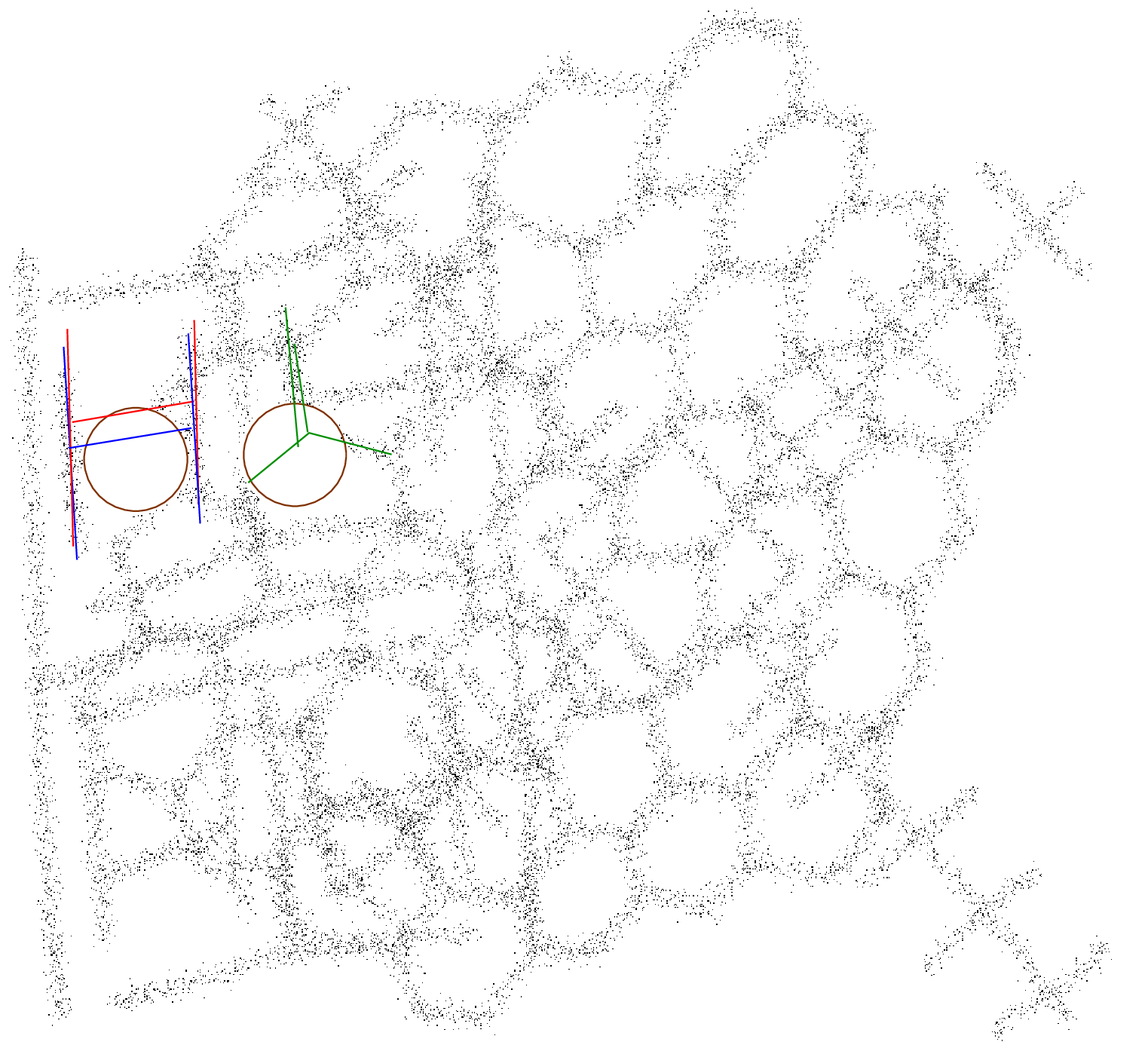}

\vskip-2pt

\longcaption{Figure 17. A hex, an Hnest and a cross. 557 symbols,
overlap index = 0.666.}

\percentsize{88}\figure{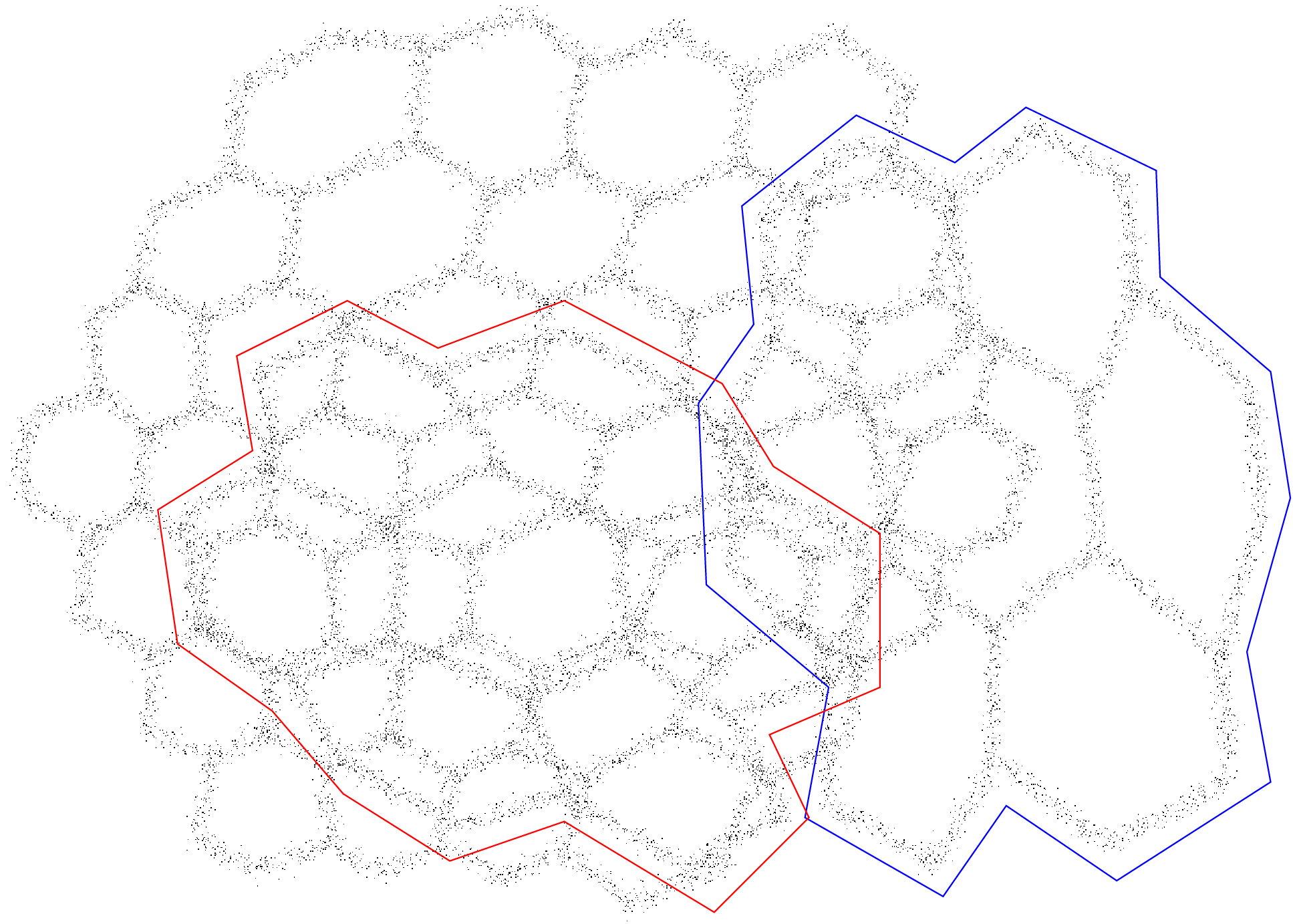}

\vskip-2pt

\longcaption{Figure 18. Three hexes. 678 symbols, overlap index =
0.432.}

\vskip-3pt

}

\newpage

\vbox{

\vskip-2pt

\percentsize{84}\figure{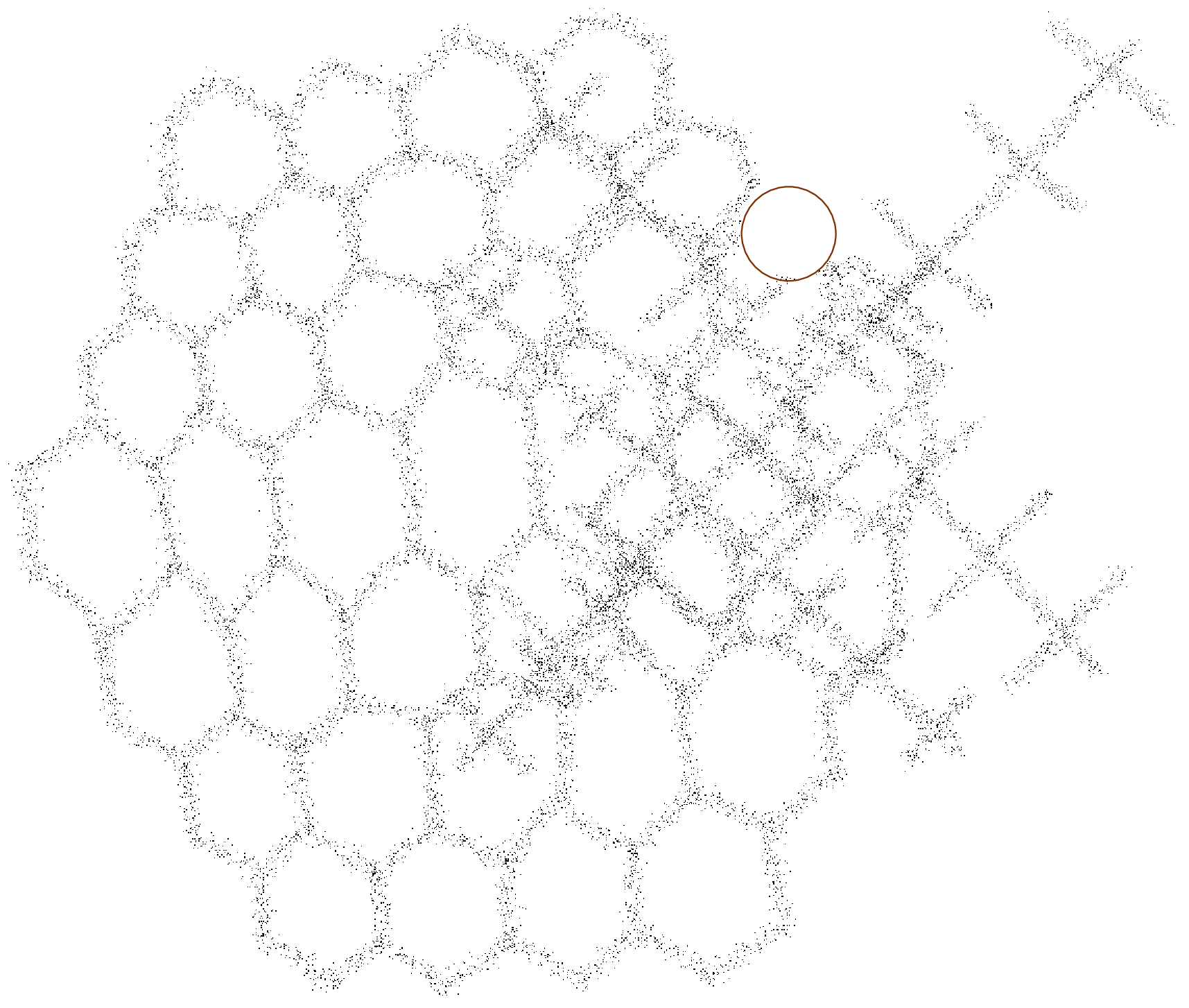}

\vskip-4pt

\longcaption{Figure 19. One hex and two crosses. 532 symbols, overlap
index = 0.579.}

\percentsize{94}\figure{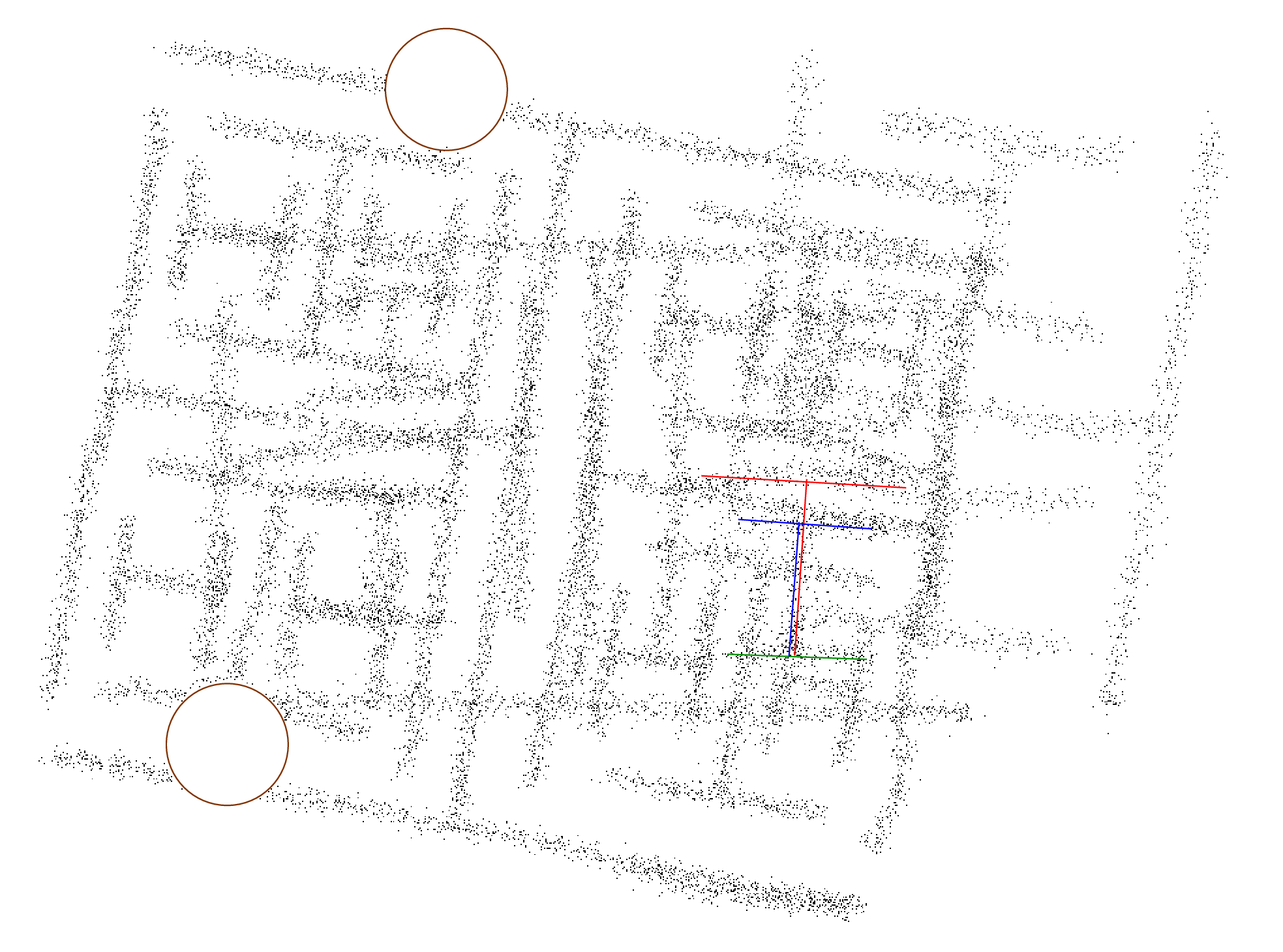}

\vskip-8pt

\longcaption{Figure 20. Three Hnests. 175 symbols, overlap index = 0.563.}

\vskip-2pt

}

\newpage

\vbox{

\vskip-8pt

\percentsize{86}\figure{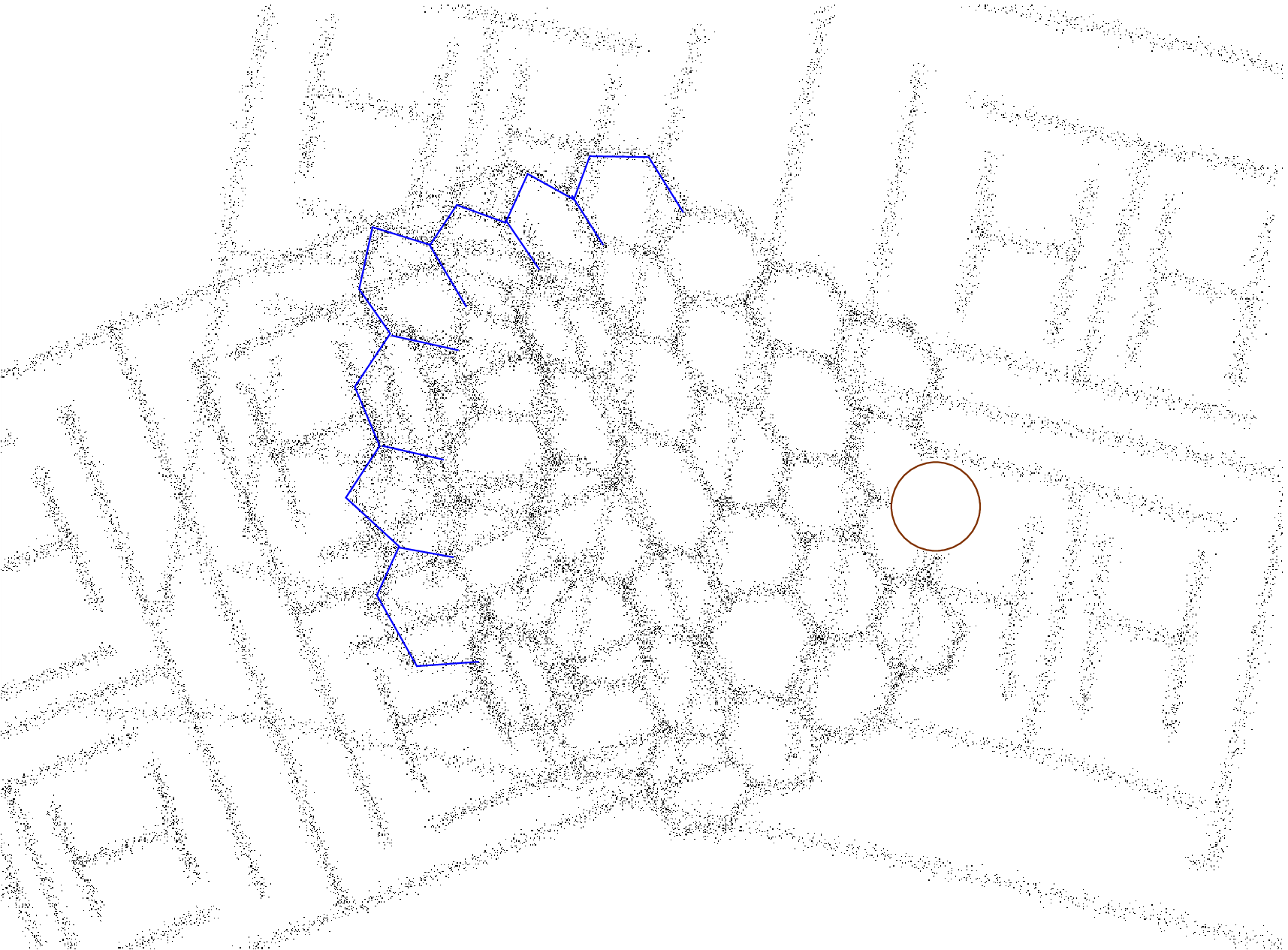}

\longcaption{Figure 21. One hex and two Hnests. 662 symbols, overlap
index = 0.387.}

\smallskip

\percentsize{85}\figure{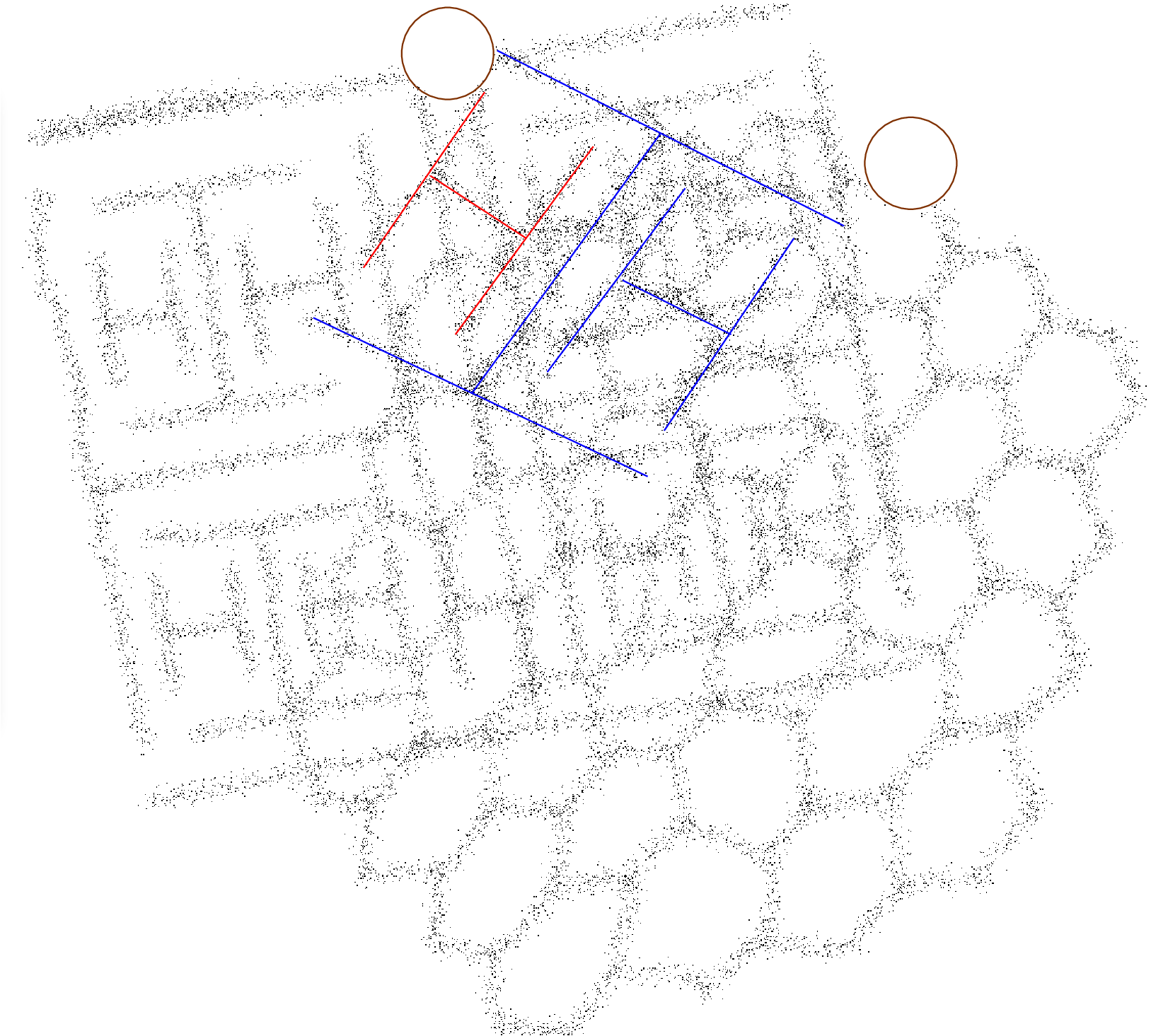}

\longcaption{Figure 22. One hex and two Hnests. 573 symbols, overlap
index = 0.502.}

\vskip-8pt

}

\newpage

\percentsize{100}\figure{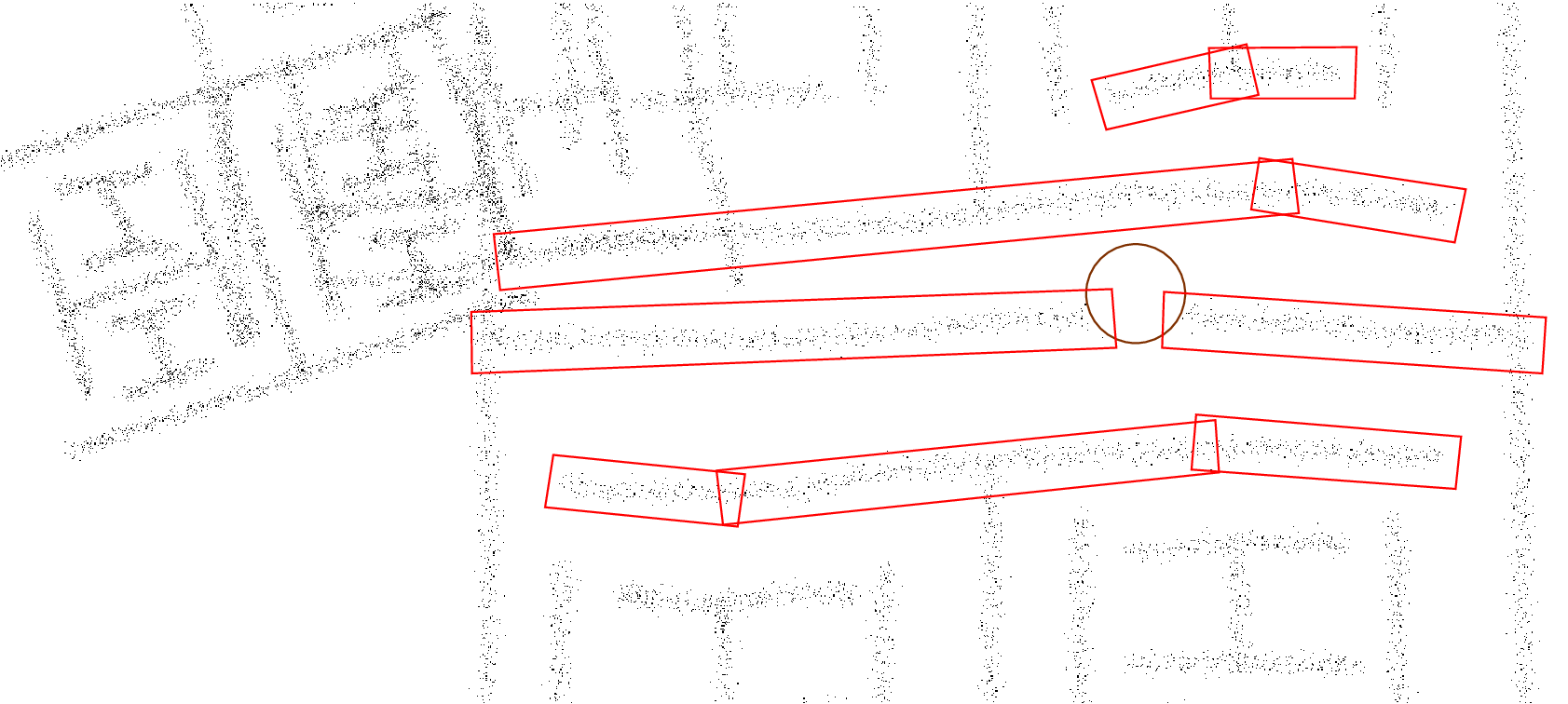}

\smallskip

\longcaption{Figure 23. A detail of an image, showing four bent
or broken lines.}

%% file: 12a.tex
\subhead{12. Conclusions}
The theoretical contributions of this paper are as follows.

\blob A new, mathematically well-founded formalism for representing the
syntactic structure of graph-like patterns declaratively, fundamentally
different from conventional graph grammars based on sets of production rules,
in which parsing is seen as the construction of a homomorphism between two
networks (a \i{pattern} and a \i{grammar}). This offers a new perspective on
the problem of representing syntactic structure in non-string patterns, and it
helps to overcome the limitations of production rules discussed in \S2; it is
also amenable to formal mathematical treatment.

\blob A technique for representing variable affine relationships using
\i{fleximaps} (which could be extended to projective transformations if
required); this is a more uniform, general, and mathematically well-founded
concept than others in the literature (\S5).

\blob A systematic treatment of affine invariance and symmetry (\S\S5.3--5.4). 
(Symmetry is a neglected topic in the literature.) 

\blob A new recognition algorithm for these grammars, which is provably correct
subject to certain qualifications (\S9.5).

\medbreak\f
There is no hope of proving favourable time-complexity bounds for the
recognition algorithm.  As pointed out in \S2, the graph parsing problem is
NP-complete, even in the absence of errors, except for highly restricted classes of
grammars. This certainly applies to my formulation of the problem too. However,
the space complexity is linear in the maximum size of the pattern network
during recognition (which may be larger than the final pattern). 

The recognition algorithm has a number of desirable features:

\blob all aspects of pattern recognition are integrated in a single process and 
act synergistically, rather than being applied as a sequence of phases;

\blob bottom-up and top-down inference are integrated; alternative
interpretations are developed in parallel;

\blob the whole of the image is processed simultaneously; there is no
\ic{traversal} of the image;

\blob because the syntactic structure of a \ic{sentence} is represented
declaratively by a network, rather than by a sequence of production rule
applications, the whole structure is available to work on throughout the
recognition process.

\medbreak\f
As a result the recognition algorithm has a combination of strong recursive
representational capabilities and general error-tolerance, not achieved before
in the literature. In particular it can

\blob represent iterative, hierarchical and nested recursive structure,
including two dimensions of iteration,

\blob recognise patterns consisting of up to 1000 symbol tokens arranged in
two-dimensional space,

\blob distinguish overlapping, indistinct or distorted symbol tokens (e.g., bent lines or
lines with pieces missing), sometimes successfully disentangling highly
cluttered images,

\blob restore missing lines or junctions in the most grammatically plausible
position.

\medskip\f
These features go some way towards substantiating my thesis (stated in \S1)
that symbol processing is not inherently brittle but is made so by the
imposition of an unwarranted sequential order on it. When this sequentiality is
discarded then robust symbol processing becomes possible.

As further work I intend to develop an algorithm that learns the fleximaps and
grammar from example images, without supervision. I believe this can be done by
extending the methods of [48].

%% file: refsa.tex
\def\ref[#1]{\medskip\f[#1] }

\ref[1] 
Aalbersberg, I.J., Engelfriet, J. \& Rozenberg, G. (1987) Restricting the complexity of regular DNLC languages. In [38], pp.\,147--166. https://doi/org/10.1007/3-540-18771-5\_51.

\ref[2] 
Aalbersberg, I.J., Rozenberg, G. \& Ehrenfeucht, A. (1986) On the membership problem for regular DNLC grammars. \i{Discrete Applied Mathematics}, {\bf 13}, no.~1, 79--85. https://doi/org/10.1016/0166-218x(86)90070-3.

\ref[3] 
Adachi, Y. (2004) Visual language design system based on context-sensitive NCE graph grammar. In  \i{Proceedings of the 2004 IEEE International Conference on Systems, Man and Cybernetics.} Vol.~1. IEEE, pp.\,502--507. https://doi/org/10.1109/icsmc.2004.1398348.

\ref[4] 
Adachi, Y., Kobayashi, S., Tsuchida, K. \& Yaku, T. (1999) An NCE context-sensitive graph grammar for visual design languages. In  \i{Proceedings 1999 IEEE Symposium on Visual Languages.} IEEE, pp.\,228--235. https://doi/org/10.1109/vl.1999.795908.

\ref[5] 
Adachi, Y. \& Nakajima, Y. (2000) A context-sensitive NCE graph grammar and its parsability. In  \i{Proceedings 2000 IEEE International Symposium on Visual Languages.} IEEE, pp.\,111--118. https://doi/org/10.1109/vl.2000.874374.

\ref[6] 
Bauderon, M. (1996) A category-theoretic approach to vertex replacement: the generation of infinite graphs. In Cuny, J., Ehrig, H., Engels, G \& Rozenberg, G. (eds) \i{Graph Grammars and Their Application to Computer Science, 5th International Workshop, 1995.} Lecture Notes in Computer Science, vol.~1073. Springer, pp.\,27--37. https://doi/org/10.1007/3-540-61228-9\_77.

\ref[7] 
Bauderon, M., Chen, R. \& Ly, O. (2008) Pullback grammars are context-free. In H.\,Ehrig, R.\,Heckel, G.\,Rozenberg \& G.\,Taentzer (eds) \i{Graph Transformations, ICGT 2008.} Lecture Notes in Computer Science, vol.~5214. Berlin: Springer, pp.\,366--378. https://doi/org/10.1007/978-3-540-87405-8\_25.

\ref[8] 
Bauderon, M., Chen, R. \& Ly, O. (2009) Context-free categorical grammars. In S.\,Bozapalidis \& G.\,Rahonis (eds) \i{Algebraic Informatics, 3rd International Conference on Algebraic Informatics, CAI 2009.} Lecture Notes in Computer Science, vol.~5725. Berlin: Springer, pp.\,160--171. https://doi/org/10.1007/978-3-642-03564-7\_10.

\ref[9] 
Bauderon, M. \& Jacquet, H. (2001) Node rewriting in graphs and hypergraphs: a categorical framework. \i{Theoretical Computer Science}, {\bf 266}, 463--487. https://doi/org/10.1016/S0304-3975(00)00200-0.

\ref[10] 
Ben-Arie, J. \& Wang, Z. (1998) Pictorial recognition of objects employing affine invariance in the frequency domain. \i{IEEE Transactions on Pattern Analysis and Machine Intelligence}, {\bf 20}, no.~6, 604--618. https://doi/org/10.1109/34.683774.

\ref[11] 
Beveridge, J.R. \& Riseman, E.M. (1997) How easy is matching 2D line models using local search? \i{IEEE Transactions on Pattern Analysis and Machine Intelligence}, {\bf 19}, no.~6, 564--579. https://doi/org/10.1109/34.601245.

\ref[12] 
Bj\"orklund, H., Bj\"orklund, J. \& Ericson, P. (2024) Tree-based generation of restricted graph languages. \i{International Journal of Foundations of Computer Science}, {\bf 351}, nos~1--2, 215--243. https://doi/org/10.1142/S0129054123480106.

\ref[13] 
Bj\"orklund, H., Drewes, F., Ericson, P. \& Starke, F. (2021) Uniform parsing for hyperedge replacement grammars. \i{Journal of Computer and System Sciences}, {\bf 118}, 1--27. https://doi/org/10.1016/j.jcss.2020.10.002.

\ref[14] 
Bj\"orklund, J., Drewes, F. \& Jonsson, A. (2023) Generation and polynomial parsing of graph languages with non-structural reentrancies. \i{Computational Linguistics}, {\bf 49}, no.~4, 841--882. https://doi/org/10.1162/coli\_a\_00488.

\ref[15] 
Blosetin, D. \& Sch\"urr, A. (1999) Computing with graphs and graph transformation. \i{Software -- Practice and Experience}, {\bf 29}, no.~3, 197--217. https://doi/org/10.1002/(SICI)1097-024X(199903)29:3<197::AID-SPE228>3.0.CO;2-F.

\ref[16] 
Brandenburg, F.J. (1988) On polynomial time graph grammars. In R.\,Cori \& M.\,Wirsing (eds) \i{Proceedings of 5th Conference on Theoretical Aspects of Computer Science.} Lecture Notes in Computer Science, vol.~294. Berlin: Springer, pp.\,227--236. https://doi/org/10.1007/BFb0035847.

\ref[17] 
Brandenburg, F.J. \& Skodinis, K. (2005) Finite graph automata for linear and boundary graph languages. \i{Theoretical Computer Science}, {\bf 332}, nos~1--3, 199--232. https://doi/org/10.1016/j.tcs.2004.09.040.

\ref[18] 
Bunke, H. \& Haller, B. (1989) A parser for context free plex grammars. In M.\,Nagl (ed.) \i{15th International Workshop on Graph-Theoretic Concepts in Computer Science.} Lecture Notes in Computer Science, vol.~411. Berlin: Springer, pp.\,136--150. https://doi/org/10.1007/3-540-52292-1\_10.

\ref[19] 
Bunke, H. \& Haller, B. (1992) Syntactic analysis of context-free plex languages for pattern recognition. In H.S.\,Baird, H.\,Bunke \& K.\,Yamamoto (eds) \i{Structured Document Image Analysis.} Berlin: Springer, pp.\,500--519. https://doi/org/10.1007/978-3-642-77281-8\_24.

\ref[20] 
Calvillo, J., Brouwer, H. \& Crocker, M.W. (2021) Semantic systematicity in connectionist language production. \i{Information}, {\bf 12}, no.~8. https://doi/org/10.3390/info12080329.

\ref[21] 
Casey, R.G. \& Lecolinet, E. (1996) A survey of methods and strategies in character segmentation. \i{IEEE Transactions on Pattern Analysis and Machine Intelligence}, {\bf 18}, no.~7, 690--706.

\ref[22] 
Celik, M. \& Yanikoglu, B. (2011) Probabilistic mathematical formula recognition using a 2D context-free graph grammar. In  \i{2011 International Conference on Document Analysis and Recognition.} IEEE Computer Society, pp.\,161--166. https://doi/org/10.1109/icdar.2011.41.

\ref[23] 
Chok, S. \& Marriott, K. (1995) Automatic construction of user interfaces from constraint multiset grammars. In  \i{Proceedings of the 11th IEEE Symposium on Visual Languages, VL '95, Darmstadt.} IEEE, pp.\,242--249. https://doi/org/10.1109/vl.1995.520815.

\ref[24] 
Chomsky, N. (1968) \i{Language and Mind.} New York: Harcourt Brace \& World.

\ref[25] 
Costagliola, G., De Lucia, A., Orefice, S. \& Tortora, G. (1998) Positional grammars: a formalism for LR-like parsing of visual languages. In K.\,Marriott \& B.\,Meyer (eds) \i{Visual Language Theory.} Berlin: Springer, pp.\,171--191.

\ref[26] 
Courcelle, B. (1987) An axiomatic definition of context-free rewriting and its application to NLC graph grammars. \i{Theoretical Computer Science}, {\bf 55}, nos~2--3, 141--181. https://doi/org/10.1016/0304-3975(87)90102-2.

\ref[27] 
Courcelle, B. (1991) The monadic second-order logic of graphs V: on closing the gap between definability and recognizability. \i{Theoretical Computer Science}, {\bf 80}, no.~2, 153--202. https://doi/org/10.1016/0304-3975(91)90387-h.

\ref[28] 
Courcelle, B. (1997) The expression of graph properties and graph transformations in monadic second-order logic. In G.\,Rozenberg (ed.) \i{Handbook of Graph Grammars and Computing by Graph Transformation, vol 1, Foundations.} Singapore: World Scientific, pp.\,313--400. https://doi/org/10.1142/9789812384720\_0005.

\ref[29] 
Das, A. \& Vemuri, R. (2009) Fuzzy logic based guidance to graph grammar framework for automated analog circuit design. In  \i{22nd International Conference on VLSI Design.} IEEE,, pp.\,445--450. https://doi/org/10.1109/vlsi.design.2009.79.

\ref[30] 
Dong, J., Chen, Q., Huang, Z., Yang, J. \& Yan, S. (2016) Parsing based on parselets: a unified deformable mixture model for human parsing. \i{IEEE Transactions on Pattern Analysis and Machine Intelligence}, {\bf 38}, no.~1, 88--101. https://doi/org/10.1109/tpami.2015.2420563.

\ref[31] 
Drewes, F., Hoffmann, B. \& Minas, M. (2019) Extending predictive shift-reduce parsing to contextual hyperedge replacement grammars. In E.\,Guerra \& F.\,Orejas (eds) \i{Graph Transformation, ICGT 2019.} Lecture Notes in Computer Science, vol.~11629. Springer, pp.\,55--72. https://doi/org/10.1007/978-3-030-23611-3\_4.

\ref[32] 
Drewes, F., Hoffmann, B. \& Minas, M. (2021) Rule-based top-down parsing for acyclic contextual hyperedge replacement grammars. In F.\,Gadducci \& T.\,Kehrer (eds) \i{Graph Transformation, ICGT 2021.} Lecture Notes in Computer Science, vol.~12741. Springer, pp.\,164--184. https://doi/org/10.1007/978-3-030-78946-6\_9.

\ref[33] 
Drewes, F., Kreowski, H.-J. \& Habel, A. (1997) Hyperedge replacement graph grammars. In G.\,Rozenberg (ed.) \i{Handbook of Graph Grammars and Computing by Graph Transformation, vol 1, Foundations.} Singapore: World Scientific, pp.\,95--162. https://doi/org/10.1142/9789812384720\_0002.

\ref[34] 
Drewes, F. \& Stade, Y. (2024) On the power of local graph expansion grammars with and without additional restrictions. \i{Theoretical Computer Science}, {\bf 1015}, 114763. https://doi/org/10.1016/j.tcs.2024.114763.

\ref[35] 
Ehrig, H. (1979) Introduction to the algebraic theory of graph grammars (a survey). In V.\,Claus, H.\,Ehrig \& G.\,Rozenberg (eds) \i{Graph-Grammars and their Application to Computer Science and Biology.} Lecture Notes in Computer Science, vol.~73. Berlin: Springer, pp.\,1--69. https://doi/org/10.1007/BFb0025714.

\ref[36] 
Ehrig, H., Korff, M. \& L\"owe, M. (1991) Tutorial introduction to the algebraic approach of graph grammars based on double and single pushouts. In [37], pp.\,24--37. https://doi/org/10.1007/BFb0017375.

\ref[37] 
Ehrig, H., Kreowski, H.J. \& Rozenberg, G. (eds) (1991) \i{Graph Grammars and Their Application to Computer Science, 4th International Workshop, Bremen, March 5-9, 1990.} Lecture Notes in Computer Science, vol.~532. Berlin: Springer. https://doi/org/10.1007/BFb0017372.

\ref[38] 
Ehrig, H., Nagl, M., Rozenberg, G. \& Rosenfeld, A. (eds) (1987) \i{Graph-Grammars and Their Application to Computer Science, 3rd International Workshop, Warrenton, Virginia, USA, December 2-6, 1986.} Lecture Notes in Computer Science, vol.~291. Berlin: Springer. https://doi/org/10.1007/3-540-18771-5.

\ref[39] 
Engelfriet, J. (1991) A characterization of context-free NCE graph languages by monadic second-order logic on trees. In Ehrig, Kreowski \etal\ (1991), pp.\,311--327. https://doi/org/10.1007/BFb0017397.

\ref[40] 
Engelfriet, J. \& Rozenberg, G. (1991) Graph grammars based on node rewriting: an introduction to NLC graph grammars. In [37], pp.\,12--23. https://doi/org/10.1007/BFb0017374.

\ref[41] 
Engelfriet, J. \& Rozenberg, G. (1997) Node replacement graph grammars. In G.\,Rozenberg (ed.) \i{Handbook of Graph Grammars and Computing by Graph Transformation, vol 1, Foundations.} Singapore: World Scientific, pp.\,1--94. https://doi/org/10.1142/9789812384720\_0001.

\ref[42] 
Fahmy, H. \& Blostein, D. (1998) A graph-rewriting paradigm for discrete relaxation: application to sheet-music recognition. \i{International Journal of Pattern Recognition and Artificial Intelligence}, {\bf 12}, no.~6, 763--799. https://doi/org/10.1142/S0218001498000439.

\ref[43] 
Ferrucci, F., Tortora, G., Tucci, M. \& Vitiello, G. (1994) A predictive parser for visual languages specified by relation grammars. In  \i{Proceedings of the 10th IEEE Symposium on Visual Languages.} Los Alamitos, CA: IEEE Computer Society Press, pp.\,245--252. https://doi/org/10.1109/vl.1994.363611.

\ref[44] 
Ferrucci, F., Tortora, G., Tucci, M. \& Vitiello, G. (1998) Relation grammars: a formalism for syntactic and semantic analysis of visual languages. In K.\,Marriott \& B.\,Meyer (eds) \i{Visual Language Theory.} Berlin: Springer, pp.\,219--243.

\ref[45] 
Fischler, M.A. \& Elschlager, R.A. (1973) The representation and matching of pictorial structures. \i{IEEE Transactions on Computers}, {\bf C-22}, no.~1, 67--92. https://doi/org/10.1109/t-c.1973.223602.

\ref[46] 
Flasi\'nski, M., Jurek, J. \& Peszek, T. (2024) Multi-derivational parsing of vague languages -- the new paradigm of syntactic pattern recognition. \i{IEEE Transactions on Pattern Analysis and Machine Intelligence}, {\bf 46}, no.~8, 5677--5691. https://doi/org/10.1109/TPAMI.2024.3367245.

\ref[47] 
Flasi\'nski, M. \& My\'sli\'nski, S. (2010) On the use of graph parsing for recognition of isolated hand postures of Polish Sign Language. \i{Pattern Recognition}, {\bf 43}, 2249--2264. https://doi/org/10.1016/j.patcog.2010.01.004.

\ref[48] 
Fletcher, P. (2001) Connectionist learning of regular graph grammars. \i{Connection Science}, {\bf 13}, no.~2, 127--188. https://doi/org/10.1080/09540090110072327.

\ref[49] 
Fletcher, P. (2004) \i{Mathematical Theory of Recursive Symbol Systems.} Technical report GEN-0401, Keele University, Computer Science Department.\allowbreak https://eprints.keele.ac.uk/id/eprint/63/.

\ref[50] 
Fletcher, P. (2023) \i{Mathematical Theory of Networks for Syntactic Pattern Recognition.} Supplement to this paper.

\ref[51] 
Fodor, J.A. \& Pylyshyn, Z.W. (1988) Connectionism and cognitive architecture: a critical analysis. \i{Cognition}, {\bf 28}, nos~1 \& 2, 3--71. https://doi/org/10.1016/0010-0277(88)90031-5.

\ref[52] 
F\"urst, L., Mernik, M. \& Mahnic, V. (2011) Improving the graph grammar parser of Rekers and Sch\"urr. \i{IET Software}, {\bf 5}, no.~2, 246--261. https://doi/org/10.1049/iet-sen.2010.0081.

\ref[53] 
Garfield, J.L. (1997) Mentalese not spoken here: computation, cognition and causation. \i{Philosophical Psychology}, {\bf 10}, no.~4, 413--435. https://doi/org/10.1080/09515089708573231.

\ref[54] 
Gilroy, S., Lopez, A. \& Maneth, S. (2017) Parsing graphs with regular graph grammars. In  \i{Proceedings of 6th Joint Conference on Lexical and Computational Semantics.} Association for Computational Linguistics, pp.\,199--208. https://doi/org/10.18653/v1/s17-1024.

\ref[55] 
Golin, E. (1991) Parsing visual languages with picture layout grammars. \i{Journal of Visual Languages and Computing}, {\bf 2}, 371--394. https://doi/org/10.1016/s1045-926x(05)80005-9.

\ref[56] 
Habel, A. (1992) \i{Hyperedge Replacement: Grammars and Languages.} Lecture Notes in Computer Science, vol.~643. Berlin: Springer. https://doi/org/10.1007/BFb0013875.

\ref[57] 
Han, F. \& Zhu, S.-C. (2009) Bottom-up/top-down image parsing with attribute grammar. \i{IEEE Transactions on Pattern Analysis and Machine Intelligence}, {\bf 31}, no.~1, 59--73. https://doi/org/10.1109/tpami.2008.65.

\ref[58] 
Hu, W. \& Zhu, S.C. (2015) Learning 3D object templates by quantizing geometry and appearance spaces. \i{IEEE Transactions on Pattern Analysis and Machine Intelligence}, {\bf 37}, no.~6, 1190--1205. https://doi/org/10.1109/tpami.2014.2362141.

\ref[59] 
Julca-Aguilar, F., Mouch\`ere, H., Viard-Gaudin, C. \& Hirata, N.S.T. (2020) A general framework for the recognition of online handwritten graphics. \i{International Journal on Document Analysis and Recognition}, {\bf 23}, no.~2, 143--160. https://doi/org/10.1007/s10032-019-00349-6.

\ref[60] 
Kaul, M. (1983) Parsing of graphs in linear time. In H.\,Ehrig, M.\,Nagl \& G.\,Rozenberg (eds) \i{2nd International Workshop on Graph Grammars and their Application to Computer Science 1982.} Lecture Notes in Computer Science, vol.~153. Berlin: Springer, pp.\,206--218. https://doi/org/10.1007/bfb0000108.

\ref[61] 
Kleyko, D., Davies, M., Frady, E.P., Kanerva, P., Kent, S.J., Olshausen, B.A., Osipov, E., Rabaey, J.M., Rachkovskij, D.A., Rahimi, A. \& Sommer, F.T. (2022) Vector symbolic architectures as a computing framework for emerging hardware. \i{Proceedings of the IEEE}, {\bf 110}, no.~10, 1538--1571. https://doi/org/10.1109/JPROC.2022.3209104.

\ref[62] 
Kong, J, Zhang, K \& Zeng, X. (2006) Spatial graph gramamrs for graphical user interfaces. \i{ACM Transactions on Computer-Human Interaction}, {\bf 13}, no.~2, 268--307. https://doi/org/10.1145/1165734.1165739.

\ref[63] 
Kong, J. \& Zhang, K. (2004) Parsing spatial graph grammars. In P.\,Bottoni, C.\,Hundhausen, S.\,Levialdi \& G.\,Tortora (eds) \i{Proceedings of the 2004 IEEE Symposium on Visual Langages and Human Centric Computing (VLHCC'04).} IEEE Computer Society, pp.\,99--101. https://doi/org/10.1109/vlhcc.2004.40.

\ref[64] 
Kosmala, A., Lavirotte, S., Pottier, L. \& Rigoll, G. (1999) On-Line Handwritten Formula Recognition using Hidden Markov Models and Context Dependent Graph Grammars. In  \i{Proceedings of 5th International Conference on Document Analysis and Recognition.} IEEE Computer Society Press, pp.\,107--110. https://doi/org/10.1109/icdar.1999.791736.

\ref[65] 
Lange, K.J. \& Welzl, E. (1987) String grammars with disconnecting, or a basic root of the difficulty in graph grammar parsing. \i{Discrete Applied Mathematics}, {\bf 16}, no.~1, 17--30. https://doi/org/10.1016/0166-218x(87)90051-5.

\ref[66] 
Lavirotte, S. \& Pottier, L. (1997) Optical formula recognition. In  \i{Proc. of 4th Int. Conf. on Document Analysis and Recognition.} Vol.~1. IEEE Computer Society, pp.\,357--361. https://doi/org/10.1109/ICDAR.1997.619871.

\ref[67] 
Li, C., Huang, L., Chen, L. \& Yu, C. (2013) BGG: A graph grammar approach for software architecture verification and reconfiguration. In  \i{Proceedings, 7th International Conference on Innovative Mobile and Internet Services in Ubiquitous Computing.} IEEE, pp.\,291--298. https://doi/org/10.1109/imis.2013.56.

\ref[68] 
Liang, X., Liu, S., Yang, J., Liu, L., Dong, J., Lin, L. \& Yan, S. (2015) Deep human parsing with active template regression. \i{IEEE Transactions on Pattern Analysis and Machine Intelligence}, {\bf 37}, no.~12, 2402--2414.

\ref[69] 
Lichtblau, U. (1991) Recognizing rooted context-free flowgraph languages in polynomial time. In Ehrig, Kreowski \& Rozenberg (1991), pp.\,538--548. https://doi/org/10.1007/BFb0017411.

\ref[70] 
Lin, L., Gong, H., Li, L. \& Wang, L. (2009) Semantic event representation and recognition using syntactic attribute graph grammar. \i{Pattern Recognition Letters}, {\bf 30}, 180--186. https://doi/org/10.1016/j.patrec.2008.02.023.

\ref[71] 
Lin, L., Wu, T., Porway, J. \& Xu, Z. (2009) A stochastic graph grammar for compositional object representation and recognition. \i{Pattern Recognition}, {\bf 42}, no.~7, 1297--1307. https://doi/org/10.1016/j.patcog.2008.10.033.

\ref[72] 
Liu, Y. \& Yang, F. (2021) EGG+: A graph grammar formalism with uncertain structure processing mechanism. \i{Journal of Logic and Computation}, {\bf 31}, no.~7, 1800--1819. https://doi/org/10.1093/logcom/exab055.

\ref[73] 
Liu, Y. \& Yang, F. (2021) An enhancement to graph grammar for the specification of edge semantics. In  \i{2021 IEEE International Conference on Progress in Informatics and Computing.} IEEE, pp.\,1--5. https://doi/org/10.1109/PIC53636.2021.9687048.

\ref[74] 
L\'opez, D. \& Pi\~naga, I. (2000) Syntactic pattern recognition by error correcting analysis on tree automata. In F.J.\,Ferri, J.M.\,I\~nesta, A.\,Amin \& P.\,Pudil (eds) \i{Advances in Pattern Recognition, SSPR/SPR 2000.} Lecture Notes in Computer Science, vol.~1876. Berlin: Springer, pp.\,133--142. https://doi/org/10.1007/3-540-44522-6\_14.

\ref[75] 
Main, M.G. \& Rozenberg, G. (1987) Fundamentals of edge-label controlled graph grammars. In [38], pp.\,411--426. https://doi/org/10.1007/3-540-18771-5\_67.

\ref[76] 
Mas, J., Llad\'os, J., Sanchez, G. \& Jorge, J.A.P. (2010) A syntactic approach based on distortion-tolerant Adjacency Grammars and a spatial-directed parser to interpret sketched diagrams. \i{Pattern Recognition}, {\bf 43}, no.~12, 4148--4164. https://doi/org/10.1016/j.patcog.2010.07.003.

\ref[77] 
Messmer, B.T. \& Bunke, H. (1998) A new algorithm for error-tolerant subgraph isomorphism detection. \i{IEEE Transactions on Pattern Analysis and Machine Intelligence}, {\bf 20}, no.~5, 493--504.

\ref[78] 
Modolo, D. \& Ferrari, V. (2018) Learning semantic part-based models from Google images. \i{IEEE Transactions on Pattern Analysis and Machine Intelligence}, {\bf 40}, no.~6, 1502--1509.

\ref[79] 
Nagl, M. (1987) Set theoretic approaches to graph grammars. In [38], pp.\,41--54. https://doi/org/10.1007/3-540-18771-5\_43.

\ref[80] 
Parizeau, M. \& Plamondon, R. (1995) A fuzzy-syntactic approach to allograph modeling for cursive script recognition. \i{IEEE Transactions on Pattern Analysis and Machine Intelligence}, {\bf 17}, no.~7, 702--712. https://doi/org/10.1109/34.391412.

\ref[81] 
Park, S., Nie, B.X. \& Zhu, S.C. (2018) Attribute and-or grammar for joint parsing of human pose, parts and attributes. \i{IEEE Transactions on Pattern Analysis and Machine Intelligence}, {\bf 40}, no.~7, 1555--1569. https://doi/org/10.1109/tpami.2017.2731842.

\ref[82] 
Ranjan, R., Patel, V.M. \& Chellappa, R. (2019) HyperFace: a deep multi-task learning framework for face detection, landmark localization, pose estimation, and gender recognition. \i{IEEE Transactions on Pattern Analysis and Machine Intelligence}, {\bf 41}, no.~1, 121--135.

\ref[83] 
Raphael, C. \& Jin, R. (2014) Optical Music Recognition on the International Music Score Library Project. In B.\,Co\"uasnon \& E.K.\,Ringger (eds) \i{Document Recognition and Retrieval XXI.} San Francisco: Soc Imaging Sci \& Technol, SPIE. https://doi/org/10.1117/12.2040247.

\ref[84] 
Rekers, J. \& Sch\"urr, A. (1997) Defining and parsing visual languages with layered graph grammars. \i{Journal of Visual Languages and Computing}, {\bf 8}, no.~1, 27--55. https://doi/org/10.1006/jvlc.1996.0027.

\ref[85] 
Rocha, J. \& Pavlidis, T. (1995) Character recognition without segmentation. \i{IEEE Transactions on Pattern Analysis and Machine Intelligence}, {\bf 17}, no.~9, 903--909.

\ref[86] 
Rozenberg, G. \& Welzl, E. (1986) Boundary NLC graph grammars -- basic definitions, normal forms, and complexity. \i{Information and Control}, {\bf 69}, nos~1--3, 136--167. https://doi/org/10.1016/s0019-9958(86)80045-6.

\ref[87] 
Sabinasz, D \& Schoner, G. (2022) A neural dynamic model perceptually grounds nested noun phrases. \i{Proceedings of the 44th Annual Meeting of the Cognitive Science Society}, {\bf 44}, 847--853. https://doi/org/10.1111/tops.12630.

\ref[88] 
S\'anchez, G., Llad\'os, J. \& Tombre, K. (2002) An error-correction graph grammar to recognize texture symbols. In D.\,Blostein \& Y.B.\,Kwon (eds) \i{Graphics Recognition Algorithms and Applications, GREC 2001.} Lecture Notes in Computer Science, vol.~2390. Berlin: Springer, pp.\,28--138. https://doi/org/10.1007/3-540-45868-9\_10.

\ref[89] 
Schuster, R. (1987) \i{Graphgrammatiken und Grapheinbettungen: Algorithmen und Komplexit\"at.} PhD thesis, Universit\"at Passau.

\ref[90] 
Shi, Z., Zeng, X., Huang, S., Qi, Z., Li, H., Hu, B., Zhang, S., Liu, Y. \& Wang, C. (2015) A method to simplify description and implementation of graph grammars. In  \i{6th International Conference on Computing, Communication and Networking Technologies, Denton, USA.} IEEE, pp.\,1--6. https://doi/org/10.1109/icccnt.2015.7395225.

\ref[91] 
Shi, Z., Zeng, X., Zou, Y., Huang, S., Li, H., Hu, B. \& Yao, Y. (2018) A temporal graph grammar formalism. \i{Journal of Visual Languages and Computing}, {\bf 47}, 62--76. https://doi/org/10.1016/j.jvlc.2018.06.003.

\ref[92] 
Si, Z. \& Zhu S.-C. (2013) Learning AND-OR templates for object recognition and detection. \i{IEEE Transactions on Pattern Analysis and Machine Intelligence}, {\bf 35}, no.~9, 2189--2205. https://doi/org/10.1109/tpami.2013.35.

\ref[93] 
Simon, L., Teboul, O., Koutsourakis, P. \& Paragios, N. (2011) Random exploration of the procedural space for single-view 3D modeling of buildings. \i{International Journal of Computer Vision}, {\bf 93}, no.~2, 253--271.

\ref[94] 
Siskind, J.M., Sherman, J.J., Pollak, I., Harper, M.P. \& Bouman, C.A. (2007) Spatial random tree grammars for modeling hierarchical structure in images with regions of arbitrary shape. \i{IEEE Transactions on Pattern Analysis and Machine Intelligence}, {\bf 29}, no.~9, 1504--1519. https://doi/org/10.1109/TPAMI.2007.1169.

\ref[95] 
Skomorowski, M. (2007) Syntactic recognition of distorted patterns by means of random graph parsing. \i{Pattern Recognition Letters}, {\bf 28}, 572--581. https://doi/org/10.1016/j.patrec.2006.10.003.

\ref[96] 
Smolensky, P. (1988) On the proper treatment of connectionism. \i{Behavioral and Brain Sciences}, {\bf 11}, no.~1, 1--23. https://doi/org/10.1017/S0140525X00052432.

\ref[97] 
Tombre, K. (1996) Structural and syntactic methods in line drawing analysis: to which extent do they work? In P.\,Perner, P.\,Wang \& A.\,Rozenfeld (eds) \i{Advances in Syntactic and Structural Pattern Recognition, SSPR '96.} Lecture Notes in Computer Science, vol.~1121. Berlin: Springer, pp.\,310--321.

\ref[98] 
Tu, Z., Chen, X., Yuille, A.L. \& Zhu, S.-C. (2005) Image parsing: unifying segmentation, detection, and recognition. \i{International Journal of Computer Vision}, {\bf 63}, no.~2, 113--140. https://doi/org/10.1007/s11263-005-6642-x.

\ref[99] 
Valveny, E. \& Marti, E. (2003) A model for image generation and symbol recognition through the deformation of lineal shapes. \i{Pattern Recognition Letters}, {\bf 24}, no.~15, 2857--2867. https://doi/org/10.1016/s0167-8655(03)00144-2.

\ref[100] 
Wang, L., Wang, Y. \& Gao, W. (2011) Mining layered grammar rules for action recognition. \i{International Journal of Computer Vision}, {\bf 93}, no.~2, 162--182. https://doi/org/10.1007/s11263-010-0393-z.

\ref[101] 
Wittenburg, K.B. (1992) Earley-style parsing for relational grammars. In  \i{Proceedings of the IEEE Workshop on Visual Languages.} Los Alamitos, CA: IEEE Computer Society Press, pp.\,192--199. https://doi/org/10.1109/wvl.1992.275765.

\ref[102] 
Wittenburg, K.B. \& Weitzman, L.M. (1998) Relational grammars: theory and practice in a visual language interface for process modeling. In K.\,Marriott \& B.\,Meyer (eds) \i{Visual Language Theory.} Berlin: Springer, pp.\,193--217.

\ref[103] 
Ye, Y.J. \& Sun, W.W. (2020) Exact yet efficient graph parsing, bi-directional locality and the constructivist hypothesis. In D.\,Jurafsky, J.\,Chai, N.\,Schluter \& J.\,Tetreault (eds) \i{58th Annual Meeting of the Association for Computational Linguistics.} Association for Computational Linguistics, pp.\,4100--4110. https://doi/org/10.18653/v1/2020.acl-main.377.

\ref[104] 
Yuan, Y., Liu, X., Dikubab, W., Liu, H., Ji, Z., Wu, Z. \& Bai, X. (2022) Syntax-Aware Network for Handwritten Mathematical Expression Recognition. In  \i{IEEE/CVF Conference on Computer Vision and Pattern Recognition (CVPR).} IEEE, pp.\,4543--4552. https://doi/org/10.1109/CVPR52688.2022.00451.

\ref[105] 
Yuille, A. L., Hallinan, P. W. \& Cohen, D. S. (1992) Feature extraction from faces using deformable templates. \i{International Journal of Computer Vision}, {\bf 8}, no.~2, 99--111. https://doi/org/10.1007/bf00127169.

\ref[106] 
Zhang, D.Q., Zhang, K. \& Cao, J. (2001) A context-sensitive graph grammar formalism for the specification of visual languages. \i{The Computer Journal}, {\bf 44}, no.~3, 187--200. https://doi/org/10.1093/comjnl/44.3.186.

\ref[107] 
Zhu, L., Chen, Y., Li, C. \& Yuille, A. (2011) Max margin learning of hierarchical configural deformable templates (HCDTs) for efficient object parsing and pose estimation. \i{International Journal of Computer Vision}, {\bf 93}, 1--21. https://doi/org/10.1007/s11263-010-0375-1.

\ref[108] 
Zhu, L., Chen, Y. \& Yuille, A. (2009) Unsupervised learning of probabilistic grammar-Markov models for object categories. \i{IEEE Transactions on Pattern Analysis and Machine Intelligence}, {\bf 31}, no.~1, 114--128. https://doi/org/10.1109/tpami.2008.67.

\ref[109] 
Zou, Y. (2024) A comparison of reserved graph grammar and edge-based graph grammar. \i{Computer Journal}, {\bf 68}, no.~2, 145--162. https://doi/org/10.1093/comjnl/bxae100.

\ref[110] 
Zou, Y., L\"u, J. \& Tao, X. (2019) Research on context of implicit context-sensitive graph grammars. \i{Journal of Computer Languages}, {\bf 51}, 241--260. https://doi/org/10.1016/j.cola.2019.01.002.

\ref[111] 
Zou, Y., Zeng, X. \& Zhu, Y. (2022) A general parsing algorithm with context matching for context-sensitive graph grammars. \i{Multimedia Tools and Applications}, {\bf 81}, 273--297. https://doi/org/10.1007/s11042-021-11076-8.